\pdfoutput=1

\documentclass[twocolumn,twocolappendix]{aastex631}
\usepackage{amsmath}
\usepackage{multirow}
\usepackage{bm}

\shorttitle{Effects of galaxy IA on WL peak statistics}
\shortauthors{Zhang et al.}
\graphicspath{{./}{figures/}}

\begin{document}

\title{Effects of galaxy intrinsic alignment on weak lensing peak statistics}

\correspondingauthor{X.K.Liu and Z.H.Fan}
\email{liuxk@ynu.edu.cn, zuhuifan@ynu.edu.cn}

\author{Tianyu Zhang}
\affiliation{South-Western Institute for Astronomy Research, Yunnan University, Kunming 650500, China}

\author{Xiangkun Liu}
\affiliation{South-Western Institute for Astronomy Research, Yunnan University, Kunming 650500, China}

\author{Chengliang Wei}
\affiliation{Purple Mountain Observatory, Chinese Academy of Sciences, Nanjing 210023, China}

\author{Guoliang Li}
\affiliation{Purple Mountain Observatory, Chinese Academy of Sciences, Nanjing 210023, China}
\affiliation{Zhejiang University-Purple Mountain Observatory Joint Research Center for Astronomy, Zhejiang University, Hangzhou 310027, China}

\author{Yu Luo}
\affiliation{Purple Mountain Observatory, Chinese Academy of Sciences, Nanjing 210023, China}

\author{Xi Kang}
\affiliation{Purple Mountain Observatory, Chinese Academy of Sciences, Nanjing 210023, China}
\affiliation{Institute for Astronomy, the School of Physics, Zhejiang University, Hangzhou 310027, China}

\author{Zuhui Fan}
\affiliation{South-Western Institute for Astronomy Research, Yunnan University, Kunming 650500, China}



\begin{abstract}

The galaxy intrinsic alignment (IA) is a dominant source of systematics in weak lensing (WL) studies. In this paper, by employing large simulations with semi-analytical galaxy formation, we investigate the IA effects
on WL peak statistics. Different simulated source galaxy samples of different redshift distributions are constructed, where both WL shear and IA signals are included. Convergence reconstruction and peak statistics are
then performed for these samples. Our results show that the IA effects on peak abundances mainly consist of two aspects. One is the additional contribution from IA to the shape noise. The other is from the 
satellite IA that can affect the peak signals from their host clusters significantly. The latter depends on the level of inclusion in a shear sample of the satellite galaxies of the clusters that contribute to WL peaks,
and thus is sensitive to the redshift distribution of source galaxies. We pay particular attention to satellite IA and adjust it artificially in the simulations to analyze the dependence of the satellite IA impacts on its strength. 
This information can potentially be incorporated into the modeling of WL peak abundances, especially for high peaks physically originated from massive clusters of galaxies, and thus
to mitigate the IA systematics on the cosmological constraints derived from WL peaks.

\end{abstract}



\section{Introduction} \label{sec:intro}
Weak gravitational lensing (WL) effects by large-scale structures induce shear and magnification signals on background sources, and are playing increasingly important roles in cosmological studies \citep{BS2001, HJ2008, FuFan, Kilbinger2015}. By measuring accurately the shapes and photometric redshifts 
of distant galaxies, WL observations have been developing in a fast pace, from the early detections of cosmic shears \citep[e.g.,][]{WaerbekeShear2000} to the demonstration observationally the feasibility of the WL probe \citep{Fu2008, Heymans2012, Kilbinger2013}, and now to the near completion of the Stage III surveys covering
a few thousand square degrees with $\sim 10^7$ to $10^8$ galaxies in the samples \citep{Kuijken2019, Abbott2021, Aihara2022}. Consequently, the delivered cosmological constraints are being improved significantly, revealing the potential $S_8$ discrepancy between the results from WL 
analyses and that extrapolated from the Planck cosmic microwave background observations \citep{Hildebrandt2020, Heymans2021, Hikage2019, Planck2020}. Here $S_8=\sigma_8(\Omega_{\rm m}/0.3)^{0.5}$ and $\Omega_{\rm m}$ and $\sigma_8$ are the present dimensionless cosmic matter density and the amplitude of the extrapolated 
linear density fluctuations over $8h^{-1}\hbox{Mpc}$ with $h$ the dimensionless Hubble constant in units of $100\hbox{ km/s/Mpc}$. On the other hand, from the current surveys, the statistical significance of the $S_8$ discrepancy is still low, and different survey analyses give rise to somewhat different
confidence levels. Thus whether this is a true problem for the $\Lambda$CDM model is still under debate \citep[e.g.,][]{Abbott2022}.

The upcoming Stage IV surveys will increase the data volume to about $10^9$ galaxies for WL analyses, leading to dramatic reductions of the statistical uncertainties in comparison with the current survey results. 
The representative projects include the ground-based Vera Rubin Observatory Legacy Survey of Space and Time
\citep[LSST;][]{Ivezic2019}, and three space-based ones of {\it Euclid} \citep{Euclid}, the Roman Space Telescope \citep[Roman;][]{WFIRST} and the China Space Station Telescope \citep[CSST;][] {CSST}. 
For them, systematics will become the dominant concerns in obtaining cosmological constraints with high accuracy.     

WL studies rely on the extraction of coherent shape distortions from galaxy samples, thus their intrinsic alignments (IA) lead to apparent contamination to WL signals \citep{Joachimi2015, Kiessling, Kirk, Troxel}. 
For cosmic shear two-point correlation (2PCF) or power spectrum analyses, IA effects have been extensively investigated and included in deriving cosmological constraints from observations \citep[e.g.,][]{Yao2020, Hildebrandt2020, Heymans2021, Abbott2022, Hikage2019}. 

To fully explore the physical information embedded in WL data, different statistics beyond two-point correlations are necessary
because of the non-Gaussian nature of large-scale structures \citep[e.g.,][]{Fu2014, Petri2015}. For that, peak statistics are important means to probe nonlinear structures \citep[e.g.,][]{vanWaer2000, Hamana2004, Dietrich2010, Fan2010, Maturi2010, Yang2011, Hamana2012, Lin2015, Yuan2018, Yuan2019, Martinet2021} and have been applied to different WL surveys 
to obtain cosmological information \citep{Shan2012, Shan2014, LiuJ2015, LiuX2015, Liu2016, Kacprzak2016, Martinet2018, Shan2018, Hamana2020, Oguri2021, Zurcher2022}. Aiming at utilizing their power in precision cosmological studies, it is important
to understand the impacts of different systematics. For IA effects on peak statistics, the investigations are not yet as detailed as that for cosmic shear 2PCF, but getting more and more attention.  

In \citet{Fan2007}, the IA contribution to the shape noise variance is analyzed, and its effects on the number of noise peaks are explored. In performing shear peak studies using Dark Energy Survey Science Verification data, \citet{Kacprzak2016} evaluate the potential impact of IA by assuming radial alignment of satellite 
galaxies in cluster halos with a scaling parameter $\gamma_{\rm scale}=0.21$ accounting for some misalignment angles \citep{SB2010}. They conclude that the IA induced bias in cosmological constraints is not significant for their analyses limited to peaks with $S/N<4$. For higher peaks however, their heights can be negatively biased considerably.

Recently, \citet{Joachim2022} carry out investigations of the impact of IA on WL statistics from aperture mass maps, including peaks, minima and the full distribution of pixel values. In their studies, an IA infusion method is taken to generate IA signals of galaxies. 
Specifically, from N-body simulations, they construct light cones for WL ray-tracing calculations. Meanwhile,  they convert the mass sheets into tidal tensors and coupled them to the intrinsic ellipticity components of galaxies 
under the nonlinear tidal alignment model \citep[NLA,][]{NLA} controlled by the coupling parameter  $A_{\rm IA}$. To better address the IA effects from galaxy positions rather than from random positions, they generate mock galaxy positions
by tracing the projected density distribution of each redshift slice assuming a linear bias parameter. Thus the galaxy clustering is included in the IA modeling, which is referred to as $\delta$-NLA \citep{Joachim2022}. 
For peaks, their studies show that IA affects high $S/N$ ones the most, with the relative number deviation caused by IA reaching about $-30\%$ and $-10\%$ for {\it Euclid}-like and KiDS-like surveys, respectively, for $A_{\rm IA}=1.5$.

In this paper, we study the IA effects on WL peaks based on large simulations with semi-analytic galaxy formation \citep{Wei}. Thus galaxies and their clustering, particularly satellite galaxies in clusters, are self-consistently included. 
From the simulated galaxies, we construct different catalogs with different redshift distributions. Furthermore, for satellite galaxies, we build different alignment samples to study systematically the dependence of the IA impacts on satellite alignment angles.

The paper is structured as follows. In \S \ref{sec:theory}, we summarize the basic theory of WL. \S \ref{sec:catalog} describes the simulations and the construction of mock galaxy catalogs. In \S \ref{sec:results} we present the results. An empirical fitting model is given in \S \ref{sec:fitting}. Summary and discussions 
are presented in \S \ref{sec:discussion}.

\section{Theory} \label{sec:theory}

In this section, we briefly summarize the theoretical background of WL, and the convergence field reconstruction used in our peak analyses. 

\subsection{Weak lensing basics} \label{sec:WL}

WL induced distortions on background sources are characterized by the following Jacobian matrix \citep{BS2001, FuFan}

\begin{equation}
\label{eq:Jacobmatrix}
    A=\begin{pmatrix}
    1-\kappa-\gamma_1 & -\gamma_2\\
    -\gamma_2 & 1-\kappa+\gamma_1
    \end{pmatrix}
\end{equation}
where the convergence $\kappa$ and the shear $\gamma_i$ are calculated from the lensing potential $\psi$ with $2\kappa=\nabla^2 \psi$ and $2\gamma_1=\partial^2\psi/\partial^2\theta_1-\partial^2\psi/\partial^2\theta_2$ and $\gamma_2=\partial^2\psi/\partial\theta_1\partial\theta_2$. 

Under the first order Born approximation, $\kappa$ is given by

\begin{equation}
\label{eq:kappa}
\begin{split}
    \kappa(\boldsymbol{\theta})=&\frac{3H_0^2\Omega_m}{2c^2}\int_0^{\chi_H} d\chi'\int^{\chi_H}_{\chi'}d\chi \\
&\bigg [p_s(\chi)\frac{r(\chi-\chi')r(\chi')}{r(\chi)a(\chi')}\bigg ]{\delta[r(\chi')\boldsymbol{\theta},\chi']},
\end{split}
\end{equation}
where $\chi$, $r$ and $a$ are the comoving radial distance, comoving angular diameter distance and the cosmic scale factor, respectively. The function $p_s$ is the source galaxy distribution function and $\delta$ is the 3-D density perturbation. 

Observationally, WL signals are extracted from the measurements of galaxy ellipticities. They are directly related to the reduced shear $\boldsymbol g=\boldsymbol \gamma/(1-\kappa)$ given by \citep{Seitz97}
\begin{equation}\label{eq:epsilon2g}
\boldsymbol \epsilon= \begin{cases}\dfrac{\boldsymbol \epsilon_s+\boldsymbol g}{1+\boldsymbol g^{*} \boldsymbol \epsilon_s}, & |g| \leq 1 \\ \dfrac{1+\boldsymbol g \boldsymbol \epsilon_s}{{\boldsymbol \epsilon_s}^*+\boldsymbol g^{*}}, & |g|>1\end{cases}.
\end{equation}
Here the bold-faced symbols represent the quantities written in the complex form, e.g., $\boldsymbol \gamma=\gamma_1+i\gamma_2$, and $\boldsymbol \epsilon$ and $\boldsymbol \epsilon_s$ are, respectively, the observed and the intrinsic ellipticities of a galaxy.

If the orientations of galaxies are intrinsically random, the expectation value of $\boldsymbol \epsilon_s$ averaged over an infinite number of galaxies should vanish, i.e., $\langle\boldsymbol \epsilon_s\rangle=0$, and thus the observed ellipticity is an unbiased estimator of the reduced shear with
\begin{equation}
\label{eq:epeqg}
    \langle\boldsymbol \epsilon\rangle=\boldsymbol g.
\end{equation}
In practice, the average can only be done over a finite number of galaxies, and therefore a residual shape noise from the intrinsic ellipticities is inevitable. Moreover, the existence of the intrinsic alignment of galaxies leads to a bias in Eq.(\ref{eq:epeqg})  as well as a contribution to the variance of the shape noise. 

\subsection{Convergence Reconstruction} \label{sec:Rec}

While WL shears are extractable quantities from observed ellipticities of galaxies, the convergence fields or their variants are more intuitively related to the projected mass distributions and thus are normally used in WL peak analyses.  

Because both $\boldsymbol \gamma$ and $\kappa$ are determined by the lensing potential, they are mutually dependent with the following relation in the Fourier space

\begin{equation}
\label{eq:kap2gam}
    \tilde {\boldsymbol \gamma}(\boldsymbol{k})=\frac{1}{\pi}\tilde{\boldsymbol D}(\boldsymbol k)\tilde{\kappa}(\boldsymbol k),
\end{equation}
with
\begin{equation}
\label{eq:D}
    \tilde{\boldsymbol D}(\boldsymbol{k})=\pi\frac{k_1^2-k_2^2+2ik_1k_2}{k_1^2+k_2^2}.
\end{equation}

As described later, in our analyses we first select a galaxy sample from the simulations with the semi-analytic galaxy formation and calculate their mock ellipticities following Eq.(\ref{eq:epsilon2g}) to generate a shear sample.  
We then use the above relation to reconstruct the convergence fields from the shear fields by employing the nonlinear iterative Kaiser-Squares reconstruction method \citep{Kaiser93,Seitz95,KS96,LiuX2015}. 
For that, we first calculate the smoothed fields $\langle{\boldsymbol{\epsilon}}\rangle$ on grid positions $\boldsymbol \theta$ by 
\begin{equation}
\label{eq:smooth}
    \langle{\boldsymbol {\epsilon}}\rangle(\boldsymbol{\theta})=\frac{\Sigma_{j}W_{{\theta}_G}(\boldsymbol{\theta}_j-\boldsymbol{\theta}){\boldsymbol {\epsilon}}(\boldsymbol{\theta}_j)}{\Sigma_{j}W_{{\theta}_G}(\boldsymbol{\theta}_{j}-\boldsymbol{\theta})}
\end{equation}
where $\boldsymbol \theta_j$ is the position of a galaxy in the shear sample, and $W$ is the smoothing function taken to be Gaussian as follows
\begin{equation}
\label{eq:Gausswindow}
    W_{\theta_{G}}(\boldsymbol{\theta})=\frac{1}{\pi\theta_G^2}\rm{exp}(-\frac{|\boldsymbol{\theta|}^2}{\theta_G^2}).
\end{equation}
We adopt the smoothing scale $\theta_G=1.5\hbox{ arcmin}$ as the fiducial value in our studies.

From $\langle{\boldsymbol{\epsilon}}\rangle$, we obtain the smoothed convergence fields $K_N$ from the reconstructions, which can be written as the sum of the WL convergence term $K$ and the shape noise term $N$, i.e., \citep[e.g.][]{vanWaer2000} 
\begin{equation}
\label{eq:KN}
    K_N(\boldsymbol{\theta})=K(\boldsymbol{\theta})+N(\boldsymbol{\theta}).
\end{equation}
In view of IA, it can affect both $K$ and $N$.

\section{Simulation catalogs} \label{sec:catalog}

In this section, we describe the simulations used in our analyses and the construction of different galaxy catalogs with different redshift distributions and different satellite IA.

\subsection{Simulations of \citet{Wei}} \label{sec:PMO}

The simulation data used in this study are from \citet{Wei}. They start from a set of N-body simulations within the ELUCID project \citep{Wang2014, Wang2016}. The simulation box size is $L_{box}=500\ \rm{h^{-1} Mpc}$ and the total number of particles is  $3072^3$.
The cosmological parameters are from WMAP9 with $\Omega_m=0.282$, $\Omega_\Lambda=0.718$, $\sigma_8=0.82$ and $h=0.697$ where $\Omega_\Lambda$ is the present dimensionless density of dark energy in the form of cosmological constant \citep{Hinshaw2013}.

From the N-body simulation, they build a full-sky light cone up to $z_{\rm max}\approx 2$ by chopping the simulation box into smaller ones with a size of $\sim 100\ \rm{h^{-1} Mpc}$ and carefully piling them up. 
Spherical ray-tracing calculations are then performed on a set of spherical mass shells with the width of $50\ \rm{h^{-1} Mpc}$ to obtain full-sky WL shear and convergence maps with the resolution of $\sim 0.43\hbox{ arcmin}$. For details, please see \citet{Wei}.

The great advantage of this set of simulations is the inclusion of the semi-analytical galaxy formation based on the model of \citet{Luo2016}, which is the modified version of \citet{FuJ2013} and \citet{Guo2013} to improve the prescription for low-mass galaxies by considering physics of gas stripping and the modeling of orphan galaxies. 
This improvement is important for properly modeling satellite galaxies. 

This semi-analytical model predicts the ratio of the bulge to total mass $B/T$ for a galaxy. This is used to divide galaxies into two classes with $B/T>0.6$ as the early type and 
$B/T<0.6$ as the late type. For early-type central galaxies, their ellipticities and orientations are assumed to be the same as their host halos calculated from the inertia tensor 
using dark matter particle distributions and then projected along the line of sight. For late-type centrals, their shapes are assigned according the angular momentum vector of the host halos. 
The 2-D axial ratio and thus the ellipticities of late-type galaxies are calculated through the projection based on the ratio of the line-of-sight angular momentum component to the total angular momentum. There the ratio of the galactic disk thickness to the diameter is set to be $0.25$ \citep{Wei}.  

For satellite galaxies, \citet{Wei} generate two sets of data with different alignments. One is to follow \citet{Joachimi2013} to set the 3-D major axis of an early-type satellite radially toward the center of its host cluster. For late-type satellites, their spin directions are assumed to be perpendicular 
to the line connecting the satellites and their centrals. In this set up, satellite galaxies are essentially radially aligned toward the cluster centrals. We refer this as highIA. The other is to assume a random distribution of satellite orientations without alignments. We name this case as lowIA. 

\begin{figure}[ht!]
\plotone{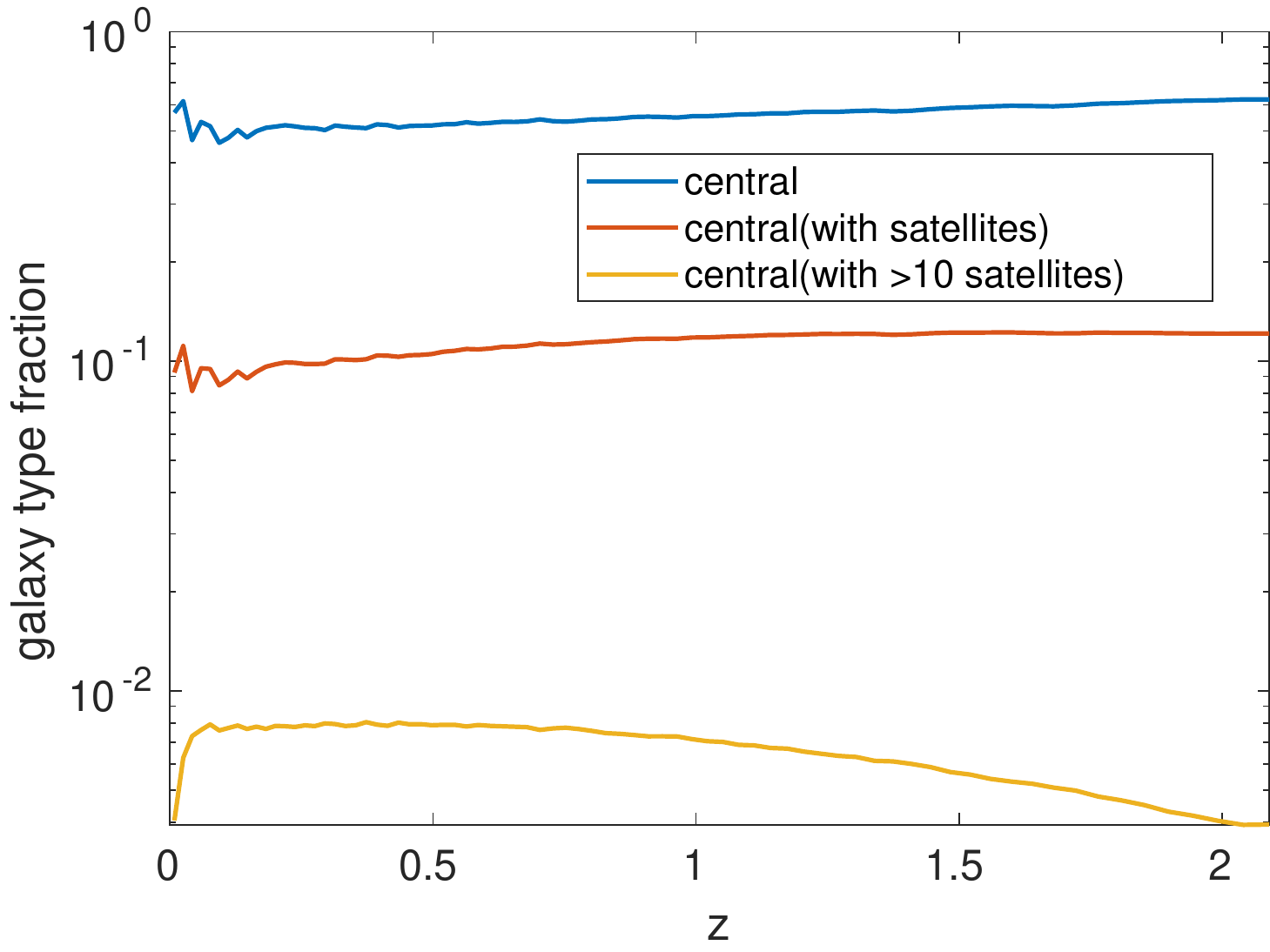}
\caption{The fractions of central galaxies. The blue, red and yellow lines are for all centrals, centrals with satellites, and centrals with more than 10 satellites, respectively. \label{fig:typeratio}}
\end{figure}

Figure \ref{fig:typeratio} presents the fractions of the total number of central galaxies (blue), centrals with satellites (red), and centrals with more than 10 satellites (yellow) to the total number of all simulated galaxies, at different redshifts. 
We can see that over the whole redshift range, centrals account for about 50\% to 60\% of the total simulated galaxies, and the rest are satellites. Among the centrals, about 20\% of them have satellites and about 2\% with more than 10 satellites. 
With the increase of the redshift, the yellow line decreases in accord with less massive clusters then. It is noted that for WL high peaks, most of them
arise from the lensing effects of massive clusters, which contain a relatively large number of satellites. From our analyses shown later, if the satellites are in a shear sample and they have strong IA signals, WL high peaks will be significantly affected. This effect depends
sensitively on the redshift distribution of a shear sample.

\subsection{Shear Catalog Construction} \label{sec:galsam}


As described in the previous section, in the simulation of \citet{Wei}, the formation and evolution of galaxies are modeled semi-analytically and their intrinsic ellipticities and orientations are also assigned.  
From the spherical ray-tracing calculations, shear and convergence maps are generated up to $z\sim 2$. Based on these, we construct different shear catalogs for analyses.  

\begin{figure}
\plotone{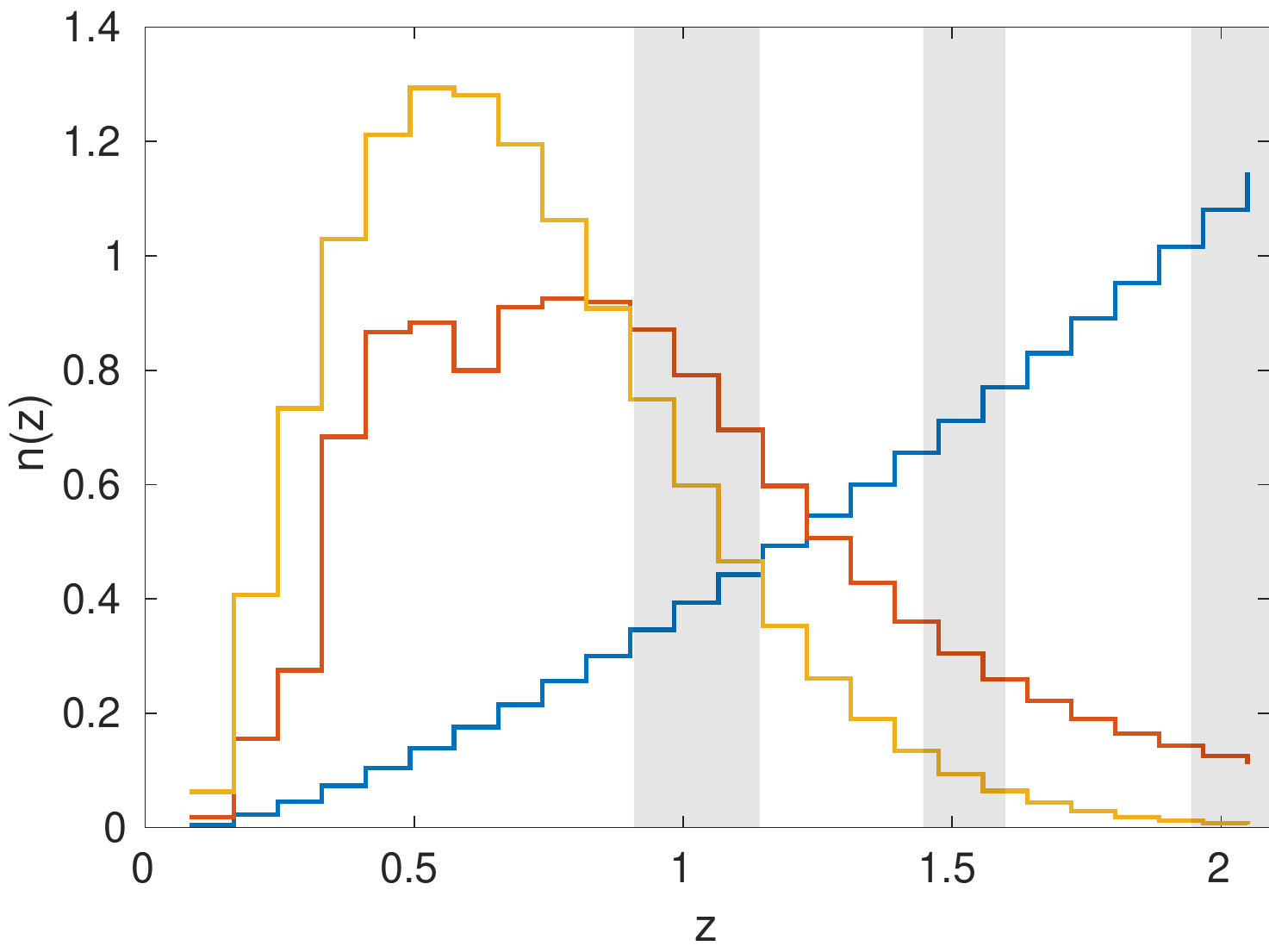}
\caption{The redshift distributions of the considered galaxy catalogs. The yellow and red histograms are for the \emph{KiDS-like} and {\it Euclid}-like samples.  The blue one is from the original galaxy sample. The three shaded regions are for redshift bins of $[0.91,1.14]$, $[1.45,1.60]$ and $[1.94,2.09]$, respectively.
\label{fig:reddis}}
\end{figure}

In total, we build five samples with different redshift distributions. One is the KiDS-like sample with the redshift distribution from \citet{Martinet2018} and the number density $n_g=8\hbox{ arcmin}^{-2}$. Another is the {\it Euclid}-like following the redshift distribution 
used in \citet{Joachim2022} and $n_g=30\hbox{ arcmin}^{-2}$. 
In addition, three samples at three different narrow redshift bins are constructed with $z=[0.91,1.14]$, $z=[1.45,1.60]$ and $z=[1.94,2.09]$ and $n_g\approx 26, 23.5$ and $23.8\hbox{ arcmin}^{-2}$, respectively.    
In Figure \ref{fig:reddis}, we show the normalized redshift distributions of our selected samples with the yellow, red and gray shaded bins for KiDS-like, {\it Euclid}-like and narrow bins, respectively. We also show the redshift distribution of the original galaxy sample in blue for comparison. 

For each catalog, we have two satellite IA settings, highIA and lowIA, from \citet{Wei}. To understand in detail
the dependence of the IA effects on the behavior of satellite alignments within clusters, we further artificially adjust their orientations to generate different
satellite IA for each shear sample.


To do this, we first identify satellite galaxies in a shear sample and their corresponding central galaxies using the IDs from \cite{Wei}.  
We then set the 3-D orientation of a satellite galaxy based on a local spherical coordinate system $(\theta_{\rm {3D}},\phi)$
in which $\theta_{\rm {3D}}$ is the zenith angle between the satellite orientation direction and the line connecting it to its central galaxy. 
Thus $\theta_{\rm{3D}}=0$ means that the satellite is radially aligned, i.e., for an early-type satellite, its long axis points toward its central, and for a late-type one, its spin is perpendicular to the radial direction. This is the highIA case.
For the case of lowIA, the satellite orientations are completely at random. To construct more IA cases, we artificially set $\theta_{\rm{3D}}$ to different fixed values. Specifically, $\theta_{\rm{3D}}=5^{\circ},10^{\circ},15^{\circ}$,..., and $45^{\circ}$ are considered. Furthermore, we also construct cases with $\theta_{\rm{3D}}=60^{\circ}$ and  $75^{\circ}$, respectively. For these two, satellite galaxies on average have different levels of 3-D tangential alignments, which might be unphysical but useful for our systematic investigations of IA effects. 

Besides fixed $\theta_{\rm{3D}}$, we also generate satellite alignments considering a dispersion of $\theta_{\rm{3D}}$. Specifically, we adopt the following probability distribution for $\theta_{\rm{3D}}$ and $\phi$
\begin{equation}
f(\theta_{\rm{3D}})d\theta_{\rm{3D}}\ d\phi=A\rm{exp}[-\frac{1}{2}(\frac{\theta_{\rm{3D}}-\theta_{0}}{\sigma_{\theta}})^2]d\theta_{\rm{3D}}\times\rm{sin}(\theta_{\rm{3D}})d\phi
\label{eq:theta_dis}
\end{equation}
where $A$ is the normalization parameter, $\theta_0$ is the average of $\theta_{\rm{3D}}$ and $\sigma_{\theta}$ is its dispersion. When $\sigma_{\theta}$ goes to infinity, this distribution is back to the 3-D uniform distribution, 
i.e., $\rm{sin}{(\theta_{\rm{3D}})}d\theta_{\rm{3D}}\ d\phi$. 
We take $\theta_0=0^{\circ}$ and consider different dispersions with $\sigma_{\theta}=10^{\circ}$, $20^{\circ}$, $30^{\circ}$ $45^{\circ}$, $60^{\circ}$ and $75^{\circ}$, respectively.
They are labeled in a format of $\hbox{0}\sigma_{\theta}$. For example, 030 stands $\sigma_{\theta}=30^{\circ}$. This set of samples correspond to the averagely radially aligned satellites with different levels of misalignment.
The above settings are for the adjustment of satellite alignments. For the central galaxies, their alignments are assigned as in \citet{Wei}. To build a reference for comparison, we further generate a noIA sample by randomly rotating both central and satellite galaxies.  

For a sample with a specific 3-D alignment setting, we then perform 2-D projections following the same approach of \citet{Joachimi2013} and \citet{Wei}. For late-type galaxies, we take the disk thickness-to-diameter ratio $r_{\rm {edge-on}}=0.25$ in accord with \citet{Wei}.
With the 2-D projected intrinsic ellipticity $\boldsymbol{\epsilon_s}$ of a galaxy, we build its observed $\boldsymbol{\epsilon}$ by using Eq.~(\ref{eq:epsilon2g}). This completes the construction of a shear catalog with an assigned alignment model.

\begin{figure}
\gridline{\plotone{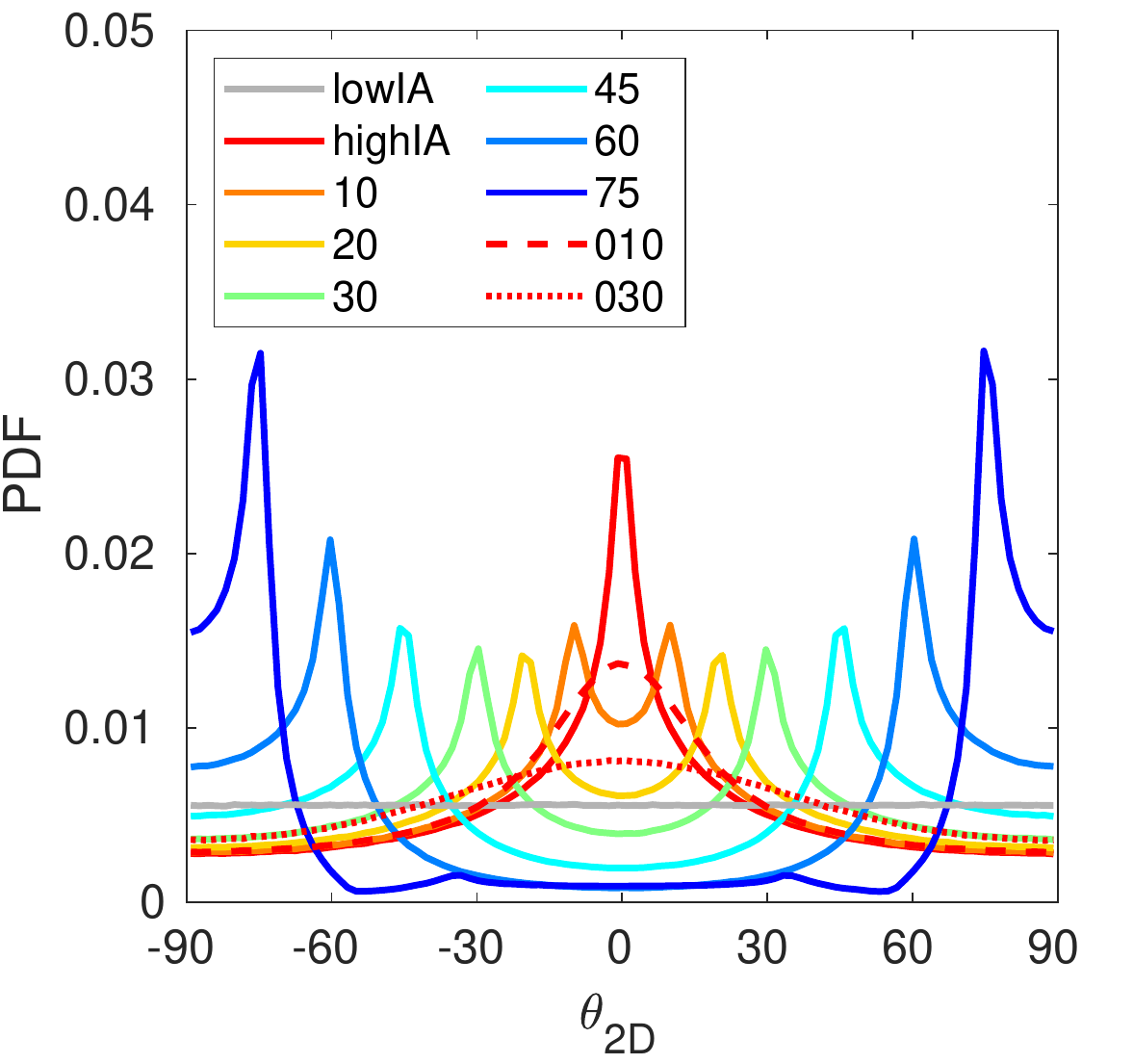}
}
\gridline{\plotone{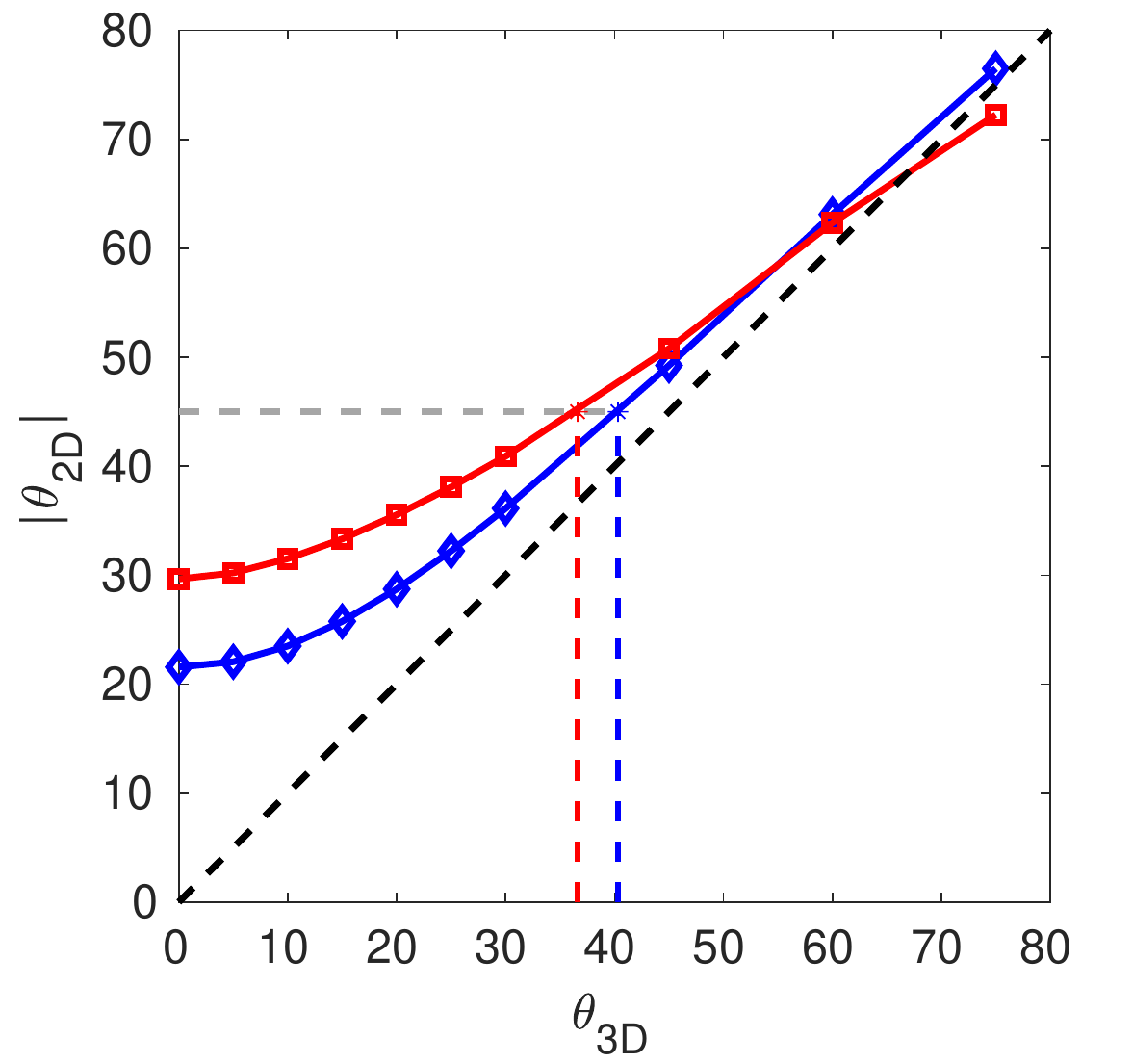}
}
\caption{Upper panel: 2D satellite orientation angle $\theta_{\rm 2D}$ distribution of different samples as measured from ${\boldsymbol{\epsilon_s}}$ without adding lensing signals. 
Lower panel: The median (blue) and mean (red) values of $|\theta_{\rm 2D}|$ with the change of $\theta_{\rm 3D}$. The grey horizontal dashed line indicates $|\theta_{\rm 2D}|=45^{\circ}$, and the red and blue vertical dashed lines
mark the corresponding values of $\theta_{\rm 3D}$ inferred from the intersections of the grey dashed line with the red and blue solid lines, respectively.
        \label{fig:2dangle}}
\end{figure}

The upper panel of Figure~\ref{fig:2dangle} shows the distribution of the angle $\theta_{\rm 2D}$ between the intrinsic 2-D projected orientation of a satellite galaxy and the radial direction to its central galaxy, 
where different lines correspond to satellite alignments with different settings as labelled in the legend. 
For example, the green solid line labelled as `30' is for the alignment model with $\theta_{\rm{3D}}=30^{\circ}$, and the line of `030' is for $\theta_0=0^{\circ}$ and $\sigma_{\theta}=30^{\circ}$. 
It is seen that for the lowIA case with randomly oriented satellite galaxies, their 2-D orientations are also random with the distribution being a flat line. For other cases, the distributions of $\theta_{\rm 2D}$ show peaks at 
$\theta_{\rm 2D}\sim \theta_{\rm{3D}}$. Adding a dispersion $\sigma_{\theta}$, the peaks are getting smoothed as expected.  

The lower panel presents the median (blue) and the average (red) of $|\theta_{\rm 2D}|$ for different $\theta_{\rm{3D}}$. It is noted that for $\theta_{\rm{3D}}<40^{\circ}$, the median value of $\theta_{\rm 2D}$ is less than $45^{\circ}$, showing that the 2-D orientations of satellite galaxies possess a level of 
radial alignment. For $\theta_{\rm{3D}}>40^{\circ}$, $\theta_{\rm 2D} (\rm {median})>45^{\circ}$, indicating a trend of tangential alignment in 2-D. The change of the 2-D orientation trend from radial to tangential at $\theta_{\rm{3D}}\approx 40^{\circ}$ rather than at $\theta_{\rm{3D}}=45^{\circ}$ is due to the projection.    
To be seen in the next section, our analysis results of the IA effects on WL peak statistics show consistent behaviors as expected from the trend change shown here.  

\section{Results\label{sec:results}}
To do peak analyses, we randomly select 156 patches from the full sky map of \cite{Wei} and each patch is $3\times3\ \rm{{deg}^{2}}$. For each considered galaxy sample with a particular satellite IA model, we perform convergence reconstruction for each patch on $1024\times 1024$ grids from galaxy ellipticities
$\boldsymbol{\epsilon}$ as described in \S\ref{sec:Rec}. We exclude $5\times\theta_G\sim43$ pixels from the edges of each convergence map in peak counting to suppress the boundary effect \citep[e.g.][]{Bartelmann95}. The final effective sky coverage is then about 1178 $\rm{{deg}^{2}}$. 
For each sample, we identify and bin peaks in terms of signal-to-noise ratio $\nu=K/\sigma_0$ with the shape noise $\sigma_0$ calculated from the corresponding noIA sample without lensing signals. This is in accord with real observations where the noise can only be estimated by randomly rotating galaxies.  

It is noted that in our setting, for different IA samples with a same redshift distribution and a same $n_g$, their galaxies have exactly the same positions and lensing signals. Thus the differences of their peak statistics solely arise from the different IA modeling.  
We therefore can reveal cleanly the impacts of IA on peak statistics.


\subsection{Dependence on the satellite IA angle\label{subsec:IAangle}}

To see the dependence of IA impacts on the satellite alignment angle, we focus on the samples with the {\it Euclid}-like redshift distribution and $n_g=30\hbox{ arcmin}^{-2}$. 

\begin{figure}
\plotone{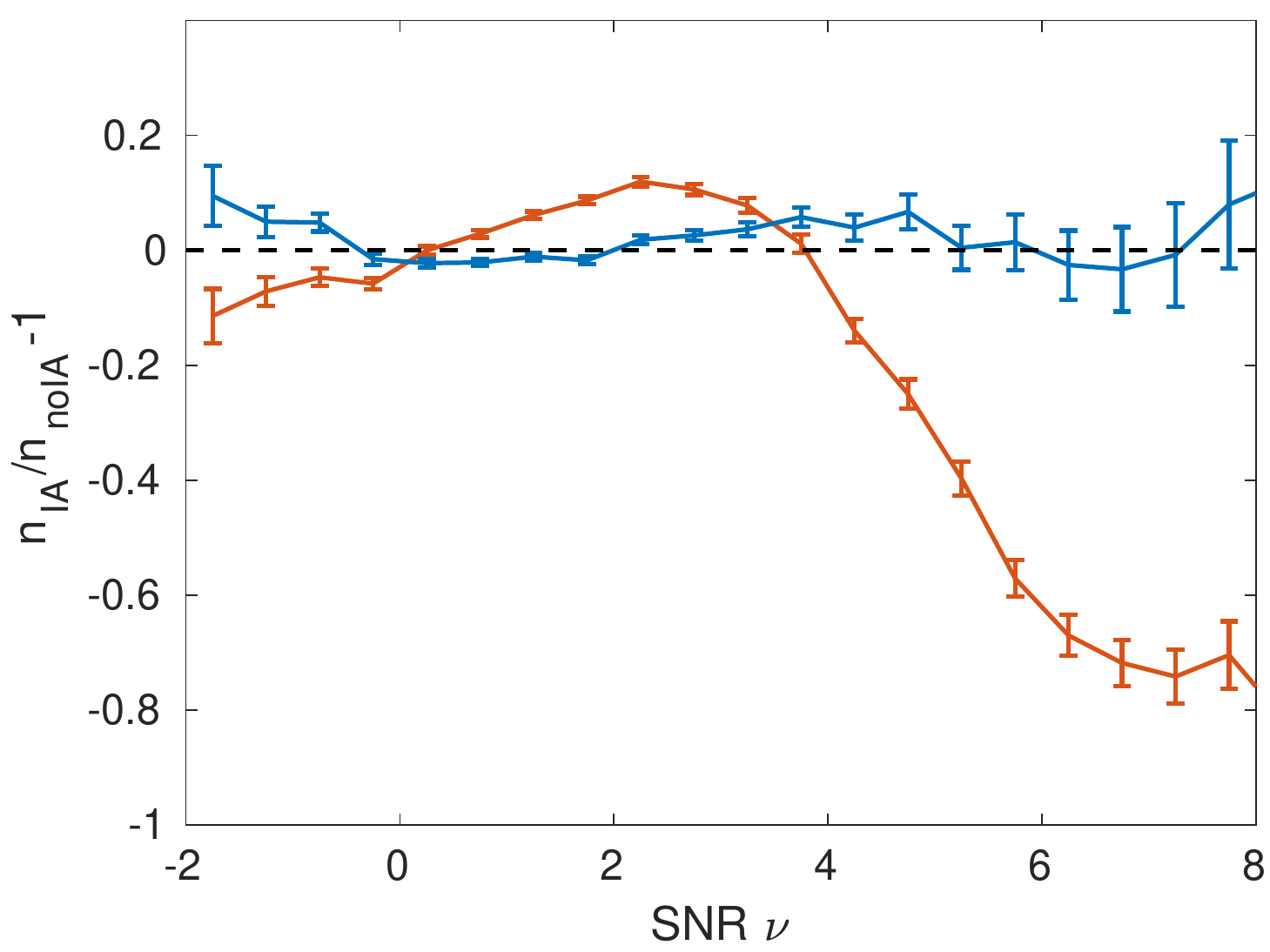}
\caption{Relative changes of peak counts for lowIA (blue) and highIA (red) samples in comparison with the noIA case for the {\it Euclid}-like sample.
\label{fig:peak_eu30}}
\end{figure}

We first present the results of highIA and lowIA in Figure \ref{fig:peak_eu30}, where the vertical axis is the relative difference of the peak numbers with respect to those of the corresponding noIA case and the error bars are Poisson errors calculated from the peak numbers in $\sim 1178 \rm{{deg}^{2}}$. 
It is seen that the two lines are very different. For lowIA, the difference from the noIA case is due to the IA of central galaxies, and we have $|n_{\rm lowIA}/n_{\rm noIA}-1|<10\%$ over the whole considered range of peaks. 
For highIA, the relative difference is much larger showing the dominant effects of satellite IA on peak statistics. For high peaks, the satellite radial IA leads to significant decreases of the peak numbers, and $n_{\rm highIA}/n_{\rm noIA}-1\sim -40\%$ and $\sim -70\%$ for $\nu\sim5$ and $6$, respectively.   

\begin{figure}
\gridline{\plotone{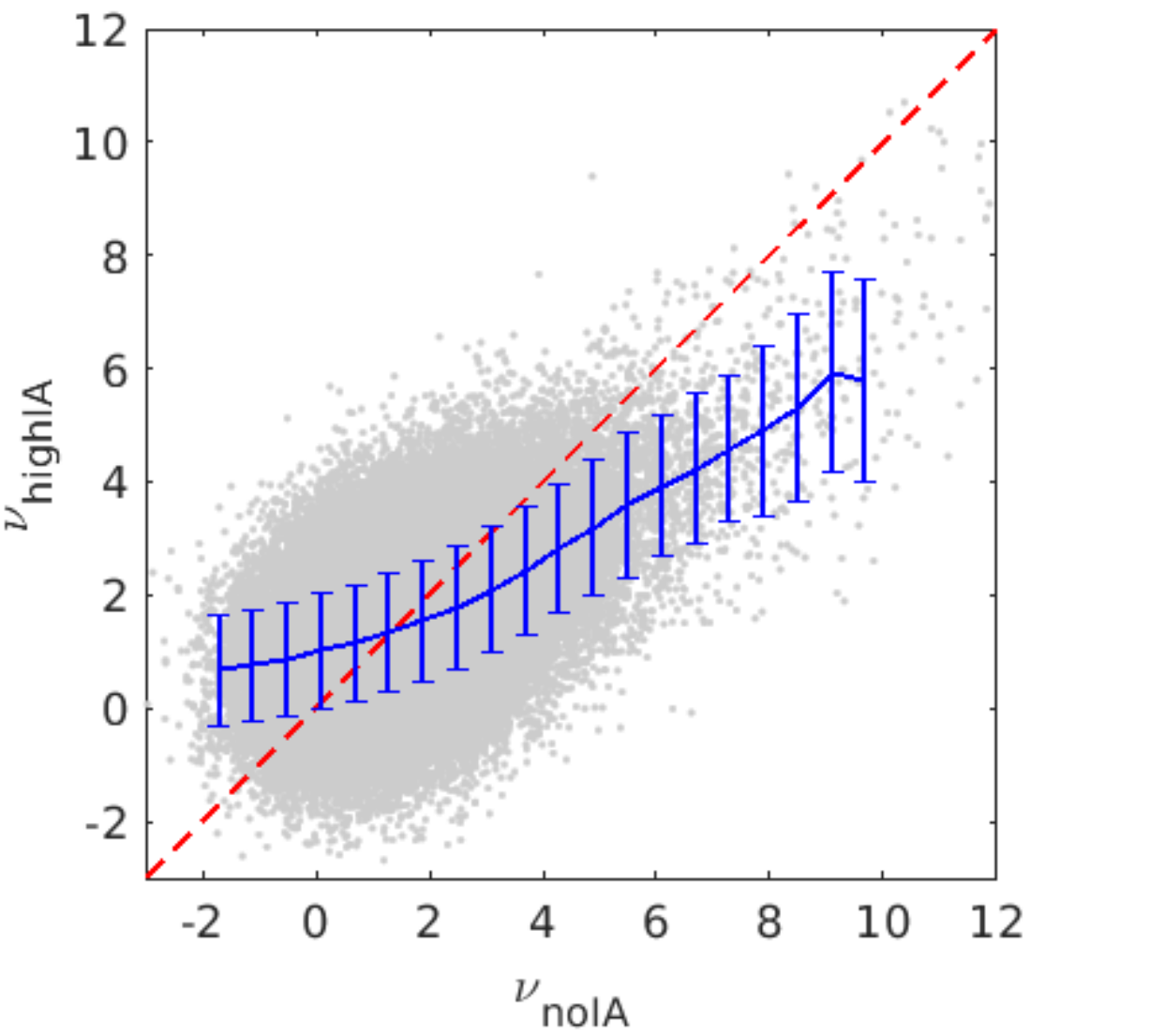}
}
\gridline{\plotone{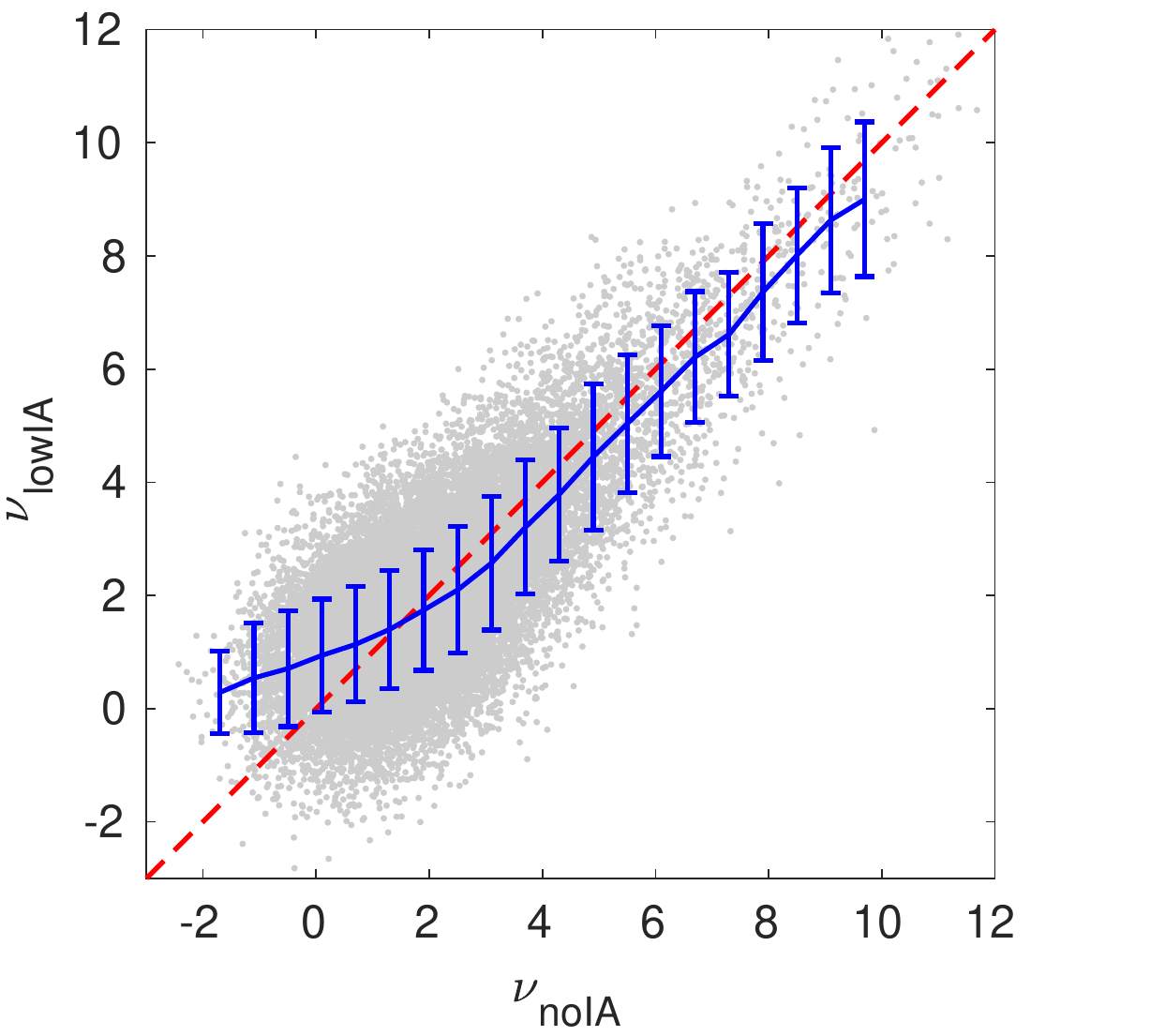}
}
\caption{Peak height comparison for matched peaks with the upper and lower panels for highIA and lowIA, respectively.
In each panel, the grey points are the scatter plots of the matched peaks, and the red dashed line indicates the 1:1 relation. The blue line shows the median peak values in the considered IA model at different bins based on $\nu_{\rm{noIA}}$ and the error bars are the $1\sigma$ dispersions.  
 \label{fig:match}}
\end{figure}

To further see the IA effects, we cross match the peaks in the two IA models with that of the corresponding noIA case. The matching radius is 17 and 5 pixels for highIA and lowIA cases respectively. 
In the highIA case, peaks are affected more for both heights and positions than that of the lowIA case, and we therefore choose a larger matching radius. In Figure \ref{fig:match}, we show the direct peak height comparison of the matched peaks with the upper and lower panels for highIA 
and lowIA, respectively. The grey points are the scatter plots of the matched peaks, and the blue lines are the median peak values of $\nu_{\rm{highIA}/\rm{lowIA}}$ of different bins based on $\nu_{\rm{noIA}}$ with the error bars showing the $1\sigma$ dispersions. 
The red dashed lines show the 1:1 relation.
We can see that in the highIA case, the high peaks are significantly lower than their noIA counterparts, while the differences are much smaller
in the lowIA case. These are in accord with the peak statistics shown in Figure \ref{fig:peak_eu30}. 

\begin{figure}
\plotone{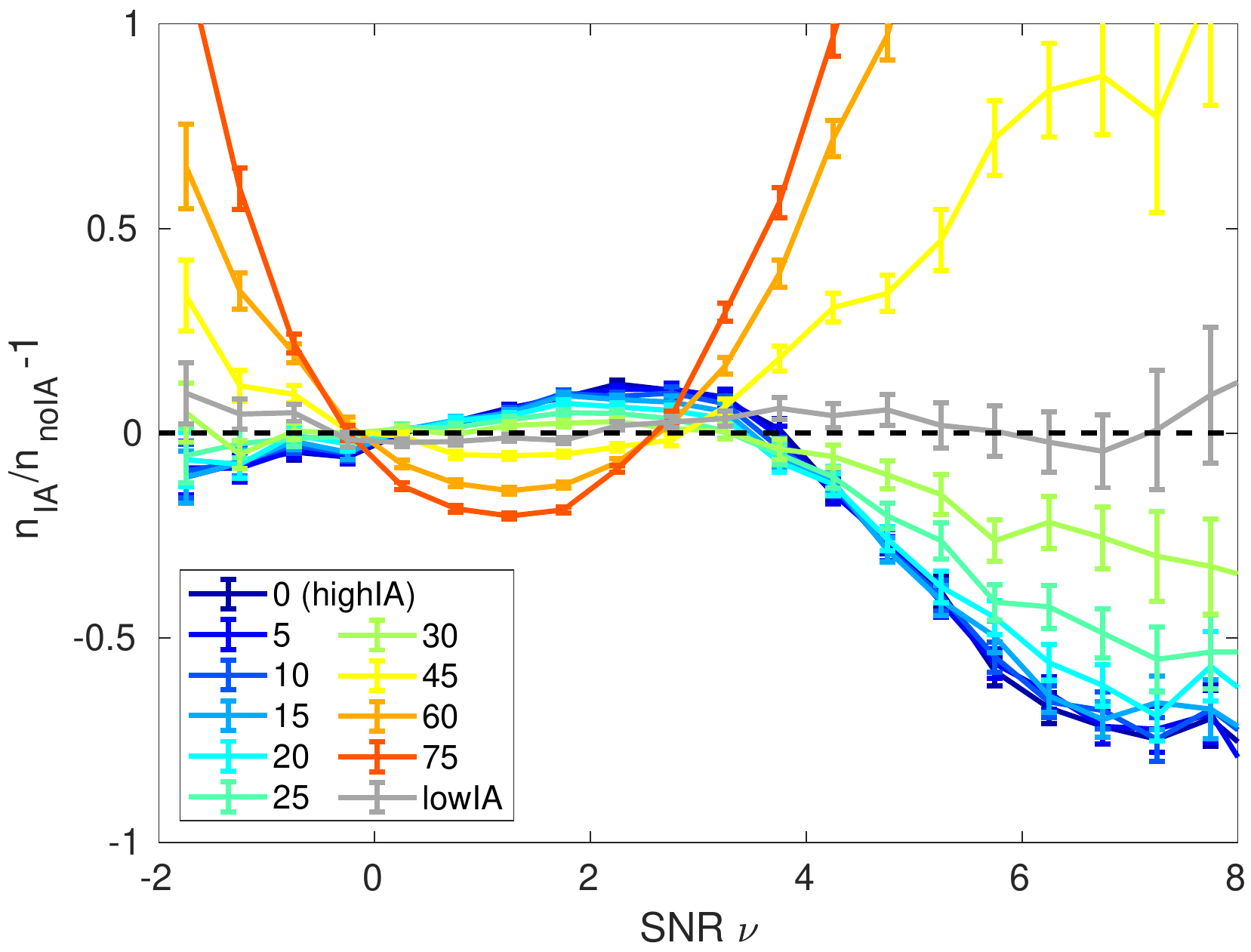}
\caption{Relative changes of peak counts with respect to the noIA case for different $\theta_{\rm{3D}}$ for the sample with the {\it Euclid}-like redshift distribution.
        \label{fig:angle_eu30}}
\end{figure}

As discussed in \citet{Wei}, the comparison with the cosmic shear 2PCF analyses from KiDS indicates that the highIA model leads to too high 2PCF signals at small scales, and the lowIA model is more consistent with the observational results. For peak statistics, its dependence on IA can be different from that of 2PCF.
For the systematic trend, we show the peak results with different satellite IA angles in the Figure \ref{fig:angle_eu30}, where different lines are for different 3-D satellite alignment angles $\theta_{\rm 3D}$ as explained in the legend. 
In consistent with the 2-D alignment angles $|\theta_{\rm 2D}|$ shown in the lower panel of Figure \ref{fig:2dangle}, when $\theta_{\rm 3D} \lesssim 40^{\circ}$, $|\theta_{\rm 2D}|\lesssim 45^{\circ}$ leading to a radial alignment tendency and resulting in a number decrease of high peaks. 
For $\theta_{\rm 3D} \gtrsim 40^{\circ}$, the number of high peaks are significantly boosted due to the tangential trend of the 2-D alignments of satellite galaxies. We also note that the IA effects on peak statistics depends sensitively on the median value of $|\theta_{\rm 2D}|$. For $\theta_{\rm 3D}<20^{\circ}$, 
the median of $|\theta_{\rm 2D}|\sim 20^{\circ}$ and changes slowly with the increase of $\theta_{\rm 3D}$. As a result, the IA effects on peak statistics from these IA models are very similar with the corresponding lines in Figure \ref{fig:angle_eu30} being very close to each other. For $\theta_{\rm 3D}>20^{\circ}$,
the slope of $|\theta_{\rm 2D}|$ in the right panel of Figure \ref{fig:2dangle} increases, and thus the differences in peak statistics between different IA models are more apparent. For $\theta_{\rm 3D}=30^{\circ}$, $n_{\rm IA}/n_{\rm noIA}-1\sim -5\%, -10\%$ and $-25\%$ for $\nu\sim 4, 5$ and $6$, respectively. 

\begin{figure}
\plotone{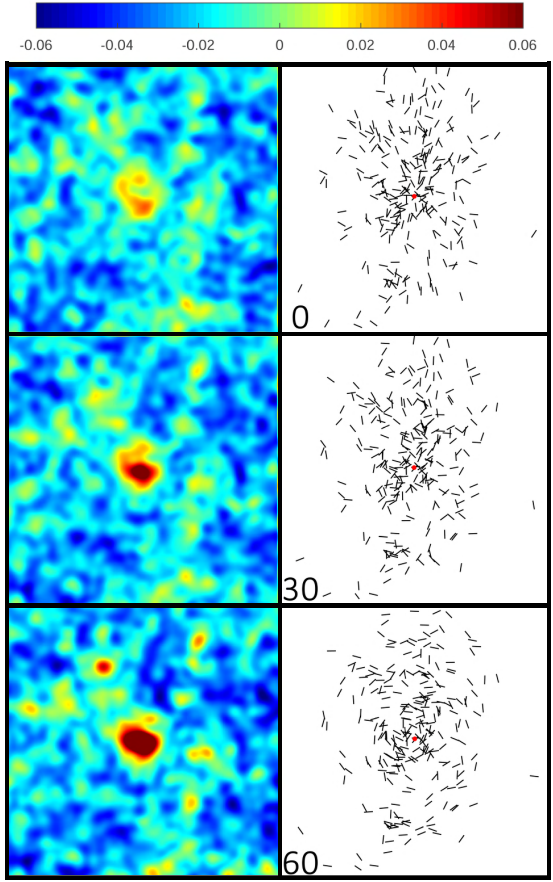}
\caption{Zoom-in convergence maps for a high peak (left) and the satellite orientations of the corresponding cluster (right) with the red dot indicating the center of the cluster.
From top to bottom, $\theta_{\rm 3D}=0^{\circ}, 30^{\circ}$ and $60^{\circ}$, respectively.
        \label{fig:kmap}}
\end{figure}

It is known that WL high peaks are typically associated with massive clusters \citep{Hamana2004, Fan2010, Yang2011, Yuan2018, Shan2018, Martinet2018, Oguri2021}. Thus the IA effects on their peak heights mainly arise from the inclusion of their satellite members in a shear catalog. 
If these member galaxies have IA signals, they can affect the peaks significantly. To illustrate this clearly, in Figure \ref{fig:kmap}, we show zoom-in convergence maps for a high peak in different satellite IA cases. For this peak, we can find a clear association with a massive cluster. 
We then identify its satellite galaxies in the shear sample, and show their orientation maps with different IA settings in the right panels. It can be seen clearly that when the satellites have perfectly radial alignments with $\theta_{\rm 3D}=0^{\circ}$, the peak from the cluster is considerably suppressed. 
With $\theta_{\rm 3D}=30^{\circ}$, the suppression is less. For $\theta_{\rm 3D}=60^{\circ}$, the IA signals have tangential trends in line with the tangential WL signals from the cluster, leading to a significant increase of the peak height.   

\subsection{Dependence on the satellie IA angle dispersion \label{subsec:dispersion}}
In the previous subsection, we present the impacts of satellite IA assuming they have a same alignment angle $\theta_{\rm{3D}}$ toward their host central galaxies. 
A more physical modeling about the satellite IA should include additionally a certain misalignment dispersion. For that, we generate satellite IA samples using the distribution of Eq. (\ref{eq:theta_dis}).
In line with other studies \citep[e.g.,][]{Knebe2008, SB2010,Kacprzak2016}, we consider the case of averagely radial alignment by taking $\theta_0=0$ with different dispersions $\sigma_{\theta}$.

\begin{figure}

\gridline{\plotone{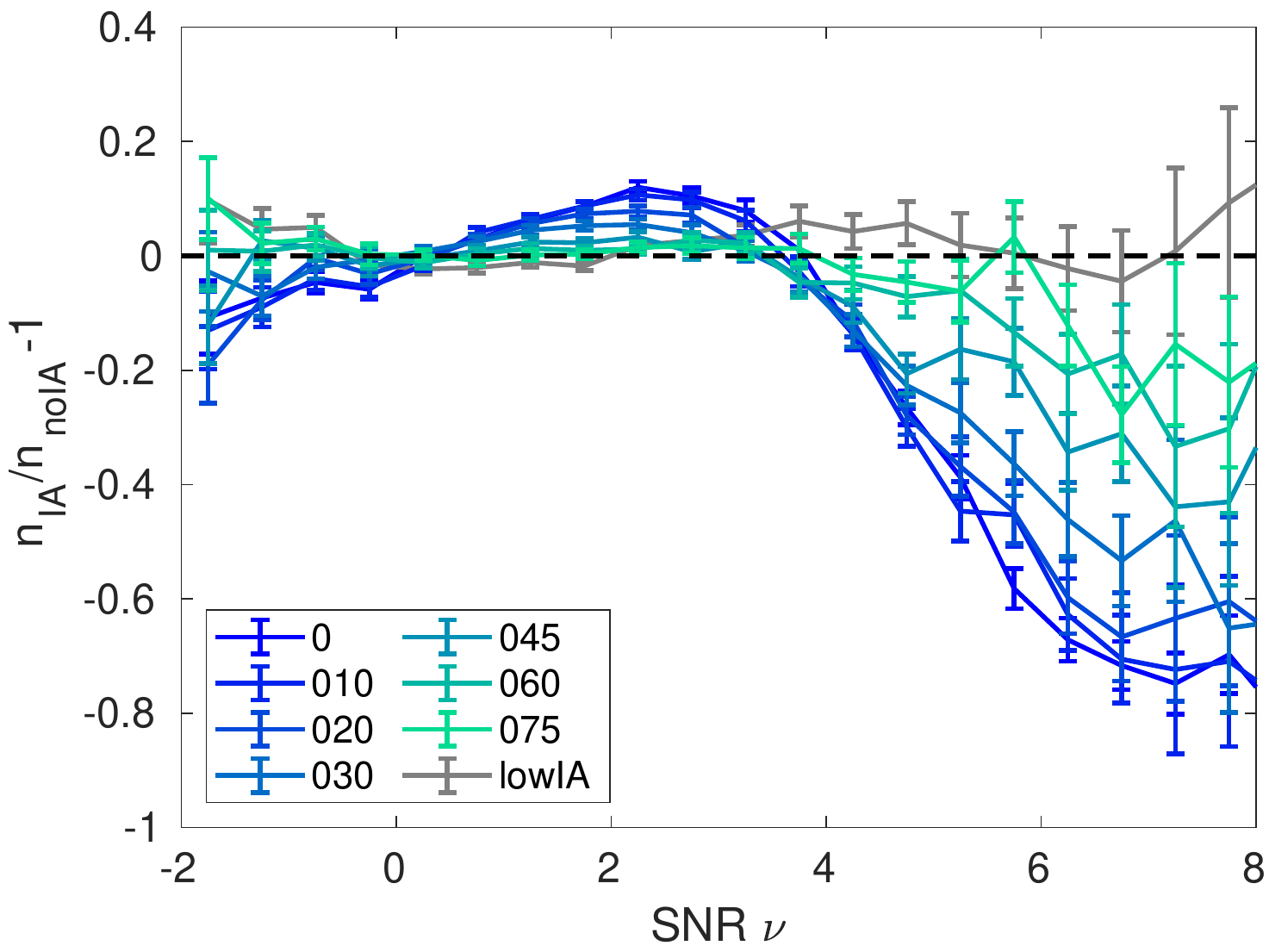}
} 
\gridline{\plotone{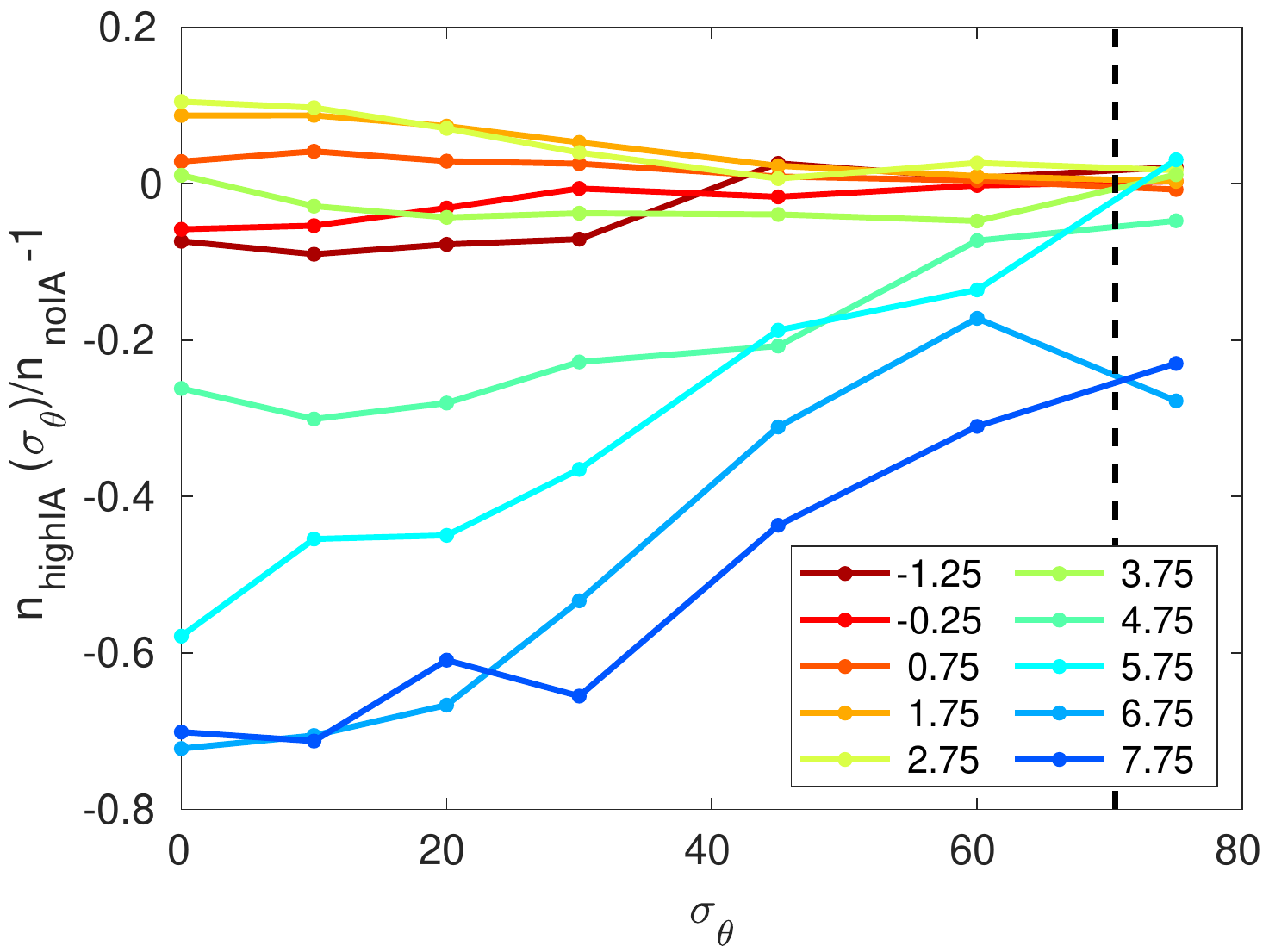}
}
\caption{Upper panel: Relative changes of peak counts for the IA samples with $\theta_0=0^{\circ}$ and different dispersions $\sigma_{\theta}$ as indicated in the legend. The lowIA results are also shown (grey).
Lower panel: The dependence of $n_{\rm IA}/n_{\rm noIA}-1$ on $\sigma_{\theta}$ with different lines for different peak heights as explained in the legend. The vertical dashed line marks the value of $\sigma_{\theta}\approx 70^{\circ}$ corresponding to $\gamma_{\rm{scale}}\approx 0.21$ as explained 
in Appendix \ref{appendix:AIAfit}.
        \label{fig:radialdispersion}}
\end{figure}

The corresponding peak analyses results for the {\it Euclid}-like sample are shown in Figure \ref{fig:radialdispersion}. The upper panel is $n_{\rm IA}/n_{\rm noIA}-1$ vs. peak height $\nu$ with different lines for different $\sigma_{\theta}$ from $0^{\circ}$ to $75^{\circ}$. 
For comparison, the result of lowIA is also shown. As expected, the IA impact decreases with the increase of the dispersion $\sigma_{\theta}$. It is noted that the results shown here have some degeneracies with the ones shown in Figure \ref{fig:angle_eu30}. 
For example, the result with $\theta_0=0^{\circ}$ and $\sigma_{\theta}=60^{\circ}$ is very similar to that of the case with a fixed $\theta_{\rm 3D}=30^{\circ}$. 

In the lower panel of Figure \ref{fig:radialdispersion}, we present the dependence of $n_{\rm IA}/n_{\rm noIA}-1$ on $\sigma_{\theta}$, where different lines are for different peak heights. For clarity, only the central values are shown without adding the error bars.
In the {\it Euclid}-like case here, $n_{\rm IA}/n_{\rm noIA}-1<10\%$ for peaks with $-1\le\nu\le 4$ for all the $\sigma_{\theta}$ values
considered. For $\nu< 0$, the IA effect leads to negeative changes of the peak numbers. For $0<\nu< 4$, the peak numbers increase with respect to the case of noIA. For peaks with $\nu>4$, their numbers are suppressed due to the satellite IA, and the suppression gets larger for higher $\nu$. 

To link the dispersion $\sigma_{\theta}$ to the scaling factor $\gamma_{\rm{scale}}$ in terms of the intrinsic aligment contribution to the cosmic shear two-point correlations \citep{SB2010}, in Appendix \ref{appendix:AIAfit}, we calculate the II correlations. 
Using correlation data on small scales where satellite IA contributes dominantly, we compute $\gamma_{\rm{scale}}$ by assuming $\hbox{II}(\sigma_{\theta})=\gamma_{\rm{scale}}^2\times \hbox{II}(\sigma_{\theta}=0)$ to obtain the relation between $\gamma_{\rm{scale}}$
and $\sigma_{\theta}$.  

By analyzing the subhalo alignments from simulations, \cite{Knebe2008} propose a distribution function about the misalignment angle of subhalos, which gives rise to $\gamma_{\rm{scale}}\sim 0.21$ \citep{SB2010}. This value has been used in 
\cite{Kacprzak2016} to evaluate the IA impacts on shear peak statistics with DES data. From Appendix \ref{appendix:AIAfit}, we can see that when $\sigma_{\theta}\sim 70^{\circ}$, $\gamma_{\rm{scale}}\sim 0.21$. 
This is indicated by the vertical dashed line in the lower panel of Figure \ref{fig:radialdispersion}. 
At this $\sigma_{\theta}$, $n_{\rm IA}/n_{\rm noIA}-1\approx 5\%$ and $25\%$ for $\nu=4.75$ and $7.75$, respectively. 

The information presented in Figure \ref{fig:radialdispersion} can be utilized in peak analyses to take into account the satellite IA impacts. In particular, the dependence of $n_{\rm IA}/n_{\rm noIA}-1$ on $\sigma_{\theta}$ can potentially allow us to
constrain $\sigma_{\theta}$ together with cosmological parameters. On the other hand, as presented in the next subsection, the IA effects on peak statistics are redshift dependent. Thus for a specific survey, its redshift distribution is needed to 
generate plots similar to Figure \ref{fig:radialdispersion}.

\subsection{The redshift dependence of the IA effects \label{subsec:redshift}}

In the above discussions, we show that satellite IAs affect peak statistics especially high peaks because of the inclusion of them in a shear sample.   
Shear samples with different redshift distributions can contain different fractions of satellites, and we therefore expect a redshift dependence of the IA impacts on peak statistics.   

\begin{figure}
\plotone{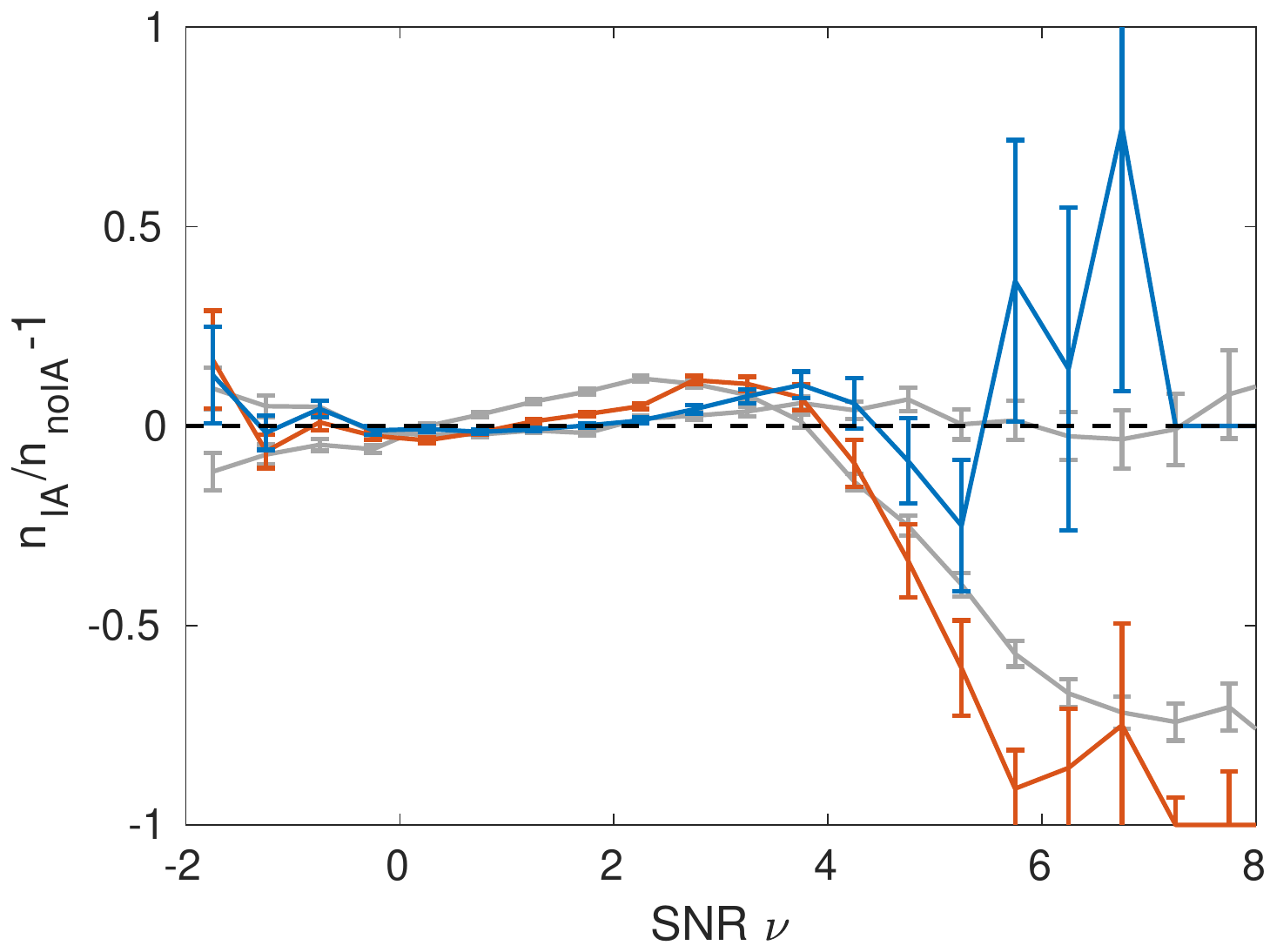}
\caption{Comparisons of the IA impacts for the {\it Euclid}-like sample (grey) and the KiDS-like sample. The results from the KiDS-like highIA and lowIA samples are shown in red and blue, respectively.\label{fig:peak_cp}}
\end{figure}

In Figure \ref{fig:peak_cp}, we show the results from the KiDS-like sample in comparison with those of the {\it Euclid}-like sample. The highIA and lowIA cases are presented. For the KiDS-like sample, $n_g=8\hbox{ arcmin}^{-2}$. Thus the shape noise is larger and the number of high peaks is smaller than that of the 
{\it Euclid}-like case, leading to larger error bars. Within the statistical uncertainties, $n_{\rm lowIA}/n_{\rm noIA}-1$ have similar behaviors in the lowIA case for the two samples. For the highIA case, the suppression of the high peak counts is stronger in the KiDS-like sample with 
$n_{\rm highIA}/n_{\rm noIA}-1\sim -60\%$ and $-90\%$ for $\nu\sim 5$ and $6$ in comparison with $\sim -40\%$ and $-70\%$ in the {\it Euclid}-like case. We note that the higher shape noise in the KiDS-like sample leads to a lower $\nu$ for a same peak. While we do see the red line shifting to the left comparing to the
grey line (highIA case) at high peak region partly due to the lower $\nu$, additional suppression in the KiDS-like case can be seen. This arises because in the KiDS-like sample, we have a larger fraction of galaxies in the redshift range of $\sim 0.2-0.6$ than that of the {\it Euclid}-like case. As 
clusters in this redshift range are abundant and contribute to WL high peaks, the higher fraction of satellite galaxies in the KiDS-like sample leads to stronger IA effects on high peak statistics. 

\begin{figure*}[]
\gridline{\fig{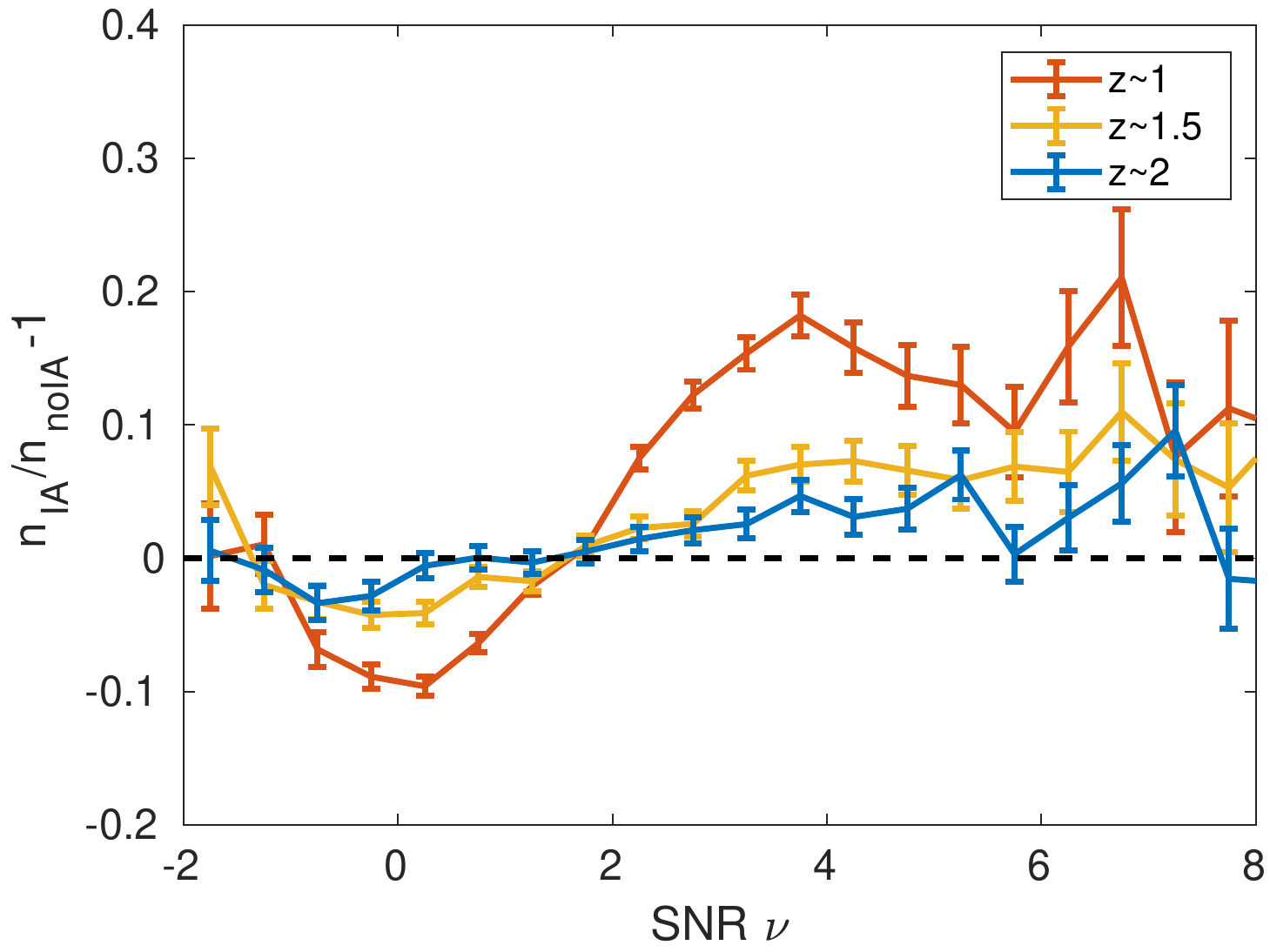}{0.4\textwidth}{}
          \fig{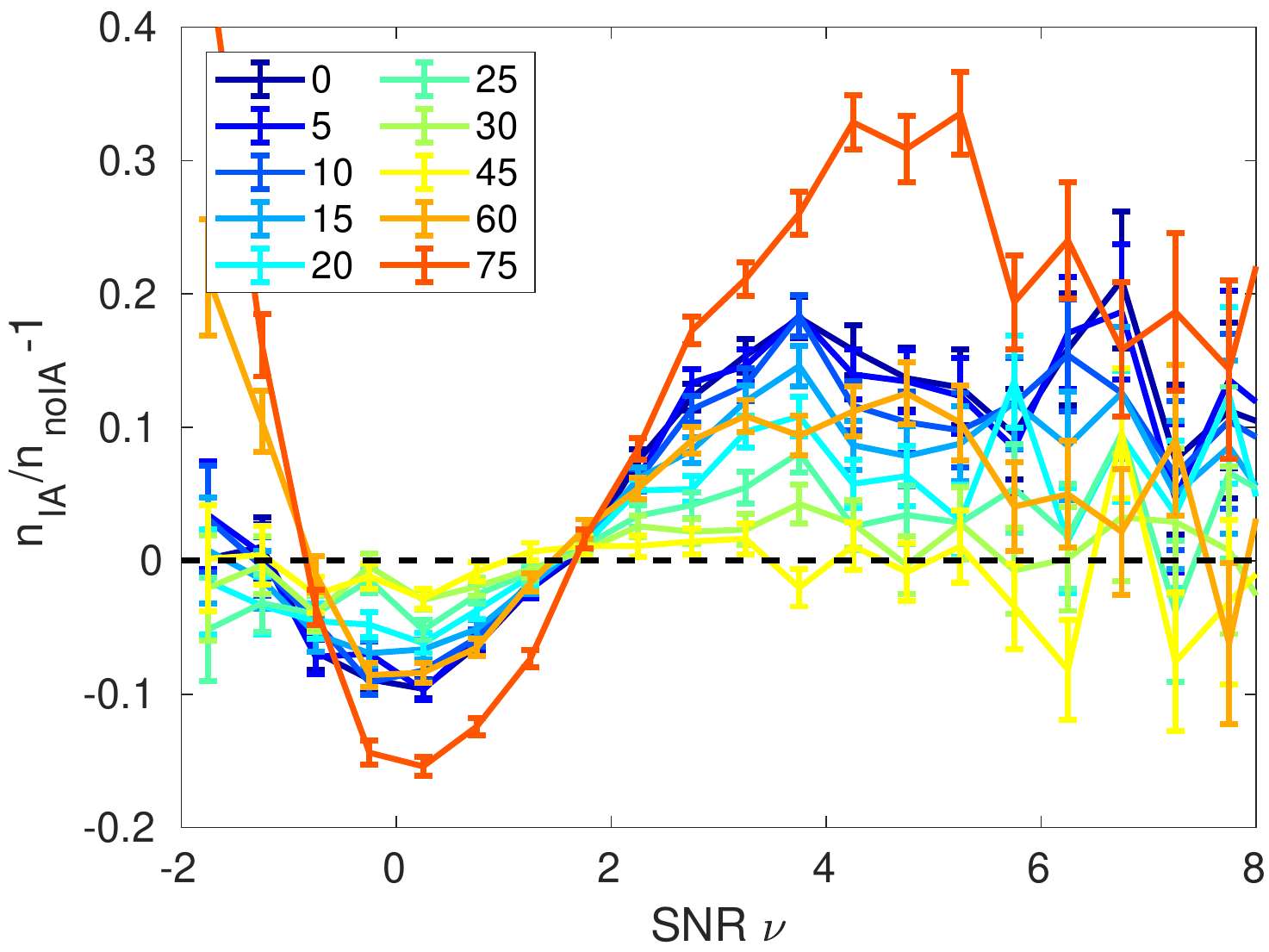}{0.4\textwidth}{}
}
\gridline{\fig{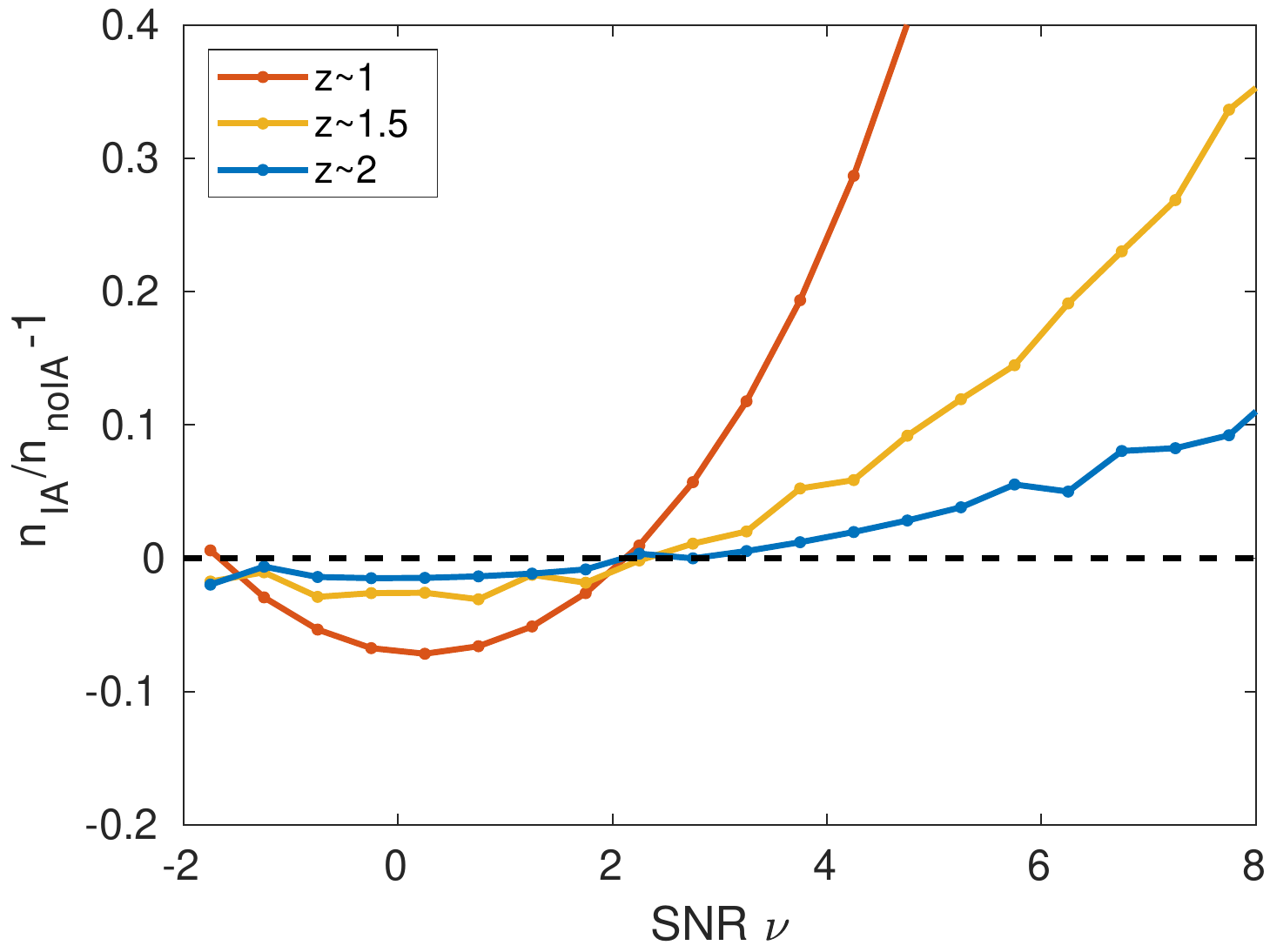}{0.4\textwidth}{}
          \fig{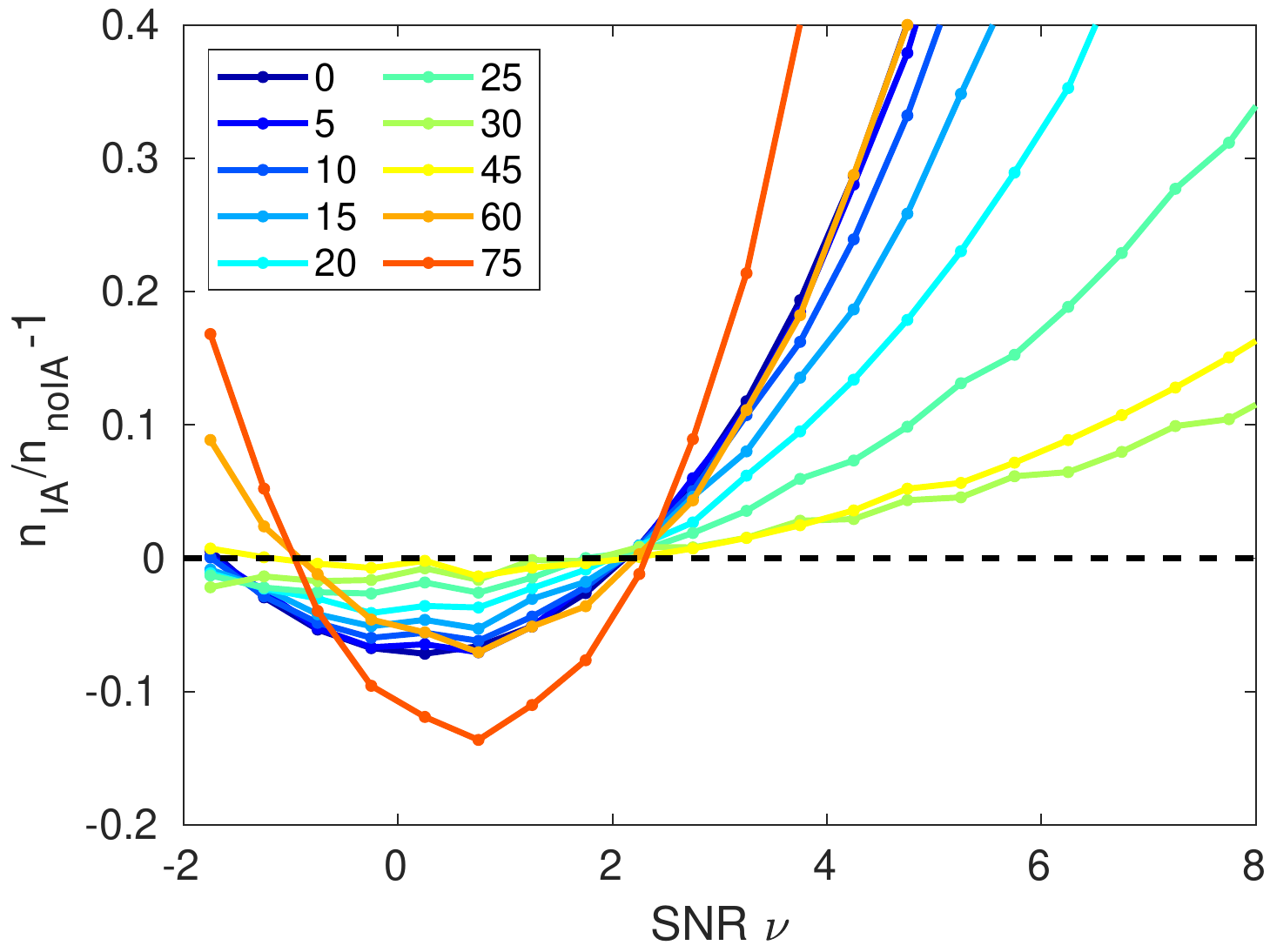}{0.4\textwidth}{}
}
\caption{Upper panels: Simulation results of the relative peak counts differences for the three narrow-redshift-bin samples of the highIA case (left) and for the $z\sim 1$ sample with different $\theta_{\rm{3D}}$ (right). 
Lower panels: The corresponding results from the Gaussian model calculations.  
\label{fig:redpk}}
\end{figure*}

To see the redshift dependence further, we present the simulation results in the upper panels of Figure \ref{fig:redpk} for the three narrow-redshift-bin samples with the satellite IA being highIA (left) and for the sample of $z\sim 1$ with different IA angle $\theta_{\rm{3D}}$. 
It is seen that the lines have markedly different trends comparing to the results of {\it Euclid}-like or KiDS-like cases. For high peak counts, there are no suppressions, instead, they increase with $n_{\rm IA}/n_{\rm noIA}-1>0$. 

The behaviors shown in Figure \ref{fig:redpk} can be explained qualitatively as follows. For sources in narrow redshift bins, the foreground clusters that contribute to WL high peaks have a negligible number of satellites in the source sample, and thus these satellite IAs do not affect the corresponding peaks. 
We therefore see no suppressions for high peak counts. On the other hand, the existence of IA induces shape noise correlations, and thus change the noise properties. This in turn can affect peak statistics \citep{Fan2007}. 

As shown in Eq.(\ref{eq:epsilon2g}), the observed ellipticities of galaxies include both their intrinsic ellipticities $\boldsymbol {\epsilon_s}$ and the lensing signals $\boldsymbol {g}$. Thus the reconstructed convergence maps contain noises arising from $\boldsymbol {\epsilon_s}$. 
The noise field $N$ can be approximated as a Gaussian random field because of the central limit theorem \citep{vanWaer2000, Fan2010}. We therefore use the Gaussian random field theory to illustrate the IA effects on noise and further on peak statistics.   

In the Appendix \ref{appendix:model}, we describe the 2-D Gaussian peak theory and present the changes of the noise parameters for different $\theta_{\rm 3D}$. From this, we do Gaussian model calculations to illustrate the IA effects on peak counts arising from their contributions to noise.

The model results are shown in the lower panels of Figure \ref{fig:redpk} for the cases corresponding to the upper panels.  
It is seen that they show qualitatively similar behaviors as the simulation results. For the three samples, the IA contributions $\sigma_{i,\rm{corr}}$ decrease with the increase of redshifts due to the
decrease of the number of clusters at high redshift (see the yellow line in Figure \ref{fig:typeratio}). Thus the relative differences of $n_{\rm IA}/n_{\rm noIA}-1$ also decrease. The dependence on $\theta_{\rm{3D}}$ is also alike qualitatively between the results
of simulations and the Gaussian model calculations. These demonstrate that the IA contributions to noise do affect peak statistics.

We need to emphasize that the Gaussian model calculations here are only for qualitative illustrations of the noise effects. For quantitative modeling, we must consider the non-Gaussian nature of the foreground structures, especially for high peaks. 
In our previous studies \citep{Fan2010, Yuan2018}, we develop a halo-based theoretical model for high peak counts in which massive halos contribute dominantly to high peaks, and the shape noise and the LSS effect other than the massive halos are modeled as Gaussian random fields.   
In our future studies, we aim to further develop the model by including the IA effects both from the noise perspective and from the inclusion of cluster satellites in shear samples.

\subsection{Dependence on the smoothing scale \label{subsec:smoothing}}

For the results shown above, our fiducial smoothing scale is taken to be $\theta_G=1.5 \hbox{ arcmin}$. To see the smoothing scale dependence, we further analyze how the IA effects change with different $\theta_G$. 

\begin{figure}
\plotone{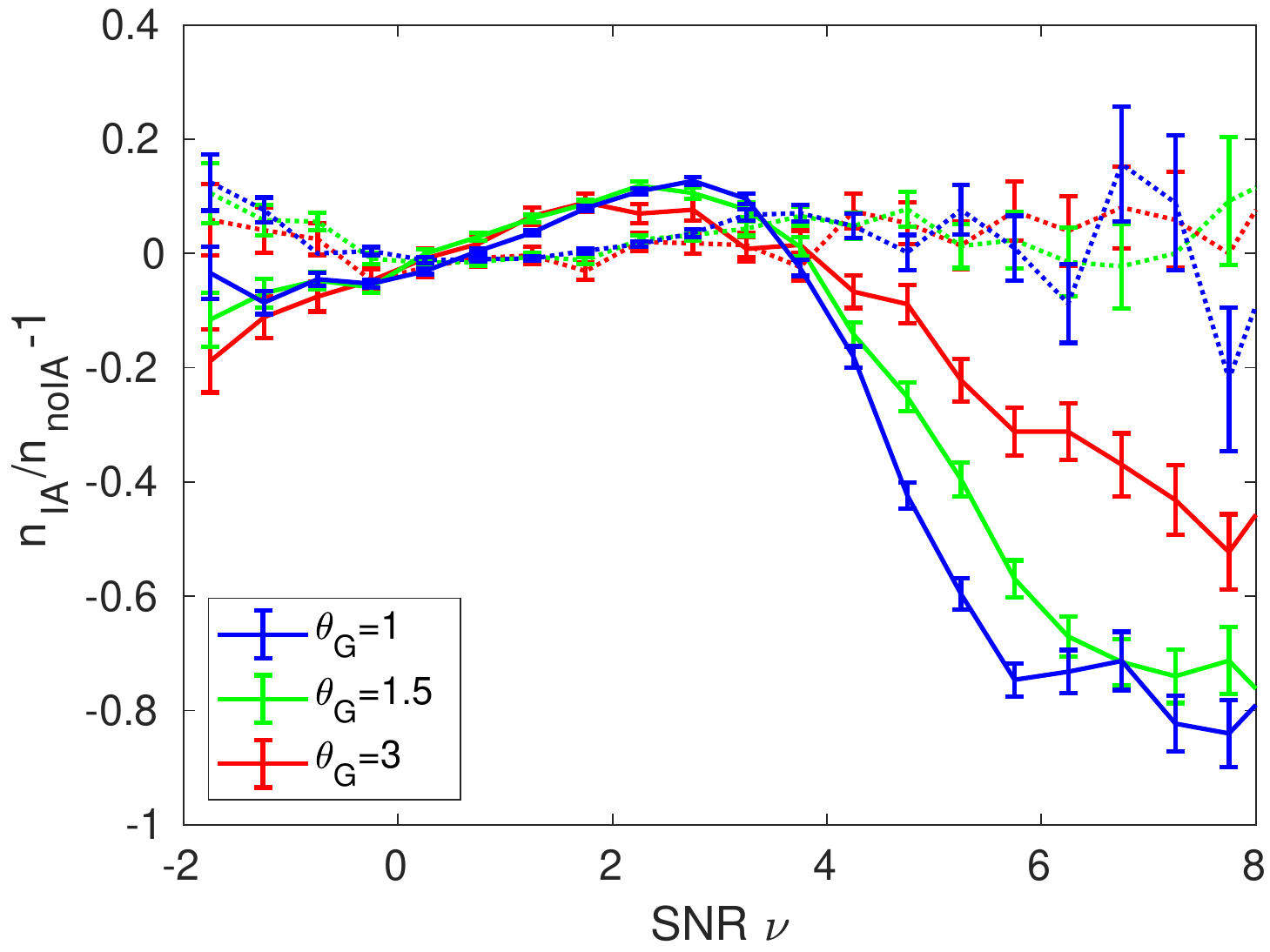}
\caption{The results with different smoothing scale $\theta_G$ as shown in the legend for the {\it Euclid}-like sample in the highIA (solid) and lowIA (dotted) cases. 
\label{fig:smooth}}
\end{figure}

Figure \ref{fig:smooth} presents the results for the {\it Euclid}-like sample in the highIA  and lowIA cases. Three values of $\theta_G$ are considered. We can see a clear dependence on $\theta_G$ in the highIA case with the IA effects less significant for larger
$\theta_G$. This can be understood as follows. Consider a massive cluster giving rise to a WL peak as the one shown in Figure \ref{fig:kmap}. Its satellite galaxies in the shear sample lead to strong IA effects on the peak. With the increase of $\theta_G$, more galaxies outside the cluster region enter the convergence reconstruction
and dilute the satellite IA effects on the peak. Statistically, we therefore have smaller $|n_{\rm highIA}/n_{\rm noIA}-1|$ for larger $\theta_G$ especially for high peaks. For $\nu\sim 5$, $n_{\rm highIA}/n_{\rm noIA}-1\sim -60\%, -40\%$ and $-15\%$ for $\theta_G=1, 1.5$, and $3 \hbox{ arcmin}$, respectively. The corresponding 
values are $\sim -80\%, -70\%$ and $-35\%$ for $\nu\sim 6$. 

\section{An empirical fitting model\label{sec:fitting}}

Our results show that WL peak counts are sensitive to IA effects. With a proper modeling, it is possible to extract IA signals, particularly satellite IA, from observed peak statistics simultaneously with the cosmological information.    

Here based on our simulation analyses, we present an empirical fitting model that allows us to build the peak counts with IA from those of noIA simulations. The model relies on the following information: the matching fractions of peaks between the cases with IA and noIA, and the average peak height shifts $\Delta\kappa$ and their scatters $\sigma_{\Delta\kappa}$
for the matched peaks. 

\begin{figure}
\plotone{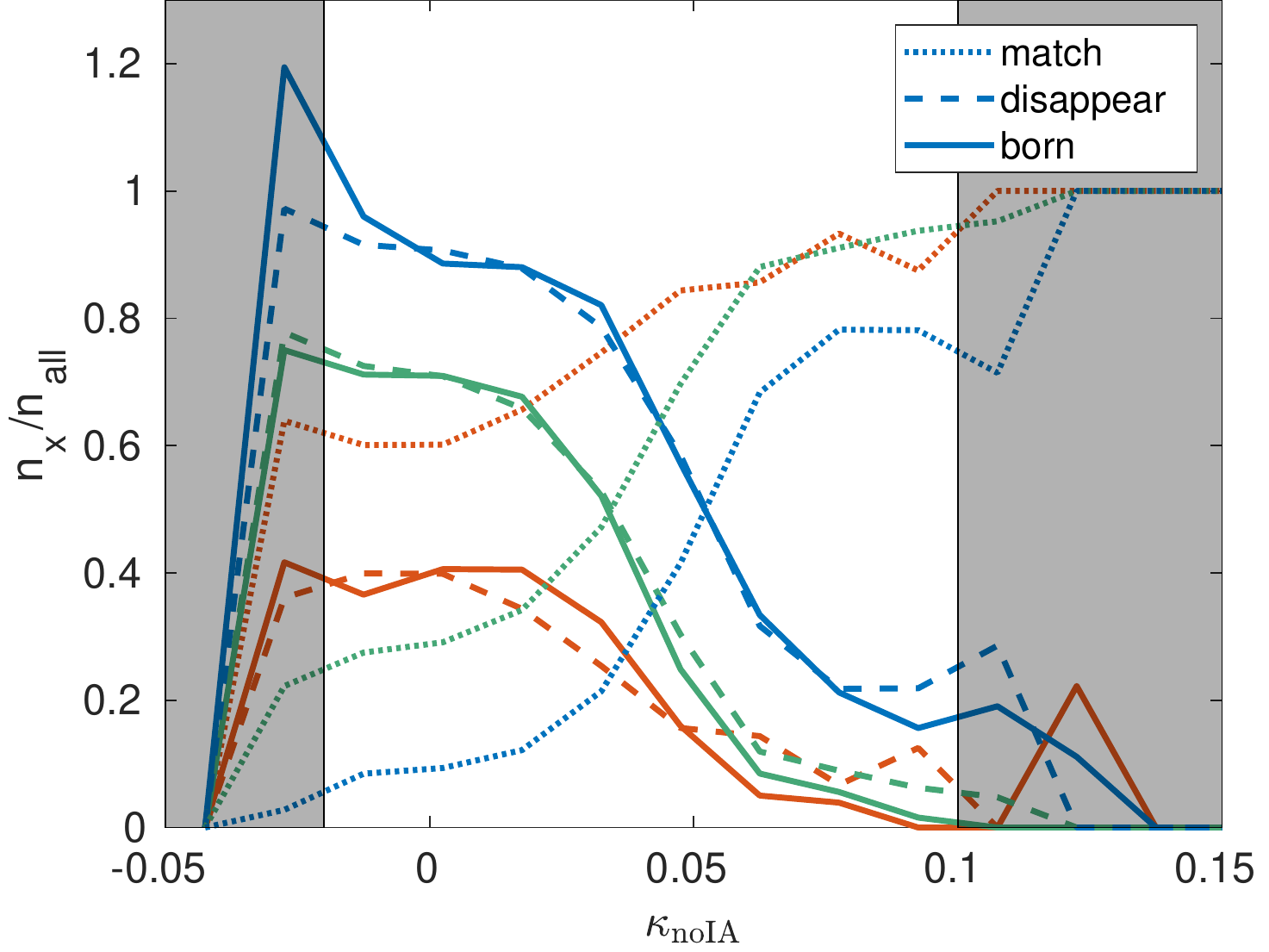}
\caption{The fractions of matched (dotted), newly born (solid), and disappeared (dashed) peaks in $\kappa_{\rm noIA}$ bins for the {\it Euclid}-like sample. The red, green and blue lines are for highIA, $\theta_{\rm 3D}=30^{\circ}$, and lowIA cases, respectively. 
Shaded regions are excluded from the analyses because of the very low peak numbers there resulting in large fluctuations.
\label{fig:matching}}
\end{figure}

Figure \ref{fig:matching} shows the matching results of peaks for the {\it Euclid}-like sample with the IA being highIA (red), $\theta_{\rm{3D}}=30^{\circ}$ (green) and lowIA (blue), respectively. The corresponding matching radius is 17, 9 and 5 pixels for the three IA cases. The dotted lines are the fractions of matched peaks. 
The solid and dashed lines are the fractions of newly born and disappeared peaks in the maps with IA comparing to those of noIA. Note that $n_{\rm all}$ here is the total number of peaks in a certain bin in the case of noIA. 

It is seen that in general, the fractions of matched peaks increase with the peak height as expected.
For the case of highIA, because of the relatively large matching radius, there are significant fractions of matched low peaks. We need to note however that low peaks largely arise from random fields including LSS and the shape noise. Thus those matched low peaks are likely just by chance to be close to each other in the cases of IA and noIA 
but not physically linked. Therefore our empirical fitting model is more meaningful for high peaks although we present the relevant results extending to low peaks. The shaded regions are excluded in the model fitting analyses because of the small numbers of peaks and thus large statistical fluctuations there.    

\begin{figure}[ht]
\gridline{\plotone{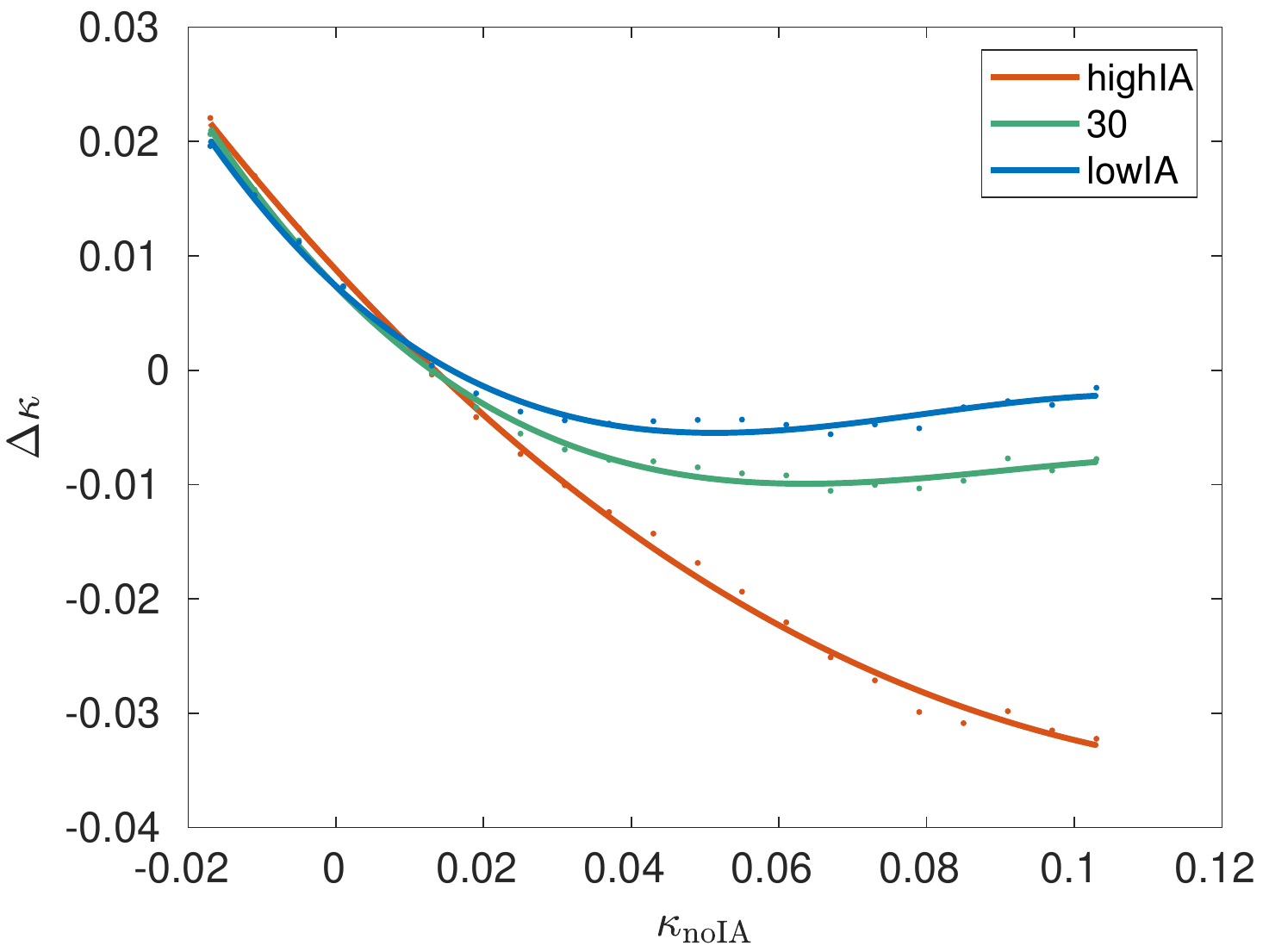}
}
\gridline{\plotone{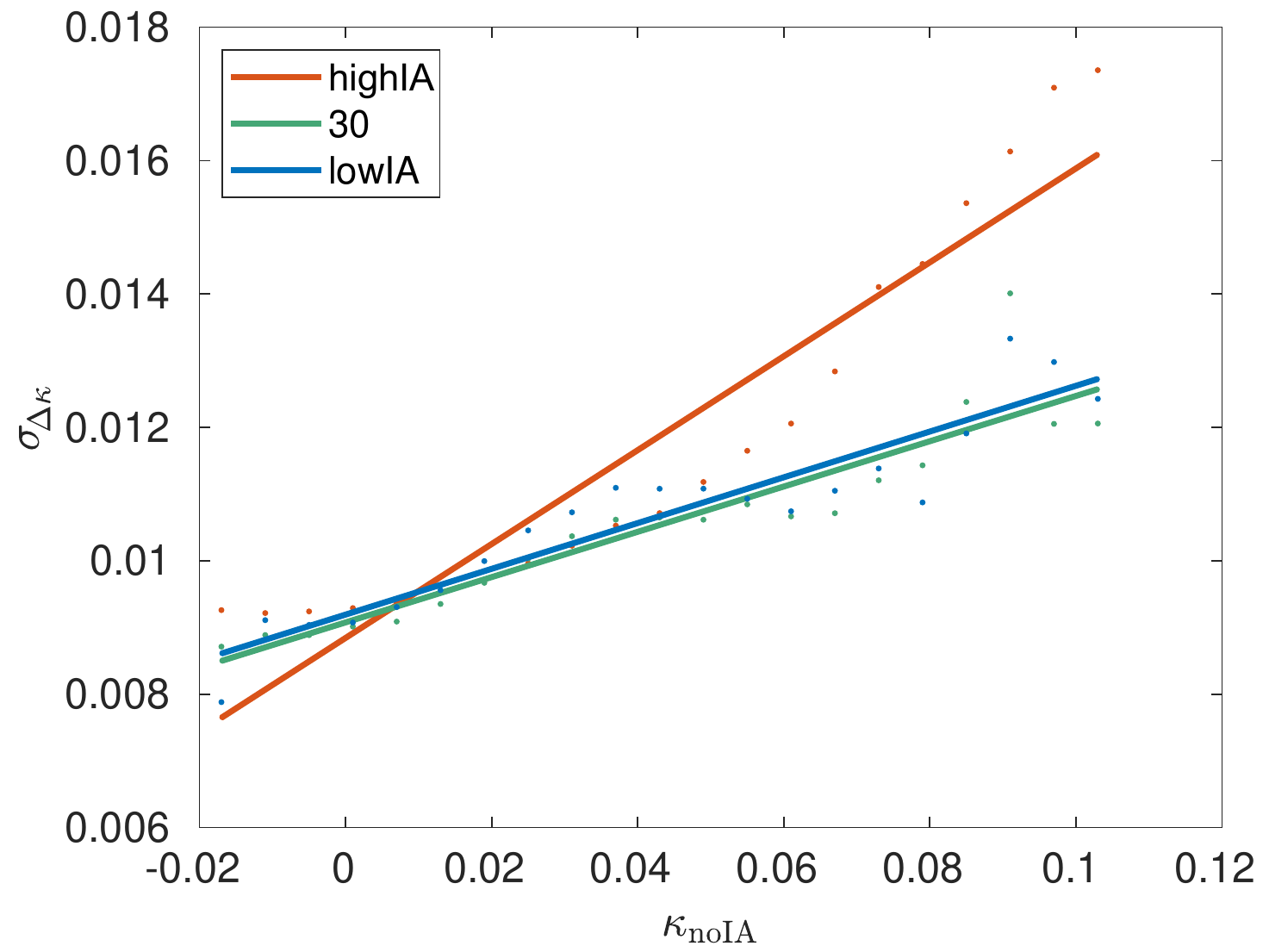}
}
\caption{The average peak height shifts $\Delta\kappa$ (upper panel) and the scatters $\sigma_{\Delta\kappa}$ (lower panel) in $\kappa_{\rm noIA}$ bins for the three considered cases: highIA (red), $30^{\circ}$ (green) and lowIA (blue). 
Dots are the results measured from the data and the solid lines are the fitting results using the parameters listed in Table \ref{tab:polyfit} as described in the main text.}
\label{fig:fitform}
\end{figure}

In Figure \ref{fig:fitform}, we show $\Delta\kappa$ and $\sigma_{\Delta\kappa}$ for the matched peaks, where the points are from simulations, and the lines are the corresponding fits. We adopt a functional form of $\Delta\kappa=A+B\kappa_{\rm{noIA}}+C\kappa_{\rm{noIA}}^2+D\kappa_{\rm{noIA}}^3$ and 
a form of $\sigma_{\Delta\kappa}=a+b \kappa_{\rm{noIA}}$ for the fits. For the three considered IA cases, we obtain the fitting parameters as listed in Table \ref{tab:polyfit}.

With the information above, we proceed to generate peak counts in the IA cases from the ones of noIA. Specifically, for each bin of $\kappa_{\rm{noIA}}$, we select a number of peaks based on the fraction of matched peaks shown in Figure \ref{fig:matching}.
For each of these peaks, we then assign it a peak height $\kappa_{\rm{IA}}$ using a Gaussian probability distribution given as 
\begin{equation}
 p(\kappa_{\rm{IA}})\propto \exp\bigg[-\frac{(\kappa_{\rm{IA}}-\kappa_{\rm{noIA}}-\Delta\kappa)^2}{2\sigma^2_{\Delta\kappa}}\bigg]\quad .
\label{eq:fitdis}
\end{equation}
For the non-matched peaks, we keep their peak heights as those in noIA. 

\begin{table}
\centering
\caption{Fitting parameters of $\Delta\kappa$ and $\sigma_{\Delta\kappa}$}
\label{tab:polyfit}
\begin{tabular}{c c c c c c}
\toprule
 & highIA & 30 & 010 & 060 & lowIA \\
\hline
{A}& 0.0088& 0.0073 & 0.0074 & 0.0086 & 0.0071\\
{B}& -0.6981& -0.6637 & -0.5947 & -0.7088 & -0.6563\\
{C}& 3.1884& 8.0604 & 8.5450 & 4.1635 & 8.6635\\
{D}& -3.2230& -29.7478 & -35.6911 & -10.4547 & -36.7084\\
{a}& 0.0088& 0.0091 & 0.0092 & 0.0090 & 0.0094\\
{b}& 0.0704& 0.0340 & 0.0343 & 0.0591 & 0.0189\\
\hline
\end{tabular}
\end{table}

\begin{figure}
\gridline{\plotone{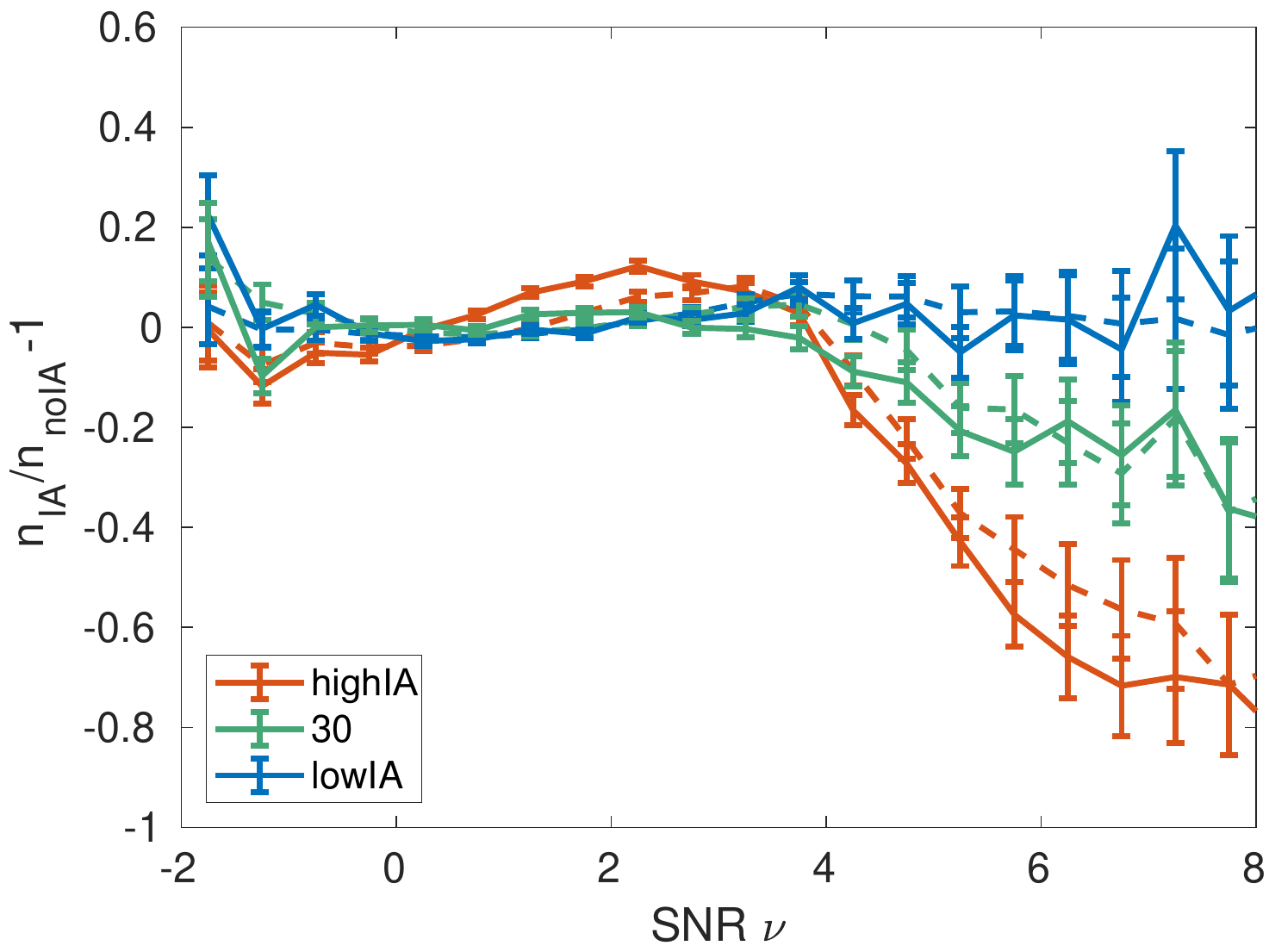}
}
\gridline{\plotone{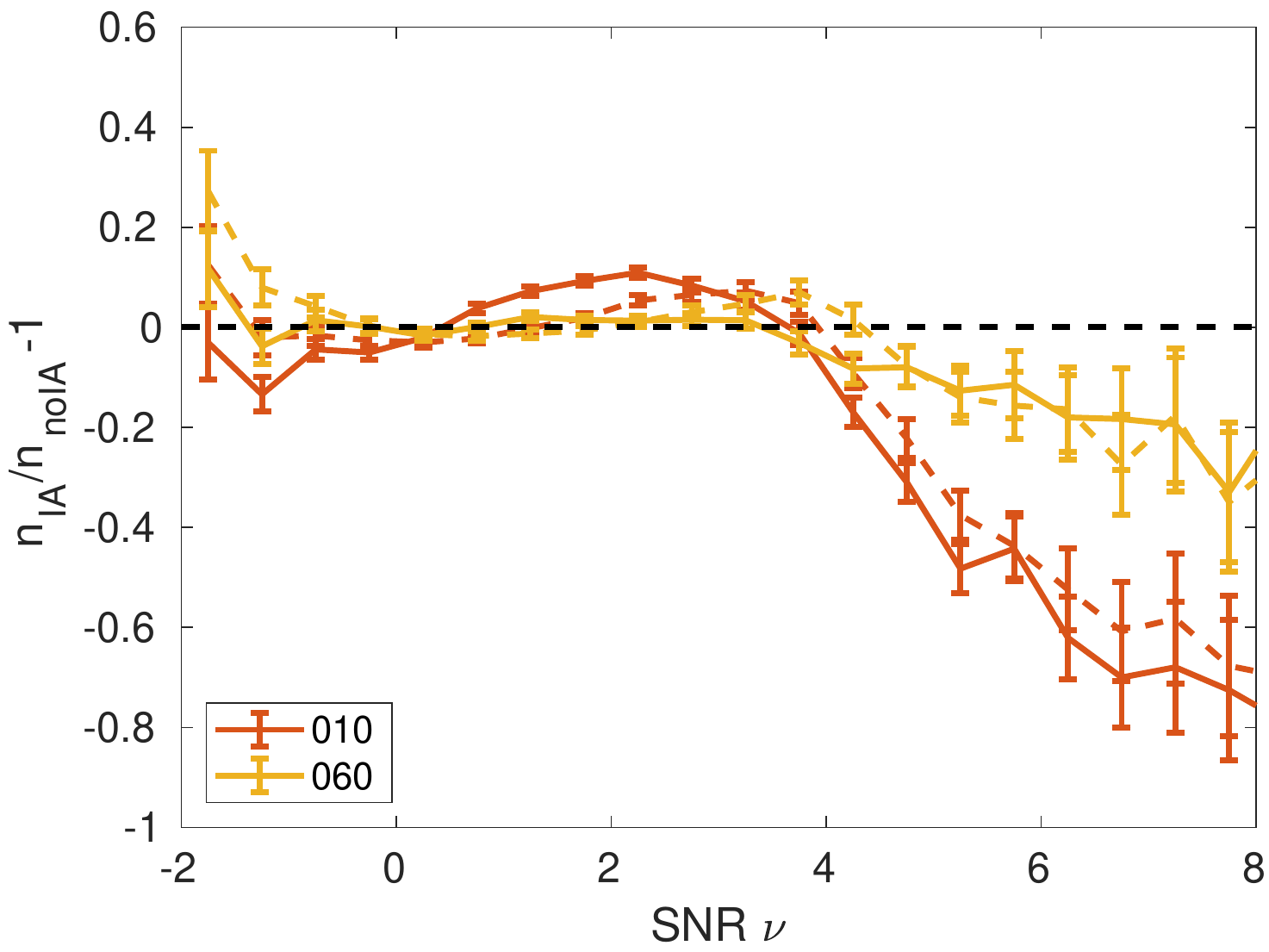}
}
\caption{The results from the empirical fitting model (dashed lines) in comparison with those from simulations (solid lines). 
The upper panel is for highIA (red), $\theta_{\rm{3D}}=30^{\circ}$ (green),
and lowIA (blue) cases, and the lower panel is for $\theta_0=0^{\circ}$ and $\sigma_{\theta}=10^{\circ}$ (red) and $60^{\circ}$ (orange), respectively.
\label{fig:pkfit}}
\end{figure}

The results are shown in the upper panel of Figure \ref{fig:pkfit}, where the solid and dashed lines are from simulations and the empirical fitting model, respectively. It is seen that within the statistical uncertainties, the fitting model can recover well the simulation results with IA for $\nu>\sim 3$.  
This provides a possible way to generate templates of peak counts with IA effects from simulations without IA. 
Following the same procedures, we also obtain fitting models for the samples with $\theta_0=0^{\circ}$ and $\sigma_{\theta}=10^{\circ}$ and $60^{\circ}$, and the results are presented in the lower panel of Figure \ref{fig:pkfit}.
For these cases, the fitting models also perform well.


\section{Summary and discussion} \label{sec:discussion}

In this study, we investigate systematically the impacts of galaxy IA on WL peak statistics using simulations with semi-analytical galaxy formation. The main conclusions are as follow.

\begin{itemize}
    \item The satellite IAs affect WL peak statistics significantly if they are included in a shear sample. When the 3-D alignment angle of satellite galaxies with respect to the radial direction toward their centrals $\theta_{\rm 3D}<\sim 40^{\circ}$, their projected 2-D orientations have a radial tendency on average, and can suppress high peak counts significantly.
    \item For averagely radially aligned satellite samples with $\theta_0=0^{\circ}$ and different misalignments represented by the dispersion $\sigma_{\theta}$, we investigate systematically 
the dependence on $\sigma_{\theta}$ of the IA impacts on peak statistics. By analyzing the II correlations, we derive a relation between $\gamma_{\rm{scale}}$ and $\sigma_{\theta}$, and find that
$\gamma_{\rm{scale}}\approx 0.21$ corresponds to $\sigma_{\theta}\approx 70^{\circ}$. At this $\sigma_{\theta}$, $n_{\rm IA}/n_{\rm noIA}-1\sim 5\%$ and $25\%$ for $\nu\sim 5$ and $8$, respectively, for the {\it Euclid}-like sample.
    \item The IA effects on WL peak counts depend on the redshift distribution of a shear sample. A sample including more satellite galaxies suffers more from the IA impacts. 
    \item For narrow-redshift-bin samples with minimal satellites from foreground clusters, the IA effects mainly arise due to their contributions to the shape noise. On the one hand, this changes the noise parameters. On the other hand it also leads to a scaling factor $\sigma_{0,\rm{noIA}}/\sigma_{0,\rm{IA}}$ in the peak S/N binning in model calculations when comparing with observed peak counts. This is because in real observations, we normally estimate the noise rms by randomly rotating galaxies in a shear sample, and thus only $\sigma_{0,\rm{noIA}}$ can be obtained and used in defining peak S/N. As a result, high peak counts are enhanced instead of suppressed in such cases. 
    \item The IA impacts also depend on the smoothing scale used in WL peak analyses. A larger smoothing scale tends to smooth out the satellite IA effects because more galaxies outside a cluster contribute to the reconstruction of the cluster convergence from the shears.
    \item We build an empirical fitting model to calculate peak counts with IA effects from the ones in noIA. This can be useful in generating templates of peak counts with IA from noIA simulations so that the IA effects can be included in deriving cosmological constraints from observed peak counts.
\end{itemize}

In \citet{Kacprzak2016}, they point out the importance of satellite IA on peak statistics, and evaluate their effects by considering radially aligned satellites with a particular scaling factor $\bar {\gamma}_{\rm scale}=0.21$. In comparison, here we construct samples with 
different satellite IA directly from simulations and perform systematic studies of the IA impacts. Because of the differences in the specific analyses, it is not straightforward to compare quantitatively our results with theirs. Qualitatively, 
similar behaviors are seen in the two studies with the IA effects stronger for higher peaks. Our systematic analyses are able to reveal the dependence of the IA impacts on the satellite alignment angles. 

For shear samples in narrow redshift bins, the strong suppression for high peak counts does not occur which is also mentioned in \citet{Kacprzak2016}. Our studies provide quantitative results of the IA effects on peak statistics for samples with different redshift distributions and demonstrate the impacts from the IA contributions to the shape noise
in the cases with source galaxies in narrow redshift bins.  

In comparison with the analyses of \citet{Joachim2022} using the infusion method to generate IA signals of galaxies based on the tidal field calculations, our settings to adjust the satellite IA to different $\theta_{\rm 3D}$ are, to a certain extent, effectively similar to adjust the overall IA amplitude parameter $A_{\rm IA}$ used in their studies.
The advantage of our simulations with semi-analytical galaxy formation is that we have galaxies of different types, particularly centrals and satellites, and their clustering is naturally included. 
We therefore are able to treat centrals and satellites separately and single out the effects of satellite IA, instead of by adjusting the overall $A_{\rm IA}$ without distinguishing different types of galaxies.  

For the IA impacts, \citet{Joachim2022} show that for tomographic peak analyses, the IA effects for the cases of source galaxies in auto-bins and cross-bins show different behaviors with the strong high peak suppression occurring in cross-bin samples. This is qualitatively consistent with our results because cross-bins contain a low redshift bin and a high redshift bin. Clusters in the low redshift bin contribute to shear signals to galaxies in the high redshift bin leading to high peaks in the case of cross-bins. Meanwhile, the satellites of these clusters are in the cross-bin sample and their IA affects high peaks significantly. On the other hand, the cases of their auto-bins are close to our narrow-redshift-bin cases where the IA effects come from the IA contributions to noise. 

We note that the inclusion of cluster satellites in a shear sample can affect WL peak signals even without IA. This arises because these satellites do not carry WL signals of their host clusters. When they are included in generating convergence or aperture mass maps, they tend to dilute the WL signals of clusters. In addition, the clustering of the satellites boost the local density of galaxies around a cluster leading to a smaller noise comparing to the average, which also affects the peak counts \citep{Kacprzak2016, Shan2018, Martinet2018, Zurcher2022}. 
If satellites have IA, they lead to additional effects on peak heights and thus peak statistics. In our analyses shown here, the positions of satellite galaxies are exactly the same in the noIA and IA cases, and thus the differences $n_{\rm IA}/n_{\rm noIA}-1$ arise solely from their IA signals. 

In real observational analyses, the dilution and the IA effects are correlated and need to be considered simultaneously. In \citet{Kacprzak2016}, they consider the two statistically by adjusting the S/N of peaks. In \citet{Shan2018}, we include the satellite dilution and the noise change due to the boosted number density around clusters in our theoretical high-peak model in deriving cosmological constraints from the observed peak counts using KiDS data. In our future studies, we will further develop the model to include the IA effects. For that, we need to carefully analyze the IA effects on WL peak signals at the level of individual clusters with different mass and at different redshifts. We note that the satellite IA effects can be regarded as strong GI effects. In addition, galaxies in a shear sample other than these satellites can also have GI signals. Together with the II term, they change the noise properties which should also be included in the modeling. The success of such a model and its application to WL observations can not only mitigate the systematic biases on cosmological parameters due to the IA effects but also allow us to extract the satellite IA information simultaneously from observed WL high peaks.

\begin{acknowledgements}
\modulolinenumbers[15]
This research is supported by the NSFC grant No.11933002. Z.H.F. also acknowledges the supports from the NSFC grant No. U1931210, the grant from the China Manned Space Projects with No. CMS-CSST-2021-A01, and a grant from CAS Interdisciplinary Innovation Team. X.K.L. is also supported by NSFC of China under Grant No. 11803028 and No. 12173033, YNU Grant No. C176220100008, and the grant from the China Manned Space Project with No. CMS-CSST-2021-B01. G.L.L. acknowledges the supports from the NSFC grant No. U1931210. C.L.W. acknowledges the supports from the NSFC grant No. 11903082 and the grant from the China Manned Space Projects with No. CMS-CSST-2021-A03.
\end{acknowledgements}

\appendix

\section{The $\sigma_{\theta}$-dependence of $\gamma_{\rm{scale}}$\label{appendix:AIAfit}}
In \cite{SB2010}, they propose to model small-scale IA by placing satellites into dark matter halos. By introducing a scaling factor $\gamma^2_{\rm{scale}}$ with respect to the II correlations from perfectly radially aligned satellites,
the misalignment of satellites can be accounted for. Adding the NLA modeling for large scales \citep{HS04, NLA} gives rise to the IA contributions to the cosmic shear 2PCF.

\begin{figure}
\plotone{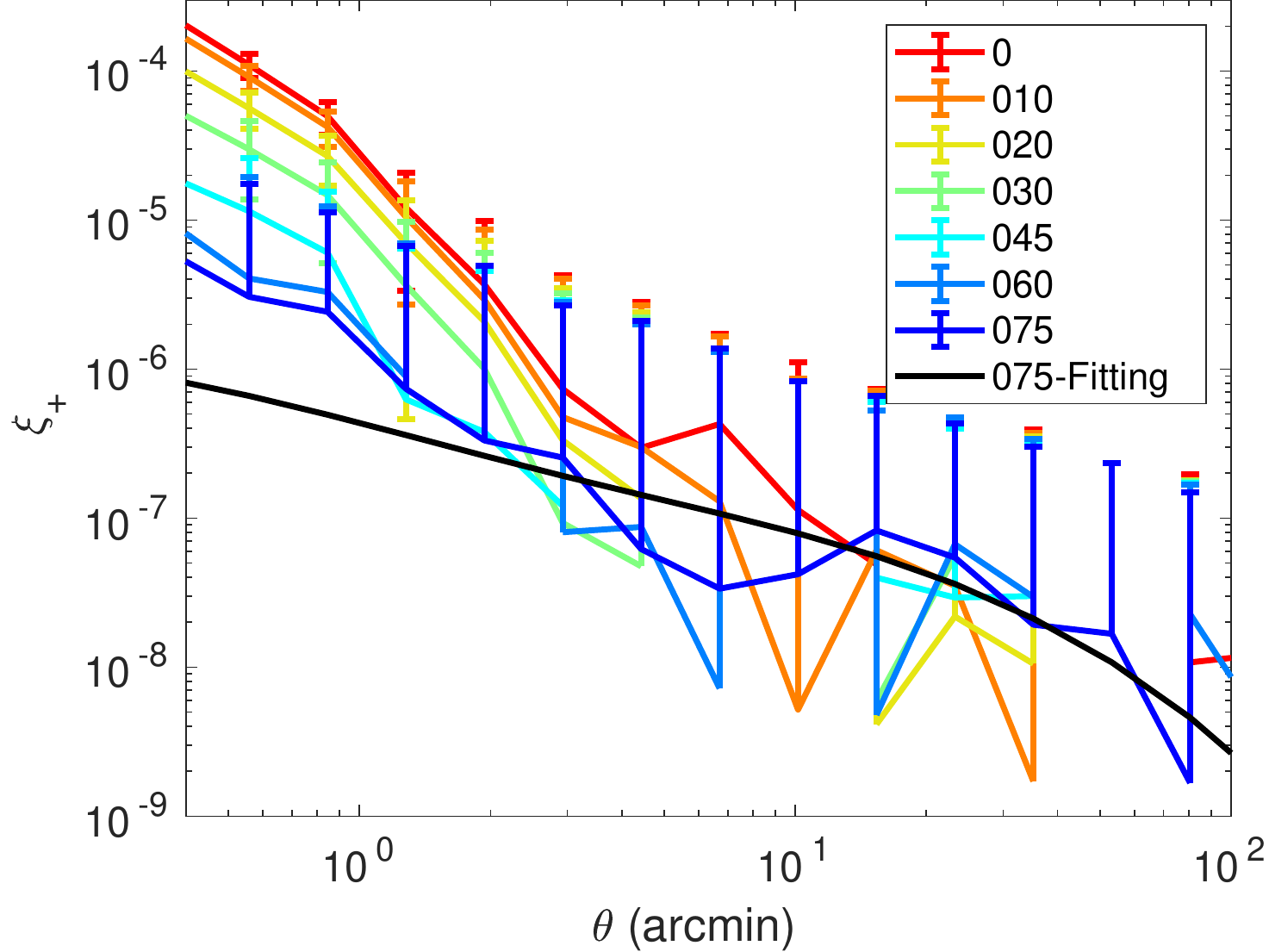}
\caption{II correlations measured from our IA samples with $\theta_0=0^{\circ}$ and different $\sigma_{\theta}$ without adding shear signals. Colored lines with error bars are the results of different $\sigma_{\theta}$ as indicated in the legend, where
e.g., 0 for $\sigma_{\theta}=0^{\circ}$ and 030 for $\sigma_{\theta}=30^{\circ}$. The black line is the theoretical NLA prediction with $A_{\rm IA}=0.98$.
\label{fig:AIAfit}}
\end{figure}

In our study here, we consider satellite IA with the average $\theta_0=0^{\circ}$ and different dispersions $\sigma_{\theta}$. This is well in line with the model of \cite{SB2010}. To derive the relation between $\gamma_{\rm{scale}}$ and $\sigma_{\theta}$,
we calculate the II correlations from our IA samples without adding shear signals. The results are shown in Figure \ref{fig:AIAfit}.
Different data points with error bars are for different $\sigma_{\theta}$.

It is seen that at the angular separation larger than about $5\hbox{ arcmin}$, the II correlations are about the same
for different $\sigma_{\theta}$ because at these scales, the central-central galaxy II correlations play dominant roles, which are the same for all the samples considered here. They can be well described by the NLA model given by \citep{HS04,NLA}
\begin{equation}
\label{eq:II}
    P_{II}(k,z)=F^2(z)P_m^{NL}(k,z).
\end{equation}
Here $P_m^{NL}(k,z)$ is the nonlinear matter power spectrum, and
\begin{equation}
\label{eq:AIA}
    F(z)=-A_{\rm IA}C_1\rho_{\rm crit}\frac{\Omega_{\rm m}}{D_+(z)}(\frac{1+z}{1+z_0})^{\eta_{\rm eff}},
\end{equation}
where $A_{\rm IA}$ is the amplitude parameter, $\rho_{\rm crit}$ is the current critical density and $D_+(z)$ is the linear growth factor.

By setting $C_1=5\times10^{-14}h^{-2}M^{-1}_\odot{\rm Mpc}^3$ and $\eta_{\rm eff}=0$ \citep{Wei} and taking into account the {\it Euclid}-like redshift distribution, we fit the NLA model to the II correlation data points at $\theta\ge 5\hbox{ arcmin}$ to obtain $A_{\rm IA}$.
The black line in Figure \ref{fig:AIAfit} is the theoretical result with $A_{\rm IA}=0.98$ fitted from the data point with $\sigma_{\theta}=75^{\circ}$. It is seen clearly while the NLA model can describe the large-scale II correlations well,
the small-scale correlations show different behaviors, in agreement with other studies \citep[e.g.,][]{SB2010,Blazek2019,Samuroff2021}.

\begin{figure}
\plotone{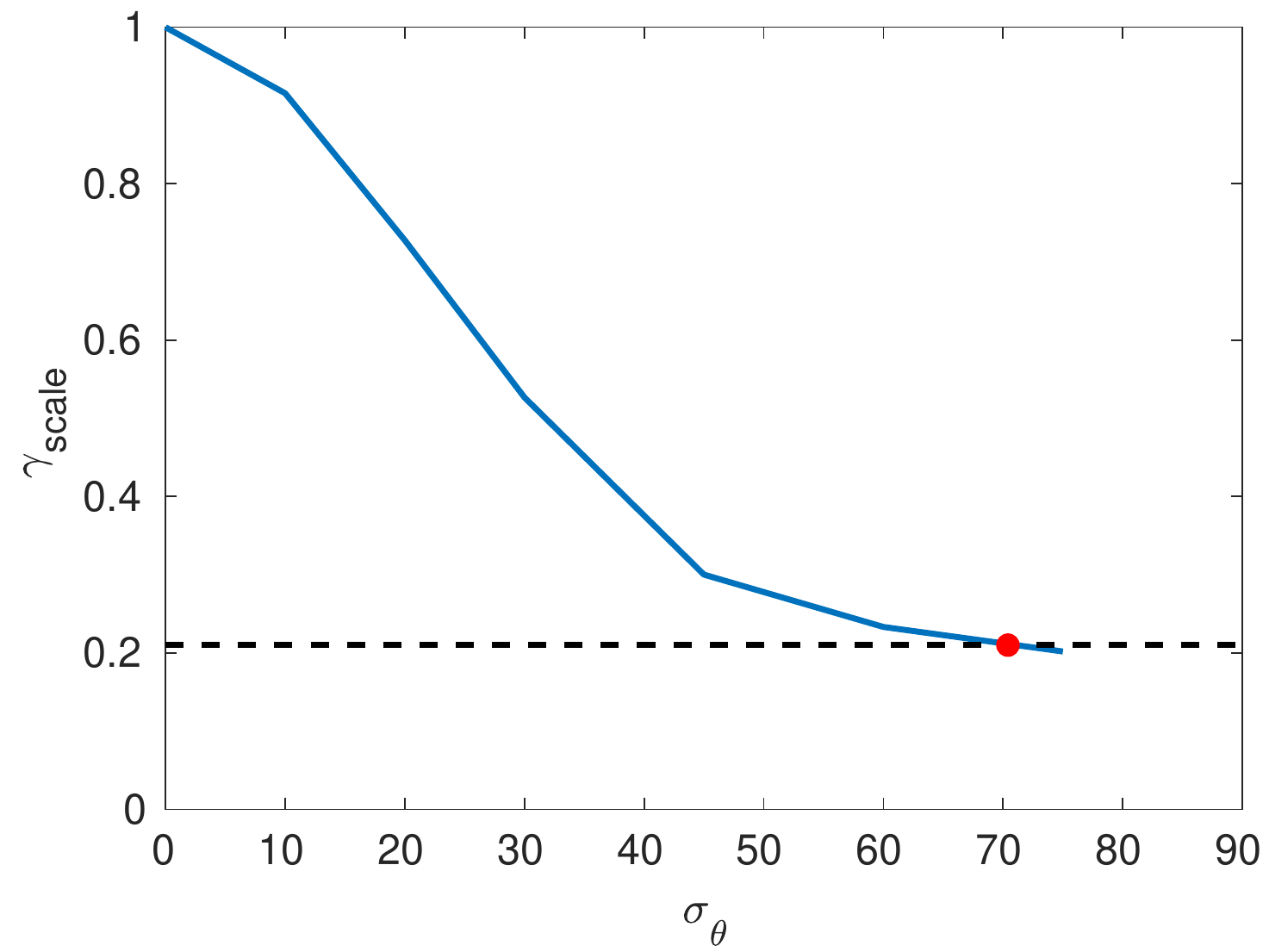}
\caption{The dependence of $\gamma_{\rm{scale}}$ on $\sigma_{\theta}$. The horizontal dashed line indicates $\gamma_{\rm{scale}}=0.21$ and the red point marks its intersection with the blue line.
\label{fig:gscale}}
\end{figure}

On small scales, the correlations are from one-halo satellite IA, and the differences are apparent for different $\sigma_{\theta}$. Following \cite{SB2010},
we compute $\gamma_{\rm{scale}}$ by assuming $\hbox{II}(\sigma_{\theta})=\gamma_{\rm{scale}}^2\times \hbox{II}(\sigma_{\theta}=0)$ and by using the II correlation data points at the angular separation less than $2\hbox{ arcmin}$. The derived relation between $\gamma_{\rm{scale}}$ and $\sigma_{\theta}$
is shown in Figure \ref{fig:gscale}. The horizontal black dashed line indicates $\gamma_{\rm{scale}}=0.21$ from \cite{Knebe2008}, and the red point shows its intersection with the blue solid line which is at $\sigma_{\theta}\approx 70^{\circ}$. We mark this $\sigma_{\theta}$ in the 
lower panel of Figure \ref{fig:radialdispersion}.

\section{The 2-D Gaussian peak model\label{appendix:model}}
For a 2-D Gaussian random field, the peak density distribution function is given by \citep{bond, vanWaer2000, Fan2007}

\begin{figure*}[ht!]
\gridline{\fig{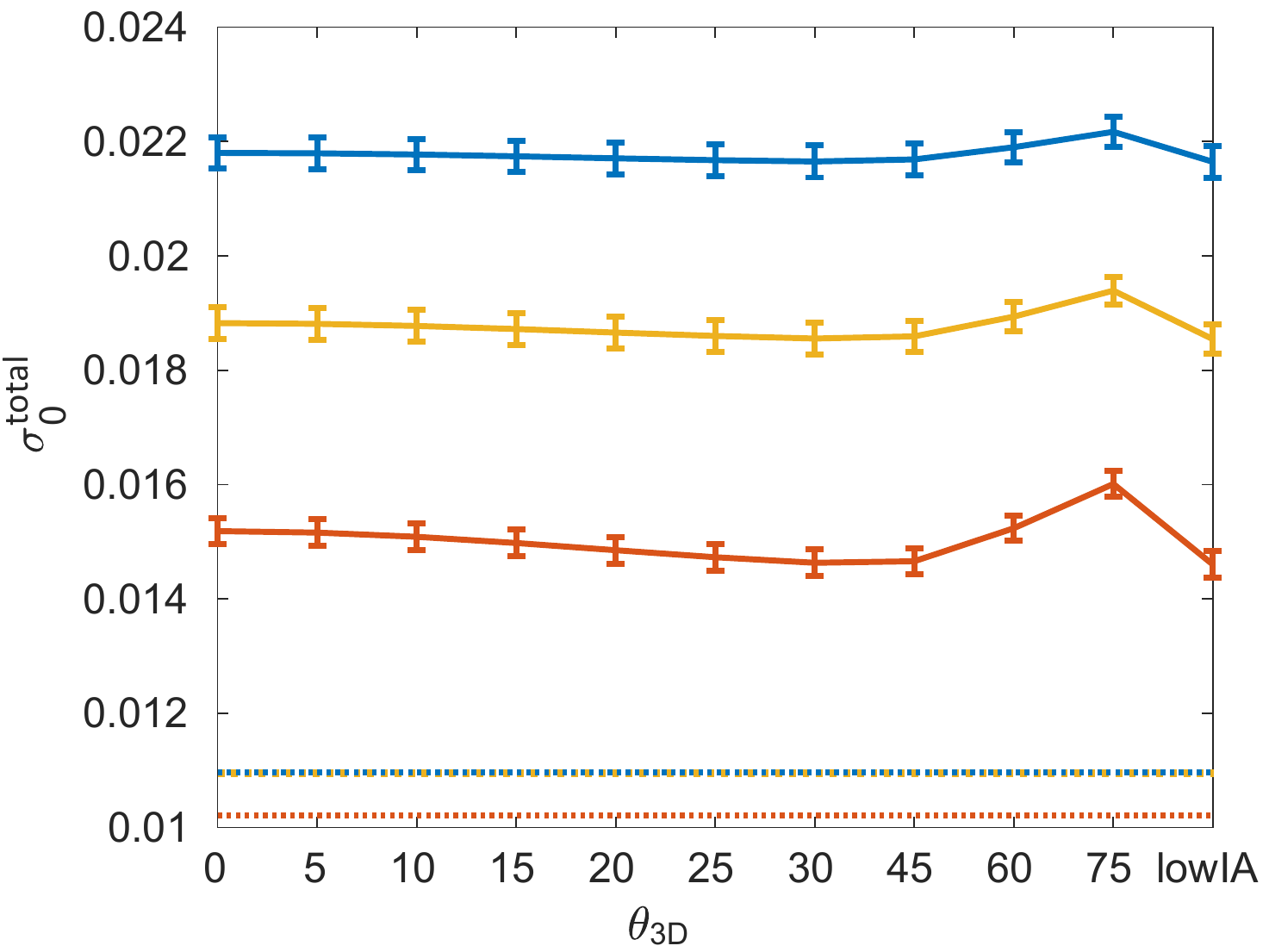}{0.32\textwidth}{}
          \fig{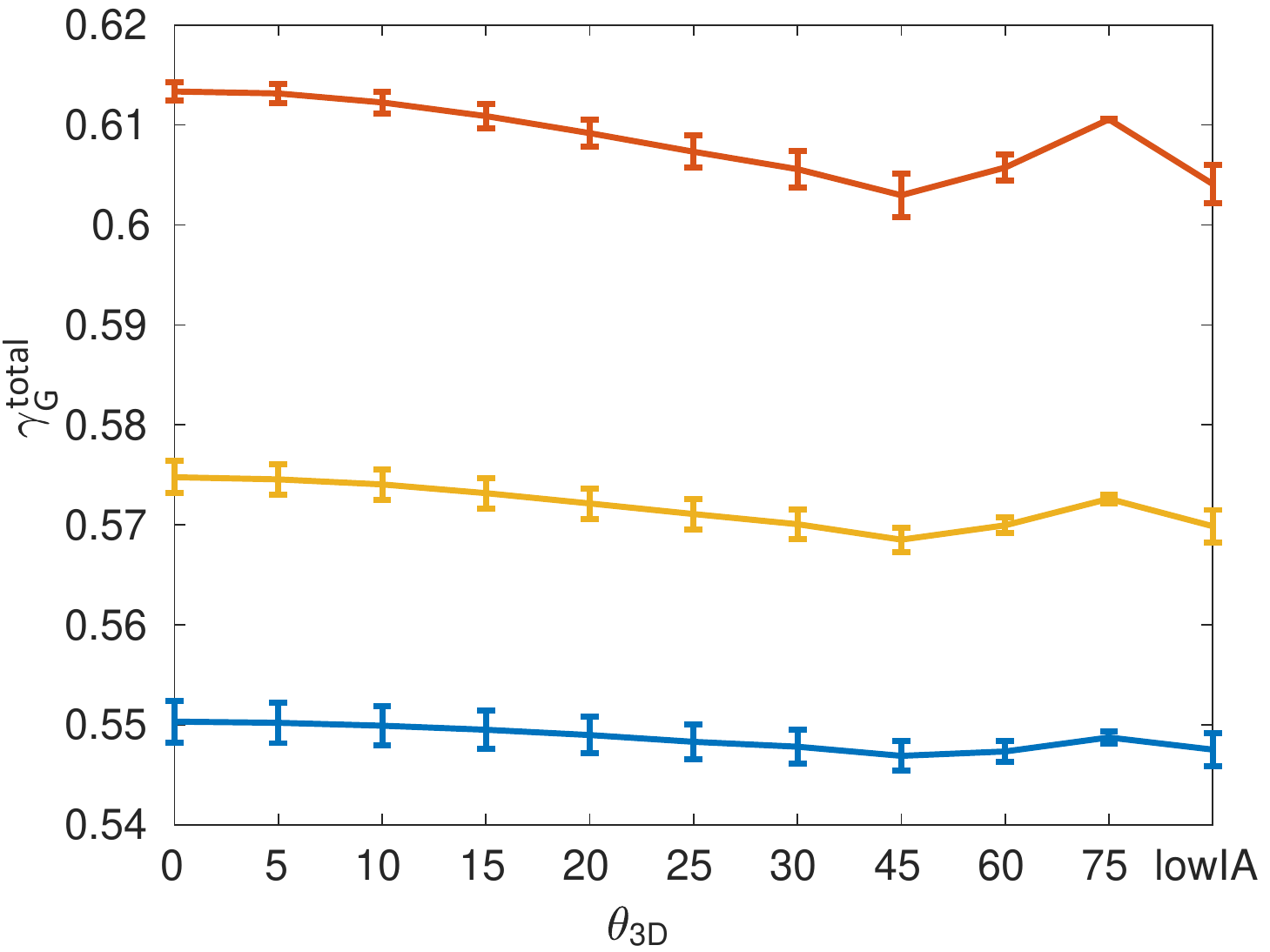}{0.32\textwidth}{}
          \fig{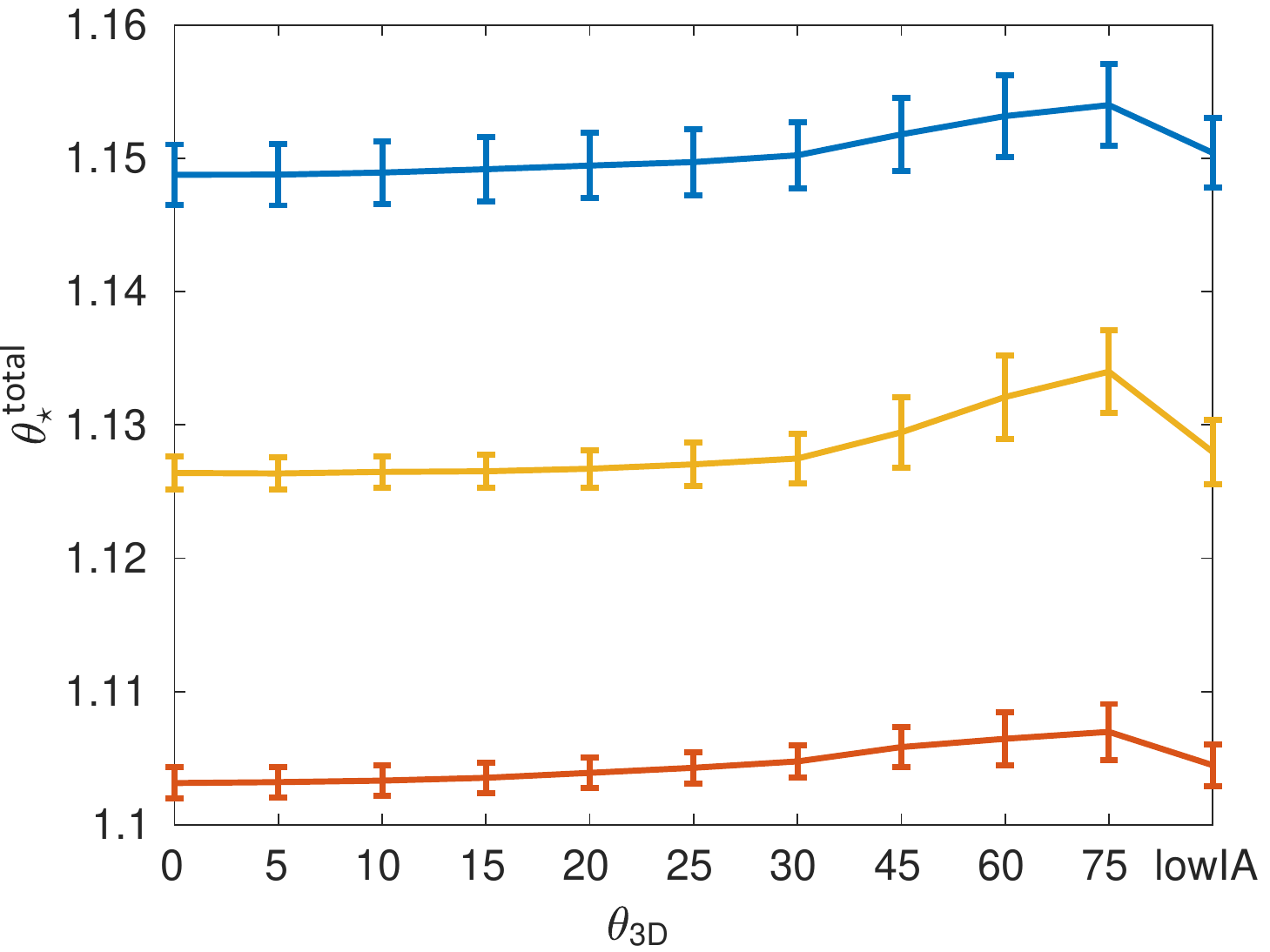}{0.32\textwidth}{}
}
\caption{Noise field parameters for the three narrow-redshift-bin samples with the red, orange and blue lines for $z\sim 1$, $\sim 1.5$ and $\sim 2$, respectively. From left to right: $\sigma^{\rm {total}}_{0}$, $\gamma^{\rm{total}}_G$, and $\theta^{\rm{total}}_\star$. The dotted lines in the left panel indicate $\sigma_{0,\rm{noIA}}$ of the three samples.
\label{fig:peakprop}}
\end{figure*}

\begin{equation}
\label{eq:bp}
    N(\nu)d\nu=\frac{1}{2\pi\theta_\star^2}\rm{exp}(-\nu^2/2)\frac{d\nu}{(2\pi)^{1/2}}G(\gamma_G,\gamma_G\nu)
\end{equation}
where $\nu=\kappa/\sigma_0$ is the signal-to-noise ratio, $\theta_\star=\sqrt{2}\sigma_1/\sigma_2$ and $\gamma_G=\sigma_1^2/(\sigma_0\sigma_2)$. The function $G$ is

\begin{equation}
\label{eq:G}
\begin{split}
    G(\gamma_G,x)=\frac{1}{2}(x^2+\frac{b^2}{2}-1){\rm erfc}(-\frac{x}{b})+\frac{xb}{2\sqrt{\pi}}{\rm exp}(-\frac{x^2}{b^2})\\+\frac{1}{2\sqrt{1+b^2}}{\rm exp}(-\frac{x^2}{1+b^2}){\rm erfc}(-\frac{x}{b\sqrt{1+b^2}}).
\end{split}
\end{equation}
with $b=[2(1-\gamma_G^2)]^{1/2}$.
The moment $\sigma_i$ of the field $N$ is given by
\begin{equation}
    \sigma_i^2=\int{d\Vec{k}k^{2i}\langle|N(k)|^2\rangle}
\end{equation}
where $|N(k)|^2$ is the power spectrum of $N$.

In the case of noIA, we have $\sigma_{0,\rm {noIA}}=\sigma_{\epsilon}/\sqrt{4\pi \theta_G^2n_g}$ under the Gaussian smoothing, and $\sigma_{0,\rm {noIA}}:\sigma_{1,\rm {noIA}}:\sigma_{2,\rm {noIA}}=1:\sqrt{2}/\theta_{G}:2\sqrt{2}\theta_{G}^2$. This leads to $\gamma_{G,\rm {noIA}}=\sqrt{2}/2$ and
$\theta_{*,\rm {noIA}}=\theta_G/\sqrt{2}$. The existence of IA gives rise to additional contributions to $\sigma_i$ with $\sigma_{i,\rm{IA}}=(\sigma^2_{i,\rm{noIA}}+\sigma^2_{i,\rm{corr}})^{1/2}$ where $\sigma^2_{i,\rm{corr}}$ is from IA. For illustrations here,
for the three narrow-redshift-bin samples, we estimate $\sigma_{i,\rm{IA}}$ directly from the galaxy ellipticities including IA but without adding lensing signals. Thus the II correlation term is accounted for, but the GI part is missed. For the considered samples in narrow redshift bins,
we expect minimal GI signals. In general for a sample with a redshift distribution, the GI contributions to noise should be included.

Besides the shape noise, to some extent, the WL effects from large-scale structures can also be described as a random field. In \citet{Yuan2018} in modeling high peak counts, we consider the massive halos as the dominant sources for high peaks. The additional WL contributions from large-scale structures
other than those massive halos (LSS) are modeled as a Gaussian random field. Although this Gaussian assumption is an approximation, it can model well the LSS contributions to high peak statistics \citep{Yuan2018}. Here, for the purpose of qualitative illustrations, we also include LSS as a Gaussian random field. Thus the total $\sigma^{\rm {total}}_{i,{\rm noIA}}$
becomes $(\sigma^2_{i,\rm{noIA}}+\sigma^2_{i,\rm{LSS}})^{1/2}$ and $\sigma^{\rm {total}}_{i,\rm{IA}}=(\sigma^2_{i,\rm{noIA}}+\sigma^2_{i,\rm{LSS}}+\sigma^2_{i,\rm{corr}})^{1/2}$, where $\sigma^2_{i,\rm{LSS}}$ are calculated following the approach described in \citet{Yuan2018}.

In Figure \ref{fig:peakprop}, we show the dependence of the noise parameters on $\theta_{\rm{3D}}$ for the three narrow-redshift-bin samples with the red, orange and blue lines for $z\sim 1$, $\sim 1.5$ and $\sim 2$, respectively. The left, middle and right panels are for $\sigma^{\rm{total}}_0$, $\gamma^{\rm{total}}_G$ and
$\theta^{\rm{total}}_*$, respectively. The dotted lines in the left panel are $\sigma_{0,\rm{noIA}}$ of the three samples, and the vertical shifts of the solid lines from their corresponding dotted lines are due to $\sigma_{0, {\rm LSS}}$, which is redshift dependent but independent of IA.

Besides the changes of the noise parameters,  in accord with real observations, we use $\sigma_{0,\rm{noIA}}$ to define $\nu$ in our simulation analyses for all the cases. To be consistent, we need to scale $\nu$ to $(\sigma_{0,\rm{noIA}}/\sigma^{\rm{total}}_{0,\rm{noIA}})\nu$ and $(\sigma_{0,\rm{noIA}}/\sigma^{\rm{total}}_{0,\rm{IA}})\nu$
in the noIA and IA cases, respectively, in calculating the peak distributions using Eq.(\ref{eq:bp}). This also affects peak statistics.

With the above information, we calculate the peak distributions for the narrow-redshift-bin samples, and the results are shown in the lower panels of Figure \ref{fig:redpk}.


\bibliography{ms}{}

\begin{thebibliography}{}
\expandafter\ifx\csname natexlab\endcsname\relax\def\natexlab#1{#1}\fi
\providecommand{\url}[1]{\href{#1}{#1}}
\providecommand{\dodoi}[1]{doi:~\href{http://doi.org/#1}{\nolinkurl{#1}}}
\providecommand{\doeprint}[1]{\href{http://ascl.net/#1}{\nolinkurl{http://ascl.net/#1}}}
\providecommand{\doarXiv}[1]{\href{https://arxiv.org/abs/#1}{\nolinkurl{https://arxiv.org/abs/#1}}}

\bibitem[{{Abbott} {et~al.}(2021){Abbott}, {Adam{\'o}w}, {Aguena}, {Allam},
  {Amon}, {Annis}, {Avila}, {Bacon}, {Banerji}, {Bechtol}, {Becker},
  {Bernstein}, {Bertin}, {Bhargava}, {Bridle}, {Brooks}, {Burke}, {Carnero
  Rosell}, {Carrasco Kind}, {Carretero}, {Castander}, {Cawthon}, {Chang},
  {Choi}, {Conselice}, {Costanzi}, {Crocce}, {da Costa}, {Davis}, {De Vicente},
  {DeRose}, {Desai}, {Diehl}, {Dietrich}, {Drlica-Wagner}, {Eckert},
  {Elvin-Poole}, {Everett}, {Evrard}, {Ferrero}, {Fert{\'e}}, {Flaugher},
  {Fosalba}, {Friedel}, {Frieman}, {Garc{\'\i}a-Bellido}, {Gaztanaga},
  {Gelman}, {Gerdes}, {Giannantonio}, {Gill}, {Gruen}, {Gruendl}, {Gschwend},
  {Gutierrez}, {Hartley}, {Hinton}, {Hollowood}, {Honscheid}, {Huterer},
  {James}, {Jeltema}, {Johnson}, {Kent}, {Kron}, {Kuehn}, {Kuropatkin},
  {Lahav}, {Li}, {Lidman}, {Lin}, {MacCrann}, {Maia}, {Manning}, {Maloney},
  {March}, {Marshall}, {Martini}, {Melchior}, {Menanteau}, {Miquel}, {Morgan},
  {Myles}, {Neilsen}, {Ogando}, {Palmese}, {Paz-Chinch{\'o}n}, {Petravick},
  {Pieres}, {Plazas}, {Pond}, {Rodriguez-Monroy}, {Romer}, {Roodman}, {Rykoff},
  {Sako}, {Sanchez}, {Santiago}, {Scarpine}, {Serrano}, {Sevilla-Noarbe},
  {Smith}, {Smith}, {Soares-Santos}, {Suchyta}, {Swanson}, {Tarle}, {Thomas},
  {To}, {Tremblay}, {Troxel}, {Tucker}, {Turner}, {Varga}, {Walker},
  {Wechsler}, {Weller}, {Wester}, {Wilkinson}, {Yanny}, {Zhang}, {Nikutta},
  {Fitzpatrick}, {Jacques}, {Scott}, {Olsen}, {Huang}, {Herrera}, {Juneau},
  {Nidever}, {Weaver}, {Adean}, {Correia}, {de Freitas}, {Freitas},
  {Singulani}, {Vila-Verde}, \& {Linea Science Server}}]{Abbott2021}
{Abbott}, T.~M.~C., {Adam{\'o}w}, M., {Aguena}, M., {et~al.} 2021, \apjs, 255,
  20, \dodoi{10.3847/1538-4365/ac00b3}

\bibitem[{{Abbott} {et~al.}(2022){Abbott}, {Aguena}, {Alarcon}, {Allam},
  {Alves}, {Amon}, {Andrade-Oliveira}, {Annis}, {Avila}, {Bacon}, {Baxter},
  {Bechtol}, {Becker}, {Bernstein}, {Bhargava}, {Birrer}, {Blazek},
  {Brandao-Souza}, {Bridle}, {Brooks}, {Buckley-Geer}, {Burke}, {Camacho},
  {Campos}, {Carnero Rosell}, {Carrasco Kind}, {Carretero}, {Castander},
  {Cawthon}, {Chang}, {Chen}, {Chen}, {Choi}, {Conselice}, {Cordero},
  {Costanzi}, {Crocce}, {da Costa}, {da Silva Pereira}, {Davis}, {Davis}, {De
  Vicente}, {DeRose}, {Desai}, {Di Valentino}, {Diehl}, {Dietrich}, {Dodelson},
  {Doel}, {Doux}, {Drlica-Wagner}, {Eckert}, {Eifler}, {Elsner}, {Elvin-Poole},
  {Everett}, {Evrard}, {Fang}, {Farahi}, {Fernandez}, {Ferrero}, {Fert{\'e}},
  {Fosalba}, {Friedrich}, {Frieman}, {Garc{\'\i}a-Bellido}, {Gatti},
  {Gaztanaga}, {Gerdes}, {Giannantonio}, {Giannini}, {Gruen}, {Gruendl},
  {Gschwend}, {Gutierrez}, {Harrison}, {Hartley}, {Herner}, {Hinton},
  {Hollowood}, {Honscheid}, {Hoyle}, {Huff}, {Huterer}, {Jain}, {James},
  {Jarvis}, {Jeffrey}, {Jeltema}, {Kovacs}, {Krause}, {Kron}, {Kuehn},
  {Kuropatkin}, {Lahav}, {Leget}, {Lemos}, {Liddle}, {Lidman}, {Lima}, {Lin},
  {MacCrann}, {Maia}, {Marshall}, {Martini}, {McCullough}, {Melchior},
  {Mena-Fern{\'a}ndez}, {Menanteau}, {Miquel}, {Mohr}, {Morgan}, {Muir},
  {Myles}, {Nadathur}, {Navarro-Alsina}, {Nichol}, {Ogando}, {Omori},
  {Palmese}, {Pandey}, {Park}, {Paz-Chinch{\'o}n}, {Petravick}, {Pieres},
  {Plazas Malag{\'o}n}, {Porredon}, {Prat}, {Raveri}, {Rodriguez-Monroy},
  {Rollins}, {Romer}, {Roodman}, {Rosenfeld}, {Ross}, {Rykoff}, {Samuroff},
  {S{\'a}nchez}, {Sanchez}, {Sanchez}, {Sanchez Cid}, {Scarpine}, {Schubnell},
  {Scolnic}, {Secco}, {Serrano}, {Sevilla-Noarbe}, {Sheldon}, {Shin}, {Smith},
  {Soares-Santos}, {Suchyta}, {Swanson}, {Tabbutt}, {Tarle}, {Thomas}, {To},
  {Troja}, {Troxel}, {Tucker}, {Tutusaus}, {Varga}, {Walker}, {Weaverdyck},
  {Wechsler}, {Weller}, {Yanny}, {Yin}, {Zhang}, {Zuntz}, \& {DES
  Collaboration}}]{Abbott2022}
{Abbott}, T.~M.~C., {Aguena}, M., {Alarcon}, A., {et~al.} 2022, \prd, 105,
  023520, \dodoi{10.1103/PhysRevD.105.023520}

\bibitem[{{Aihara} {et~al.}(2022){Aihara}, {AlSayyad}, {Ando}, {Armstrong},
  {Bosch}, {Egami}, {Furusawa}, {Furusawa}, {Harasawa}, {Harikane}, {Hsieh},
  {Ikeda}, {Ito}, {Iwata}, {Kodama}, {Koike}, {Kokubo}, {Komiyama}, {Li},
  {Liang}, {Lin}, {Lupton}, {Lust}, {MacArthur}, {Mawatari}, {Mineo},
  {Miyatake}, {Miyazaki}, {More}, {Morishima}, {Murayama}, {Nakajima},
  {Nakata}, {Nishizawa}, {Oguri}, {Okabe}, {Okura}, {Ono}, {Osato}, {Ouchi},
  {Pan}, {Plazas Malag{\'o}n}, {Price}, {Reed}, {Rykoff}, {Shibuya},
  {Simunovic}, {Strauss}, {Sugimori}, {Suto}, {Suzuki}, {Takada}, {Takagi},
  {Takata}, {Takita}, {Tanaka}, {Tang}, {Taranu}, {Terai}, {Toba}, {Turner},
  {Uchiyama}, {Vijarnwannaluk}, {Waters}, {Yamada}, {Yamamoto}, \&
  {Yamashita}}]{Aihara2022}
{Aihara}, H., {AlSayyad}, Y., {Ando}, M., {et~al.} 2022, \pasj,
  \dodoi{10.1093/pasj/psab122}

\bibitem[{{Bartelmann}(1995)}]{Bartelmann95}
{Bartelmann}, M. 1995, \aap, 303, 643.
\newblock \doarXiv{astro-ph/9412051}

\bibitem[{{Bartelmann} \& {Schneider}(2001)}]{BS2001}
{Bartelmann}, M., \& {Schneider}, P. 2001, \physrep, 340, 291,
  \dodoi{10.1016/S0370-1573(00)00082-X}

\bibitem[{{Bond} \& {Efstathiou}(1987)}]{bond}
{Bond}, J.~R., \& {Efstathiou}, G. 1987, \mnras, 226, 655,
  \dodoi{10.1093/mnras/226.3.655}

\bibitem[{{Bridle} \& {King}(2007)}]{NLA}
{Bridle}, S., \& {King}, L. 2007, New Journal of Physics, 9, 444,
  \dodoi{10.1088/1367-2630/9/12/444}

\bibitem[{{Dietrich} \& {Hartlap}(2010)}]{Dietrich2010}
{Dietrich}, J.~P., \& {Hartlap}, J. 2010, \mnras, 402, 1049,
  \dodoi{10.1111/j.1365-2966.2009.15948.x}

\bibitem[{{Fan} {et~al.}(2010){Fan}, {Shan}, \& {Liu}}]{Fan2010}
{Fan}, Z., {Shan}, H., \& {Liu}, J. 2010, \apj, 719, 1408,
  \dodoi{10.1088/0004-637X/719/2/1408}

\bibitem[{{Fan}(2007)}]{Fan2007}
{Fan}, Z.~H. 2007, \apj, 669, 10, \dodoi{10.1086/521182}

\bibitem[{{Fu} {et~al.}(2013){Fu}, {Kauffmann}, {Huang}, {Yates}, {Moran},
  {Heckman}, {Dav{\'e}}, {Guo}, \& {Henriques}}]{FuJ2013}
{Fu}, J., {Kauffmann}, G., {Huang}, M.-l., {et~al.} 2013, \mnras, 434, 1531,
  \dodoi{10.1093/mnras/stt1117}

\bibitem[{{Fu} {et~al.}(2008){Fu}, {Semboloni}, {Hoekstra}, {Kilbinger}, {van
  Waerbeke}, {Tereno}, {Mellier}, {Heymans}, {Coupon}, {Benabed}, {Benjamin},
  {Bertin}, {Dor{\'e}}, {Hudson}, {Ilbert}, {Maoli}, {Marmo}, {McCracken}, \&
  {M{\'e}nard}}]{Fu2008}
{Fu}, L., {Semboloni}, E., {Hoekstra}, H., {et~al.} 2008, \aap, 479, 9,
  \dodoi{10.1051/0004-6361:20078522}

\bibitem[{{Fu} {et~al.}(2014){Fu}, {Kilbinger}, {Erben}, {Heymans},
  {Hildebrandt}, {Hoekstra}, {Kitching}, {Mellier}, {Miller}, {Semboloni},
  {Simon}, {Van Waerbeke}, {Coupon}, {Harnois-D{\'e}raps}, {Hudson}, {Kuijken},
  {Rowe}, {Schrabback}, {Vafaei}, \& {Velander}}]{Fu2014}
{Fu}, L., {Kilbinger}, M., {Erben}, T., {et~al.} 2014, \mnras, 441, 2725,
  \dodoi{10.1093/mnras/stu754}

\bibitem[{{Fu} \& {Fan}(2014)}]{FuFan}
{Fu}, L.-P., \& {Fan}, Z.-H. 2014, Research in Astronomy and Astrophysics, 14,
  1061, \dodoi{10.1088/1674-4527/14/9/002}

\bibitem[{{Gong} {et~al.}(2019){Gong}, {Liu}, {Cao}, {Chen}, {Fan}, {Li}, {Li},
  {Li}, {Zhang}, \& {Zhan}}]{CSST}
{Gong}, Y., {Liu}, X., {Cao}, Y., {et~al.} 2019, \apj, 883, 203,
  \dodoi{10.3847/1538-4357/ab391e}

\bibitem[{{Guo} {et~al.}(2013){Guo}, {White}, {Angulo}, {Henriques}, {Lemson},
  {Boylan-Kolchin}, {Thomas}, \& {Short}}]{Guo2013}
{Guo}, Q., {White}, S., {Angulo}, R.~E., {et~al.} 2013, \mnras, 428, 1351,
  \dodoi{10.1093/mnras/sts115}

\bibitem[{{Hamana} {et~al.}(2012){Hamana}, {Oguri}, {Shirasaki}, \&
  {Sato}}]{Hamana2012}
{Hamana}, T., {Oguri}, M., {Shirasaki}, M., \& {Sato}, M. 2012, \mnras, 425,
  2287, \dodoi{10.1111/j.1365-2966.2012.21582.x}

\bibitem[{{Hamana} {et~al.}(2020){Hamana}, {Shirasaki}, \& {Lin}}]{Hamana2020}
{Hamana}, T., {Shirasaki}, M., \& {Lin}, Y.-T. 2020, \pasj, 72, 78,
  \dodoi{10.1093/pasj/psaa068}

\bibitem[{{Hamana} {et~al.}(2004){Hamana}, {Takada}, \& {Yoshida}}]{Hamana2004}
{Hamana}, T., {Takada}, M., \& {Yoshida}, N. 2004, \mnras, 350, 893,
  \dodoi{10.1111/j.1365-2966.2004.07691.x}

\bibitem[{{Harnois-D{\'e}raps} {et~al.}(2022){Harnois-D{\'e}raps}, {Martinet},
  \& {Reischke}}]{Joachim2022}
{Harnois-D{\'e}raps}, J., {Martinet}, N., \& {Reischke}, R. 2022, \mnras, 509,
  3868, \dodoi{10.1093/mnras/stab3222}

\bibitem[{{Heymans} {et~al.}(2012){Heymans}, {Van Waerbeke}, {Miller}, {Erben},
  {Hildebrandt}, {Hoekstra}, {Kitching}, {Mellier}, {Simon}, {Bonnett},
  {Coupon}, {Fu}, {Harnois D{\'e}raps}, {Hudson}, {Kilbinger}, {Kuijken},
  {Rowe}, {Schrabback}, {Semboloni}, {van Uitert}, {Vafaei}, \&
  {Velander}}]{Heymans2012}
{Heymans}, C., {Van Waerbeke}, L., {Miller}, L., {et~al.} 2012, \mnras, 427,
  146, \dodoi{10.1111/j.1365-2966.2012.21952.x}

\bibitem[{{Heymans} {et~al.}(2021){Heymans}, {Tr{\"o}ster}, {Asgari}, {Blake},
  {Hildebrandt}, {Joachimi}, {Kuijken}, {Lin}, {S{\'a}nchez}, {van den Busch},
  {Wright}, {Amon}, {Bilicki}, {de Jong}, {Crocce}, {Dvornik}, {Erben},
  {Fortuna}, {Getman}, {Giblin}, {Glazebrook}, {Hoekstra}, {Joudaki},
  {Kannawadi}, {K{\"o}hlinger}, {Lidman}, {Miller}, {Napolitano}, {Parkinson},
  {Schneider}, {Shan}, {Valentijn}, {Verdoes Kleijn}, \& {Wolf}}]{Heymans2021}
{Heymans}, C., {Tr{\"o}ster}, T., {Asgari}, M., {et~al.} 2021, \aap, 646, A140,
  \dodoi{10.1051/0004-6361/202039063}

\bibitem[{{Hikage} {et~al.}(2019){Hikage}, {Oguri}, {Hamana}, {More},
  {Mandelbaum}, {Takada}, {K{\"o}hlinger}, {Miyatake}, {Nishizawa}, {Aihara},
  {Armstrong}, {Bosch}, {Coupon}, {Ducout}, {Ho}, {Hsieh}, {Komiyama},
  {Lanusse}, {Leauthaud}, {Lupton}, {Medezinski}, {Mineo}, {Miyama},
  {Miyazaki}, {Murata}, {Murayama}, {Shirasaki}, {Sif{\'o}n}, {Simet},
  {Speagle}, {Spergel}, {Strauss}, {Sugiyama}, {Tanaka}, {Utsumi}, {Wang}, \&
  {Yamada}}]{Hikage2019}
{Hikage}, C., {Oguri}, M., {Hamana}, T., {et~al.} 2019, \pasj, 71, 43,
  \dodoi{10.1093/pasj/psz010}

\bibitem[{{Hildebrandt} {et~al.}(2020){Hildebrandt}, {K{\"o}hlinger}, {van den
  Busch}, {Joachimi}, {Heymans}, {Kannawadi}, {Wright}, {Asgari}, {Blake},
  {Hoekstra}, {Joudaki}, {Kuijken}, {Miller}, {Morrison}, {Tr{\"o}ster},
  {Amon}, {Archidiacono}, {Brieden}, {Choi}, {de Jong}, {Erben}, {Giblin},
  {Mead}, {Peacock}, {Radovich}, {Schneider}, {Sif{\'o}n}, \&
  {Tewes}}]{Hildebrandt2020}
{Hildebrandt}, H., {K{\"o}hlinger}, F., {van den Busch}, J.~L., {et~al.} 2020,
  \aap, 633, A69, \dodoi{10.1051/0004-6361/201834878}

\bibitem[{{Hinshaw} {et~al.}(2013){Hinshaw}, {Larson}, {Komatsu}, {Spergel},
  {Bennett}, {Dunkley}, {Nolta}, {Halpern}, {Hill}, {Odegard}, {Page}, {Smith},
  {Weiland}, {Gold}, {Jarosik}, {Kogut}, {Limon}, {Meyer}, {Tucker}, {Wollack},
  \& {Wright}}]{Hinshaw2013}
{Hinshaw}, G., {Larson}, D., {Komatsu}, E., {et~al.} 2013, \apjs, 208, 19,
  \dodoi{10.1088/0067-0049/208/2/19}

\bibitem[{{Hirata} \& {Seljak}(2004)}]{HS04}
{Hirata}, C.~M., \& {Seljak}, U. 2004, \prd, 70, 063526,
  \dodoi{10.1103/PhysRevD.70.063526}

\bibitem[{{Hoekstra} \& {Jain}(2008)}]{HJ2008}
{Hoekstra}, H., \& {Jain}, B. 2008, Annual Review of Nuclear and Particle
  Science, 58, 99, \dodoi{10.1146/annurev.nucl.58.110707.171151}

\bibitem[{{Ivezi{\'c}} {et~al.}(2019){Ivezi{\'c}}, {Kahn}, {Tyson}, {Abel},
  {Acosta}, {Allsman}, {Alonso}, {AlSayyad}, {Anderson}, {Andrew}, {Angel},
  {Angeli}, {Ansari}, {Antilogus}, {Araujo}, {Armstrong}, {Arndt}, {Astier},
  {Aubourg}, {Auza}, {Axelrod}, {Bard}, {Barr}, {Barrau}, {Bartlett}, {Bauer},
  {Bauman}, {Baumont}, {Bechtol}, {Bechtol}, {Becker}, {Becla}, {Beldica},
  {Bellavia}, {Bianco}, {Biswas}, {Blanc}, {Blazek}, {Blandford}, {Bloom},
  {Bogart}, {Bond}, {Booth}, {Borgland}, {Borne}, {Bosch}, {Boutigny},
  {Brackett}, {Bradshaw}, {Brandt}, {Brown}, {Bullock}, {Burchat}, {Burke},
  {Cagnoli}, {Calabrese}, {Callahan}, {Callen}, {Carlin}, {Carlson},
  {Chandrasekharan}, {Charles-Emerson}, {Chesley}, {Cheu}, {Chiang}, {Chiang},
  {Chirino}, {Chow}, {Ciardi}, {Claver}, {Cohen-Tanugi}, {Cockrum}, {Coles},
  {Connolly}, {Cook}, {Cooray}, {Covey}, {Cribbs}, {Cui}, {Cutri}, {Daly},
  {Daniel}, {Daruich}, {Daubard}, {Daues}, {Dawson}, {Delgado}, {Dellapenna},
  {de Peyster}, {de Val-Borro}, {Digel}, {Doherty}, {Dubois},
  {Dubois-Felsmann}, {Durech}, {Economou}, {Eifler}, {Eracleous}, {Emmons},
  {Fausti Neto}, {Ferguson}, {Figueroa}, {Fisher-Levine}, {Focke}, {Foss},
  {Frank}, {Freemon}, {Gangler}, {Gawiser}, {Geary}, {Gee}, {Geha}, {Gessner},
  {Gibson}, {Gilmore}, {Glanzman}, {Glick}, {Goldina}, {Goldstein}, {Goodenow},
  {Graham}, {Gressler}, {Gris}, {Guy}, {Guyonnet}, {Haller}, {Harris},
  {Hascall}, {Haupt}, {Hernandez}, {Herrmann}, {Hileman}, {Hoblitt}, {Hodgson},
  {Hogan}, {Howard}, {Huang}, {Huffer}, {Ingraham}, {Innes}, {Jacoby}, {Jain},
  {Jammes}, {Jee}, {Jenness}, {Jernigan}, {Jevremovi{\'c}}, {Johns}, {Johnson},
  {Johnson}, {Jones}, {Juramy-Gilles}, {Juri{\'c}}, {Kalirai}, {Kallivayalil},
  {Kalmbach}, {Kantor}, {Karst}, {Kasliwal}, {Kelly}, {Kessler}, {Kinnison},
  {Kirkby}, {Knox}, {Kotov}, {Krabbendam}, {Krughoff}, {Kub{\'a}nek},
  {Kuczewski}, {Kulkarni}, {Ku}, {Kurita}, {Lage}, {Lambert}, {Lange},
  {Langton}, {Le Guillou}, {Levine}, {Liang}, {Lim}, {Lintott}, {Long},
  {Lopez}, {Lotz}, {Lupton}, {Lust}, {MacArthur}, {Mahabal}, {Mandelbaum},
  {Markiewicz}, {Marsh}, {Marshall}, {Marshall}, {May}, {McKercher}, {McQueen},
  {Meyers}, {Migliore}, {Miller}, {Mills}, {Miraval}, {Moeyens}, {Moolekamp},
  {Monet}, {Moniez}, {Monkewitz}, {Montgomery}, {Morrison}, {Mueller},
  {Muller}, {Mu{\~n}oz Arancibia}, {Neill}, {Newbry}, {Nief}, {Nomerotski},
  {Nordby}, {O'Connor}, {Oliver}, {Olivier}, {Olsen}, {O'Mullane}, {Ortiz},
  {Osier}, {Owen}, {Pain}, {Palecek}, {Parejko}, {Parsons}, {Pease},
  {Peterson}, {Peterson}, {Petravick}, {Libby Petrick}, {Petry},
  {Pierfederici}, {Pietrowicz}, {Pike}, {Pinto}, {Plante}, {Plate}, {Plutchak},
  {Price}, {Prouza}, {Radeka}, {Rajagopal}, {Rasmussen}, {Regnault}, {Reil},
  {Reiss}, {Reuter}, {Ridgway}, {Riot}, {Ritz}, {Robinson}, {Roby}, {Roodman},
  {Rosing}, {Roucelle}, {Rumore}, {Russo}, {Saha}, {Sassolas}, {Schalk},
  {Schellart}, {Schindler}, {Schmidt}, {Schneider}, {Schneider}, {Schoening},
  {Schumacher}, {Schwamb}, {Sebag}, {Selvy}, {Sembroski}, {Seppala}, {Serio},
  {Serrano}, {Shaw}, {Shipsey}, {Sick}, {Silvestri}, {Slater}, {Smith},
  {Smith}, {Sobhani}, {Soldahl}, {Storrie-Lombardi}, {Stover}, {Strauss},
  {Street}, {Stubbs}, {Sullivan}, {Sweeney}, {Swinbank}, {Szalay}, {Takacs},
  {Tether}, {Thaler}, {Thayer}, {Thomas}, {Thornton}, {Thukral}, {Tice},
  {Trilling}, {Turri}, {Van Berg}, {Vanden Berk}, {Vetter}, {Virieux},
  {Vucina}, {Wahl}, {Walkowicz}, {Walsh}, {Walter}, {Wang}, {Wang}, {Warner},
  {Wiecha}, {Willman}, {Winters}, {Wittman}, {Wolff}, {Wood-Vasey}, {Wu},
  {Xin}, {Yoachim}, \& {Zhan}}]{Ivezic2019}
{Ivezi{\'c}}, {\v{Z}}., {Kahn}, S.~M., {Tyson}, J.~A., {et~al.} 2019, \apj,
  873, 111, \dodoi{10.3847/1538-4357/ab042c}

\bibitem[{{Joachimi} {et~al.}(2011){Joachimi}, {Mandelbaum}, {Abdalla}, \&
  {Bridle}}]{Joachimi2011}
{Joachimi}, B., {Mandelbaum}, R., {Abdalla}, F.~B., \& {Bridle}, S.~L. 2011,
  \aap, 527, A26, \dodoi{10.1051/0004-6361/201015621}

\bibitem[{{Joachimi} {et~al.}(2013){Joachimi}, {Semboloni}, {Hilbert}, {Bett},
  {Hartlap}, {Hoekstra}, \& {Schneider}}]{Joachimi2013}
{Joachimi}, B., {Semboloni}, E., {Hilbert}, S., {et~al.} 2013, \mnras, 436,
  819, \dodoi{10.1093/mnras/stt1618}

\bibitem[{{Joachimi} {et~al.}(2015){Joachimi}, {Cacciato}, {Kitching},
  {Leonard}, {Mandelbaum}, {Sch{\"a}fer}, {Sif{\'o}n}, {Hoekstra}, {Kiessling},
  {Kirk}, \& {Rassat}}]{Joachimi2015}
{Joachimi}, B., {Cacciato}, M., {Kitching}, T.~D., {et~al.} 2015, \ssr, 193, 1,
  \dodoi{10.1007/s11214-015-0177-4}

\bibitem[{{Kacprzak} {et~al.}(2016){Kacprzak}, {Kirk}, {Friedrich}, {Amara},
  {Refregier}, {Marian}, {Dietrich}, {Suchyta}, {Aleksi{\'c}}, {Bacon},
  {Becker}, {Bonnett}, {Bridle}, {Chang}, {Eifler}, {Hartley}, {Huff},
  {Krause}, {MacCrann}, {Melchior}, {Nicola}, {Samuroff}, {Sheldon}, {Troxel},
  {Weller}, {Zuntz}, {Abbott}, {Abdalla}, {Armstrong}, {Benoit-L{\'e}vy},
  {Bernstein}, {Bernstein}, {Bertin}, {Brooks}, {Burke}, {Carnero Rosell},
  {Carrasco Kind}, {Carretero}, {Castander}, {Crocce}, {D'Andrea}, {da Costa},
  {Desai}, {Diehl}, {Evrard}, {Neto}, {Flaugher}, {Fosalba}, {Frieman},
  {Gerdes}, {Goldstein}, {Gruen}, {Gruendl}, {Gutierrez}, {Honscheid}, {Jain},
  {James}, {Jarvis}, {Kuehn}, {Kuropatkin}, {Lahav}, {Lima}, {March},
  {Marshall}, {Martini}, {Miller}, {Miquel}, {Mohr}, {Nichol}, {Nord},
  {Plazas}, {Romer}, {Roodman}, {Rykoff}, {Sanchez}, {Scarpine}, {Schubnell},
  {Sevilla-Noarbe}, {Smith}, {Soares-Santos}, {Sobreira}, {Swanson}, {Tarle},
  {Thomas}, {Vikram}, {Walker}, {Zhang}, \& {DES Collaboration}}]{Kacprzak2016}
{Kacprzak}, T., {Kirk}, D., {Friedrich}, O., {et~al.} 2016, \mnras, 463, 3653,
  \dodoi{10.1093/mnras/stw2070}

\bibitem[{{Kaiser} \& {Squires}(1993)}]{Kaiser93}
{Kaiser}, N., \& {Squires}, G. 1993, \apj, 404, 441, \dodoi{10.1086/172297}

\bibitem[{{Kiessling} {et~al.}(2015){Kiessling}, {Cacciato}, {Joachimi},
  {Kirk}, {Kitching}, {Leonard}, {Mandelbaum}, {Sch{\"a}fer}, {Sif{\'o}n},
  {Brown}, \& {Rassat}}]{Kiessling}
{Kiessling}, A., {Cacciato}, M., {Joachimi}, B., {et~al.} 2015, \ssr, 193, 67,
  \dodoi{10.1007/s11214-015-0203-6}

\bibitem[{{Kilbinger}(2015)}]{Kilbinger2015}
{Kilbinger}, M. 2015, Reports on Progress in Physics, 78, 086901,
  \dodoi{10.1088/0034-4885/78/8/086901}

\bibitem[{{Kilbinger} {et~al.}(2013){Kilbinger}, {Fu}, {Heymans}, {Simpson},
  {Benjamin}, {Erben}, {Harnois-D{\'e}raps}, {Hoekstra}, {Hildebrandt},
  {Kitching}, {Mellier}, {Miller}, {Van Waerbeke}, {Benabed}, {Bonnett},
  {Coupon}, {Hudson}, {Kuijken}, {Rowe}, {Schrabback}, {Semboloni}, {Vafaei},
  \& {Velander}}]{Kilbinger2013}
{Kilbinger}, M., {Fu}, L., {Heymans}, C., {et~al.} 2013, \mnras, 430, 2200,
  \dodoi{10.1093/mnras/stt041}

\bibitem[{{Kirk} {et~al.}(2015){Kirk}, {Brown}, {Hoekstra}, {Joachimi},
  {Kitching}, {Mandelbaum}, {Sif{\'o}n}, {Cacciato}, {Choi}, {Kiessling},
  {Leonard}, {Rassat}, \& {Sch{\"a}fer}}]{Kirk}
{Kirk}, D., {Brown}, M.~L., {Hoekstra}, H., {et~al.} 2015, \ssr, 193, 139,
  \dodoi{10.1007/s11214-015-0213-4}

\bibitem[{{Kuijken} {et~al.}(2019){Kuijken}, {Heymans}, {Dvornik},
  {Hildebrandt}, {de Jong}, {Wright}, {Erben}, {Bilicki}, {Giblin}, {Shan},
  {Getman}, {Grado}, {Hoekstra}, {Miller}, {Napolitano}, {Paolilo}, {Radovich},
  {Schneider}, {Sutherland}, {Tewes}, {Tortora}, {Valentijn}, \& {Verdoes
  Kleijn}}]{Kuijken2019}
{Kuijken}, K., {Heymans}, C., {Dvornik}, A., {et~al.} 2019, \aap, 625, A2,
  \dodoi{10.1051/0004-6361/201834918}

\bibitem[{{Laureijs} {et~al.}(2011){Laureijs}, {Amiaux}, {Arduini},
  {Augu{\`e}res}, {Brinchmann}, {Cole}, {Cropper}, {Dabin}, {Duvet}, {Ealet},
  {Garilli}, {Gondoin}, {Guzzo}, {Hoar}, {Hoekstra}, {Holmes}, {Kitching},
  {Maciaszek}, {Mellier}, {Pasian}, {Percival}, {Rhodes}, {Saavedra Criado},
  {Sauvage}, {Scaramella}, {Valenziano}, {Warren}, {Bender}, {Castander},
  {Cimatti}, {Le F{\`e}vre}, {Kurki-Suonio}, {Levi}, {Lilje}, {Meylan},
  {Nichol}, {Pedersen}, {Popa}, {Rebolo Lopez}, {Rix}, {Rottgering},
  {Zeilinger}, {Grupp}, {Hudelot}, {Massey}, {Meneghetti}, {Miller}, {Paltani},
  {Paulin-Henriksson}, {Pires}, {Saxton}, {Schrabback}, {Seidel}, {Walsh},
  {Aghanim}, {Amendola}, {Bartlett}, {Baccigalupi}, {Beaulieu}, {Benabed},
  {Cuby}, {Elbaz}, {Fosalba}, {Gavazzi}, {Helmi}, {Hook}, {Irwin}, {Kneib},
  {Kunz}, {Mannucci}, {Moscardini}, {Tao}, {Teyssier}, {Weller}, {Zamorani},
  {Zapatero Osorio}, {Boulade}, {Foumond}, {Di Giorgio}, {Guttridge}, {James},
  {Kemp}, {Martignac}, {Spencer}, {Walton}, {Bl{\"u}mchen}, {Bonoli},
  {Bortoletto}, {Cerna}, {Corcione}, {Fabron}, {Jahnke}, {Ligori}, {Madrid},
  {Martin}, {Morgante}, {Pamplona}, {Prieto}, {Riva}, {Toledo}, {Trifoglio},
  {Zerbi}, {Abdalla}, {Douspis}, {Grenet}, {Borgani}, {Bouwens}, {Courbin},
  {Delouis}, {Dubath}, {Fontana}, {Frailis}, {Grazian}, {Koppenh{\"o}fer},
  {Mansutti}, {Melchior}, {Mignoli}, {Mohr}, {Neissner}, {Noddle}, {Poncet},
  {Scodeggio}, {Serrano}, {Shane}, {Starck}, {Surace}, {Taylor},
  {Verdoes-Kleijn}, {Vuerli}, {Williams}, {Zacchei}, {Altieri}, {Escudero
  Sanz}, {Kohley}, {Oosterbroek}, {Astier}, {Bacon}, {Bardelli}, {Baugh},
  {Bellagamba}, {Benoist}, {Bianchi}, {Biviano}, {Branchini}, {Carbone},
  {Cardone}, {Clements}, {Colombi}, {Conselice}, {Cresci}, {Deacon}, {Dunlop},
  {Fedeli}, {Fontanot}, {Franzetti}, {Giocoli}, {Garcia-Bellido}, {Gow},
  {Heavens}, {Hewett}, {Heymans}, {Holland}, {Huang}, {Ilbert}, {Joachimi},
  {Jennins}, {Kerins}, {Kiessling}, {Kirk}, {Kotak}, {Krause}, {Lahav}, {van
  Leeuwen}, {Lesgourgues}, {Lombardi}, {Magliocchetti}, {Maguire}, {Majerotto},
  {Maoli}, {Marulli}, {Maurogordato}, {McCracken}, {McLure}, {Melchiorri},
  {Merson}, {Moresco}, {Nonino}, {Norberg}, {Peacock}, {Pello}, {Penny},
  {Pettorino}, {Di Porto}, {Pozzetti}, {Quercellini}, {Radovich}, {Rassat},
  {Roche}, {Ronayette}, {Rossetti}, {Sartoris}, {Schneider}, {Semboloni},
  {Serjeant}, {Simpson}, {Skordis}, {Smadja}, {Smartt}, {Spano}, {Spiro},
  {Sullivan}, {Tilquin}, {Trotta}, {Verde}, {Wang}, {Williger}, {Zhao},
  {Zoubian}, \& {Zucca}}]{Euclid}
{Laureijs}, R., {Amiaux}, J., {Arduini}, S., {et~al.} 2011, arXiv e-prints,
  arXiv:1110.3193.
\newblock \doarXiv{1110.3193}

\bibitem[{{Lin} \& {Kilbinger}(2015)}]{Lin2015}
{Lin}, C.-A., \& {Kilbinger}, M. 2015, \aap, 583, A70,
  \dodoi{10.1051/0004-6361/201526659}

\bibitem[{{Liu} {et~al.}(2016){Liu}, {Li}, {Zhao}, {Chiu}, {Fang}, {Pan},
  {Wang}, {Du}, {Yuan}, {Fu}, \& {Fan}}]{Liu2016}
{Liu}, X., {Li}, B., {Zhao}, G.-B., {et~al.} 2016, \prl, 117, 051101,
  \dodoi{10.1103/PhysRevLett.117.051101}

\bibitem[{{LiuJ} {et~al.}(2015){LiuJ}, {Petri}, {Haiman}, {Hui}, {Kratochvil},
  \& {May}}]{LiuJ2015}
{LiuJ}, J., {Petri}, A., {Haiman}, Z., {et~al.} 2015, \prd, 91, 063507,
  \dodoi{10.1103/PhysRevD.91.063507}

\bibitem[{{LiuX} {et~al.}(2015){LiuX}, {Pan}, {Li}, {Shan}, {Wang}, {Fu},
  {Fan}, {Kneib}, {Leauthaud}, {Van Waerbeke}, {Makler}, {Moraes}, {Erben}, \&
  {Charbonnier}}]{LiuX2015}
{LiuX}, X., {Pan}, C., {Li}, R., {et~al.} 2015, \mnras, 450, 2888,
  \dodoi{10.1093/mnras/stv784}

\bibitem[{{Luo} {et~al.}(2016){Luo}, {Kang}, {Kauffmann}, \& {Fu}}]{Luo2016}
{Luo}, Y., {Kang}, X., {Kauffmann}, G., \& {Fu}, J. 2016, \mnras, 458, 366,
  \dodoi{10.1093/mnras/stw268}

\bibitem[{{Martinet} {et~al.}(2021){Martinet}, {Harnois-D{\'e}raps}, {Jullo},
  \& {Schneider}}]{Martinet2021}
{Martinet}, N., {Harnois-D{\'e}raps}, J., {Jullo}, E., \& {Schneider}, P. 2021,
  \aap, 646, A62, \dodoi{10.1051/0004-6361/202039679}

\bibitem[{{Martinet} {et~al.}(2018){Martinet}, {Schneider}, {Hildebrandt},
  {Shan}, {Asgari}, {Dietrich}, {Harnois-D{\'e}raps}, {Erben}, {Grado},
  {Heymans}, {Hoekstra}, {Klaes}, {Kuijken}, {Merten}, \&
  {Nakajima}}]{Martinet2018}
{Martinet}, N., {Schneider}, P., {Hildebrandt}, H., {et~al.} 2018, \mnras, 474,
  712, \dodoi{10.1093/mnras/stx2793}

\bibitem[{{Maturi} {et~al.}(2010){Maturi}, {Angrick}, {Pace}, \&
  {Bartelmann}}]{Maturi2010}
{Maturi}, M., {Angrick}, C., {Pace}, F., \& {Bartelmann}, M. 2010, \aap, 519,
  A23, \dodoi{10.1051/0004-6361/200912866}

\bibitem[{{Oguri} {et~al.}(2021){Oguri}, {Miyazaki}, {Li}, {Luo}, {Mitsuishi},
  {Miyatake}, {More}, {Nishizawa}, {Okabe}, {Ota}, {Plazas Malag{\'o}n}, \&
  {Utsumi}}]{Oguri2021}
{Oguri}, M., {Miyazaki}, S., {Li}, X., {et~al.} 2021, \pasj, 73, 817,
  \dodoi{10.1093/pasj/psab047}

\bibitem[{{Petri} {et~al.}(2015){Petri}, {Liu}, {Haiman}, {May}, {Hui}, \&
  {Kratochvil}}]{Petri2015}
{Petri}, A., {Liu}, J., {Haiman}, Z., {et~al.} 2015, \prd, 91, 103511,
  \dodoi{10.1103/PhysRevD.91.103511}

\bibitem[{{Planck Collaboration} {et~al.}(2020){Planck Collaboration},
  {Aghanim}, {Akrami}, {Ashdown}, {Aumont}, {Baccigalupi}, {Ballardini},
  {Banday}, {Barreiro}, {Bartolo}, {Basak}, {Battye}, {Benabed}, {Bernard},
  {Bersanelli}, {Bielewicz}, {Bock}, {Bond}, {Borrill}, {Bouchet}, {Boulanger},
  {Bucher}, {Burigana}, {Butler}, {Calabrese}, {Cardoso}, {Carron},
  {Challinor}, {Chiang}, {Chluba}, {Colombo}, {Combet}, {Contreras}, {Crill},
  {Cuttaia}, {de Bernardis}, {de Zotti}, {Delabrouille}, {Delouis}, {Di
  Valentino}, {Diego}, {Dor{\'e}}, {Douspis}, {Ducout}, {Dupac}, {Dusini},
  {Efstathiou}, {Elsner}, {En{\ss}lin}, {Eriksen}, {Fantaye}, {Farhang},
  {Fergusson}, {Fernandez-Cobos}, {Finelli}, {Forastieri}, {Frailis},
  {Fraisse}, {Franceschi}, {Frolov}, {Galeotta}, {Galli}, {Ganga},
  {G{\'e}nova-Santos}, {Gerbino}, {Ghosh}, {Gonz{\'a}lez-Nuevo}, {G{\'o}rski},
  {Gratton}, {Gruppuso}, {Gudmundsson}, {Hamann}, {Handley}, {Hansen},
  {Herranz}, {Hildebrandt}, {Hivon}, {Huang}, {Jaffe}, {Jones}, {Karakci},
  {Keih{\"a}nen}, {Keskitalo}, {Kiiveri}, {Kim}, {Kisner}, {Knox},
  {Krachmalnicoff}, {Kunz}, {Kurki-Suonio}, {Lagache}, {Lamarre}, {Lasenby},
  {Lattanzi}, {Lawrence}, {Le Jeune}, {Lemos}, {Lesgourgues}, {Levrier},
  {Lewis}, {Liguori}, {Lilje}, {Lilley}, {Lindholm}, {L{\'o}pez-Caniego},
  {Lubin}, {Ma}, {Mac{\'\i}as-P{\'e}rez}, {Maggio}, {Maino}, {Mandolesi},
  {Mangilli}, {Marcos-Caballero}, {Maris}, {Martin}, {Martinelli},
  {Mart{\'\i}nez-Gonz{\'a}lez}, {Matarrese}, {Mauri}, {McEwen}, {Meinhold},
  {Melchiorri}, {Mennella}, {Migliaccio}, {Millea}, {Mitra},
  {Miville-Desch{\^e}nes}, {Molinari}, {Montier}, {Morgante}, {Moss}, {Natoli},
  {N{\o}rgaard-Nielsen}, {Pagano}, {Paoletti}, {Partridge}, {Patanchon},
  {Peiris}, {Perrotta}, {Pettorino}, {Piacentini}, {Polastri}, {Polenta},
  {Puget}, {Rachen}, {Reinecke}, {Remazeilles}, {Renzi}, {Rocha}, {Rosset},
  {Roudier}, {Rubi{\~n}o-Mart{\'\i}n}, {Ruiz-Granados}, {Salvati}, {Sandri},
  {Savelainen}, {Scott}, {Shellard}, {Sirignano}, {Sirri}, {Spencer},
  {Sunyaev}, {Suur-Uski}, {Tauber}, {Tavagnacco}, {Tenti}, {Toffolatti},
  {Tomasi}, {Trombetti}, {Valenziano}, {Valiviita}, {Van Tent}, {Vibert},
  {Vielva}, {Villa}, {Vittorio}, {Wandelt}, {Wehus}, {White}, {White},
  {Zacchei}, \& {Zonca}}]{Planck2020}
{Planck Collaboration}, {Aghanim}, N., {Akrami}, Y., {et~al.} 2020, \aap, 641,
  A6, \dodoi{10.1051/0004-6361/201833910}

\bibitem[{{Schneider} \& {Bridle}(2010)}]{SB2010}
{Schneider}, M.~D., \& {Bridle}, S. 2010, \mnras, 402, 2127,
  \dodoi{10.1111/j.1365-2966.2009.15956.x}

\bibitem[{{Seitz} \& {Schneider}(1995)}]{Seitz95}
{Seitz}, C., \& {Schneider}, P. 1995, \aap, 297, 287.
\newblock \doarXiv{astro-ph/9408050}

\bibitem[{{Seitz} \& {Schneider}(1997)}]{Seitz97}
---. 1997, \aap, 318, 687.
\newblock \doarXiv{astro-ph/9601079}

\bibitem[{{Shan} {et~al.}(2012){Shan}, {Kneib}, {Tao}, {Fan}, {Jauzac},
  {Limousin}, {Massey}, {Rhodes}, {Thanjavur}, \& {McCracken}}]{Shan2012}
{Shan}, H., {Kneib}, J.-P., {Tao}, C., {et~al.} 2012, \apj, 748, 56,
  \dodoi{10.1088/0004-637X/748/1/56}

\bibitem[{{Shan} {et~al.}(2018){Shan}, {Liu}, {Hildebrandt}, {Pan}, {Martinet},
  {Fan}, {Schneider}, {Asgari}, {Harnois-D{\'e}raps}, {Hoekstra}, {Wright},
  {Dietrich}, {Erben}, {Getman}, {Grado}, {Heymans}, {Klaes}, {Kuijken},
  {Merten}, {Puddu}, {Radovich}, \& {Wang}}]{Shan2018}
{Shan}, H., {Liu}, X., {Hildebrandt}, H., {et~al.} 2018, \mnras, 474, 1116,
  \dodoi{10.1093/mnras/stx2837}

\bibitem[{{Shan} {et~al.}(2014){Shan}, {Kneib}, {Comparat}, {Jullo},
  {Charbonnier}, {Erben}, {Makler}, {Moraes}, {Van Waerbeke}, {Courbin},
  {Meylan}, {Tao}, \& {Taylor}}]{Shan2014}
{Shan}, H.~Y., {Kneib}, J.-P., {Comparat}, J., {et~al.} 2014, \mnras, 442,
  2534, \dodoi{10.1093/mnras/stu1040}

\bibitem[{{Spergel} {et~al.}(2015){Spergel}, {Gehrels}, {Baltay}, {Bennett},
  {Breckinridge}, {Donahue}, {Dressler}, {Gaudi}, {Greene}, {Guyon}, {Hirata},
  {Kalirai}, {Kasdin}, {Macintosh}, {Moos}, {Perlmutter}, {Postman},
  {Rauscher}, {Rhodes}, {Wang}, {Weinberg}, {Benford}, {Hudson}, {Jeong},
  {Mellier}, {Traub}, {Yamada}, {Capak}, {Colbert}, {Masters}, {Penny},
  {Savransky}, {Stern}, {Zimmerman}, {Barry}, {Bartusek}, {Carpenter}, {Cheng},
  {Content}, {Dekens}, {Demers}, {Grady}, {Jackson}, {Kuan}, {Kruk}, {Melton},
  {Nemati}, {Parvin}, {Poberezhskiy}, {Peddie}, {Ruffa}, {Wallace}, {Whipple},
  {Wollack}, \& {Zhao}}]{WFIRST}
{Spergel}, D., {Gehrels}, N., {Baltay}, C., {et~al.} 2015, arXiv e-prints.
\newblock \doarXiv{1503.03757}

\bibitem[{{Squires} \& {Kaiser}(1996)}]{KS96}
{Squires}, G., \& {Kaiser}, N. 1996, \apj, 473, 65, \dodoi{10.1086/178127}

\bibitem[{{Troxel} \& {Ishak}(2015)}]{Troxel}
{Troxel}, M.~A., \& {Ishak}, M. 2015, \physrep, 558, 1,
  \dodoi{10.1016/j.physrep.2014.11.001}

\bibitem[{{van Waerbeke}(2000)}]{vanWaer2000}
{van Waerbeke}, L. 2000, \mnras, 313, 524,
  \dodoi{10.1046/j.1365-8711.2000.03259.x}

\bibitem[{{Van Waerbeke} {et~al.}(2000){Van Waerbeke}, {Mellier}, {Erben},
  {Cuillandre}, {Bernardeau}, {Maoli}, {Bertin}, {McCracken}, {Le F{\`e}vre},
  {Fort}, {Dantel-Fort}, {Jain}, \& {Schneider}}]{WaerbekeShear2000}
{Van Waerbeke}, L., {Mellier}, Y., {Erben}, T., {et~al.} 2000, \aap, 358, 30.
\newblock \doarXiv{astro-ph/0002500}

\bibitem[{{Wang} {et~al.}(2014){Wang}, {Mo}, {Yang}, {Jing}, \&
  {Lin}}]{Wang2014}
{Wang}, H., {Mo}, H.~J., {Yang}, X., {Jing}, Y.~P., \& {Lin}, W.~P. 2014, \apj,
  794, 94, \dodoi{10.1088/0004-637X/794/1/94}

\bibitem[{{Wang} {et~al.}(2016){Wang}, {Mo}, {Yang}, {Zhang}, {Shi}, {Jing},
  {Liu}, {Li}, {Kang}, \& {Gao}}]{Wang2016}
{Wang}, H., {Mo}, H.~J., {Yang}, X., {et~al.} 2016, \apj, 831, 164,
  \dodoi{10.3847/0004-637X/831/2/164}

\bibitem[{{Wei} {et~al.}(2018){Wei}, {Li}, {Kang}, {Luo}, {Xia}, {Wang},
  {Yang}, {Wang}, {Jing}, {Mo}, {Lin}, {Wang}, {Li}, {Lu}, {Zhang}, {Lim},
  {Tweed}, \& {Cui}}]{Wei}
{Wei}, C., {Li}, G., {Kang}, X., {et~al.} 2018, \apj, 853, 25,
  \dodoi{10.3847/1538-4357/aaa40d}

\bibitem[{{Yang} {et~al.}(2011){Yang}, {Kratochvil}, {Wang}, {Lim}, {Haiman},
  \& {May}}]{Yang2011}
{Yang}, X., {Kratochvil}, J.~M., {Wang}, S., {et~al.} 2011, \prd, 84, 043529,
  \dodoi{10.1103/PhysRevD.84.043529}

\bibitem[{{Yao} {et~al.}(2020){Yao}, {Shan}, {Zhang}, {Kneib}, \&
  {Jullo}}]{Yao2020}
{Yao}, J., {Shan}, H., {Zhang}, P., {Kneib}, J.-P., \& {Jullo}, E. 2020, \apj,
  904, 135, \dodoi{10.3847/1538-4357/abc175}

\bibitem[{{Yuan} {et~al.}(2018){Yuan}, {Liu}, {Pan}, {Wang}, \&
  {Fan}}]{Yuan2018}
{Yuan}, S., {Liu}, X., {Pan}, C., {Wang}, Q., \& {Fan}, Z. 2018, \apj, 857,
  112, \dodoi{10.3847/1538-4357/aab900}

\bibitem[{{Yuan} {et~al.}(2019){Yuan}, {Pan}, {Liu}, {Wang}, \&
  {Fan}}]{Yuan2019}
{Yuan}, S., {Pan}, C., {Liu}, X., {Wang}, Q., \& {Fan}, Z. 2019, \apj, 884,
  164, \dodoi{10.3847/1538-4357/ab40a5}

\bibitem[{{Z{\"u}rcher} {et~al.}(2022){Z{\"u}rcher}, {Fluri}, {Sgier},
  {Kacprzak}, {Gatti}, {Doux}, {Whiteway}, {R{\'e}fr{\'e}gier}, {Chang},
  {Jeffrey}, {Jain}, {Lemos}, {Bacon}, {Alarcon}, {Amon}, {Bechtol}, {Becker},
  {Bernstein}, {Campos}, {Chen}, {Choi}, {Davis}, {Derose}, {Dodelson},
  {Elsner}, {Elvin-Poole}, {Everett}, {Ferte}, {Gruen}, {Harrison}, {Huterer},
  {Jarvis}, {Leget}, {Maccrann}, {Mccullough}, {Muir}, {Myles}, {Navarro
  Alsina}, {Pandey}, {Prat}, {Raveri}, {Rollins}, {Roodman}, {Sanchez},
  {Secco}, {Sheldon}, {Shin}, {Troxel}, {Tutusaus}, {Yin}, {Aguena}, {Allam},
  {Andrade-Oliveira}, {Annis}, {Bertin}, {Brooks}, {Burke}, {Carnero Rosell},
  {Carrasco Kind}, {Carretero}, {Castander}, {Cawthon}, {Conselice},
  {Costanzi}, {da Costa}, {da Silva Pereira}, {Davis}, {De Vicente}, {Desai},
  {Diehl}, {Dietrich}, {Doel}, {Eckert}, {Evrard}, {Ferrero}, {Flaugher},
  {Fosalba}, {Friedel}, {Frieman}, {Garcia-Bellido}, {Gaztanaga}, {Gerdes},
  {Giannantonio}, {Gruendl}, {Gschwend}, {Gutierrez}, {Hinton}, {Hollowood},
  {Honscheid}, {Hoyle}, {James}, {Kuehn}, {Kuropatkin}, {Lahav}, {Lidman},
  {Lima}, {Maia}, {Marshall}, {Melchior}, {Menanteau}, {Miquel}, {Morgan},
  {Palmese}, {Paz-Chinchon}, {Pieres}, {Plazas Malag{\'o}n}, {Reil}, {Rodriguez
  Monroy}, {Romer}, {Sanchez}, {Scarpine}, {Schubnell}, {Serrano}, {Sevilla},
  {Smith}, {Suchyta}, {Tarle}, {Thomas}, {To}, {Varga}, {Weller}, {Wilkinson},
  \& {DES Collaboration}}]{Zurcher2022}
{Z{\"u}rcher}, D., {Fluri}, J., {Sgier}, R., {et~al.} 2022, \mnras, 511, 2075,
  \dodoi{10.1093/mnras/stac078}

\end{thebibliography}
\bibliographystyle{aasjournal}



\end{document}